\documentclass[a4paper,11pt]{article}
\pdfoutput=1
\usepackage{geometry}
\usepackage{jheppub}
\usepackage[utf8]{inputenc}

\usepackage{dsfont}
\usepackage{simplewick}

\usepackage{mathtools,amsmath,amssymb}
\usepackage{graphicx}

\usepackage{lmodern}
\usepackage[T1]{fontenc}
\usepackage{microtype}

\usepackage{hyperref}

\usepackage{xspace}

\usepackage{array}

\usepackage{picture}
\usepackage{calc}

\usepackage{subfigure}

\usepackage{booktabs}

\usepackage{xspace}

\usepackage{placeins}

\DeclareMathOperator{\phaneq}{\phantom{{}=}}

\DeclareMathOperator{\tr}{tr}

\makeatletter
\renewcommand*\env@matrix[1][\arraystretch]{%
  \edef\arraystretch{#1}%
  \hskip -\arraycolsep
  \let\@ifnextchar\new@ifnextchar
  \array{*\c@MaxMatrixCols c}}
\makeatother

\makeatletter
\def\mmatrix{\@ifnextchar[{\@amwith}{\@mwithout}}
\def\@amwith[#1]{\@ifnextchar[{\@mwwith[#1]}{\@mwith[#1]}}
\def\@mwwith[#1][#2]#3{
\begingroup\setlength{\arraycolsep}{#2pt}
    \begin{pmatrix}[#1]
        #3
    \end{pmatrix}
\endgroup
}
\def\@mwith[#1]#2{
    \begin{pmatrix}[#1]
        #2
    \end{pmatrix}
}
\def\@mwithout#1{
    \begin{pmatrix}
        #1
    \end{pmatrix}
}
\makeatother

\makeatletter
\DeclareRobustCommand*{\bfseries}{%
  \not@math@alphabet\bfseries\mathbf
  \fontseries\bfdefault\selectfont
  \boldmath
}

\makeatother

\usepackage{lipsum}
\makeatletter
\if@titlepage
  \renewenvironment{abstract}{%
      \titlepage
      \null\vfil
      \@beginparpenalty\@lowpenalty
      \begin{center}%
        \bfseries \abstractname
        \@endparpenalty\@M
      \end{center}}%
     {\par\vfil\null\endtitlepage}
\else
  \renewenvironment{abstract}{%
      \if@twocolumn
        \section*{\abstractname}%
      \else
        \small
        \begin{center}%
          {\bfseries \abstractname\vspace{-.5em}\vspace{\z@}}%
        \end{center}%
        \quotation
      \fi}
      {\if@twocolumn\else\endquotation\fi}
\fi
\makeatother

\renewcommand{\digamma}{\Psi}

\xspaceaddexceptions{]\}}

\newcommand{\pcl}{\ensuremath{\phi^\text{cl}}\xspace}
\newcommand{\pt}{\ensuremath{\tilde \phi}\xspace}
\newcommand{\ad}{\ensuremath{\mathrm{ad}\xspace}}

\newcommand{\su}[1]{\ensuremath{\mathfrak{su}(#1)\xspace}}

\newcommand{\gym}{\ensuremath{g_{\mathrm{YM}}\xspace}}
\newcommand{\diff}{\ensuremath{\mathrm{d}\xspace}}
\newcommand{\arctanh}{\ensuremath{\mathrm{arctanh}\xspace}}
        
\newcommand{\measy}{\ensuremath{m_{\text{easy}}}}

\title{A Quantum Check of Non-Supersymmetric AdS/dCFT}
\author{Aleix Gimenez Grau, Charlotte Kristjansen, Matthias Volk and Matthias Wilhelm}

\begin{document}

\begingroup\parindent0pt
\begin{flushright}\footnotesize
\end{flushright}
\vspace*{4em}
\centering
\begingroup\LARGE
\bf
A Quantum Check of Non-Supersymmetric AdS/dCFT 
\par\endgroup
\vspace{2.5em}
\begingroup\large{\bf Aleix Gimenez Grau, Charlotte Kristjansen, Matthias Volk and Matthias Wilhelm}
\par\endgroup
\vspace{1em}
\begingroup\itshape
Niels Bohr Institute, Copenhagen University,\\
Blegdamsvej 17, 2100 Copenhagen \O{}, Denmark\\

\par\endgroup
\vspace{1em}
\begingroup\ttfamily
aleix.gimenez@desy.de,$^*$
kristjan@nbi.ku.dk, 
mvolk@nbi.ku.dk, 
matthias.wilhelm@nbi.ku.dk \\
\par\endgroup
\vspace{2.5em}
\endgroup

\begin{abstract}
\noindent
Via a challenging field-theory computation, we confirm a supergravity prediction for the non-supersymmetric
D3-D7 probe-brane system with probe geometry $AdS_4\times S^2\times S^2$, stabilized by fluxes. Supergravity predicts, in a 
certain double-scaling limit,  the value of the one-point functions of chiral primaries of the dual defect version of ${\cal N}=4$
SYM theory, where the fluxes translate into $SO(3)\times SO(3)$-symmetric, Lie-algebra-valued vacuum expectation values for all six scalar fields.  Using a generalization of the technique based on fuzzy spherical harmonics
developed for the related D3-D5 probe-brane system, we diagonalize the resulting mass matrix of the field theory. Subsequently,
we calculate the planar one-loop correction to the vacuum expectation values of the scalars in dimensional reduction and find 
that it is UV finite and non-vanishing. We then proceed to calculating the one-loop correction to the planar one-point function of 
any single-trace scalar operator and explicitly evaluate this correction for a 1/2-BPS operator of length $L$ at two leading orders
in the double-scaling limit, finding exact agreement with the supergravity prediction.

\noindent
\end{abstract}


\vfill
{\footnotesize ${}^*$Correspondence address after November 1, 2018. }

\thispagestyle{empty}

\newpage
\hrule
\setcounter{tocdepth}{2}
\tableofcontents
\afterTocSpace
\hrule
\afterTocRuleSpace

\section{Introduction and Summary}
\label{sec:introduction}

Introducing defects such as boundaries or interfaces in conformal field theories (CFTs) does not only make these theories more adapt to experimental situations in condensed matter systems but also constitutes a natural 
step in exploring the limits of applicability of modern approaches to  quantum field theory such as duality, integrability and the conformal bootstrap program, see e.g.\ \cite{Andrei:2018die}.
From the latter perspective, various defect versions of the four-dimensional maximally supersymmetric Yang-Mills (${\cal N}=4$ SYM) theory constitute particularly interesting arenas for investigation. 

An example of such a defect CFT is the field theory dual to the D3-D5 probe-brane setup with $k$ units of background gauge-field flux~\cite{Constable:1999ac,Karch:2000gx}, see \cite{deLeeuw:2017cop} for a review.  The presence of the flux translates into the rank of the gauge group of the defect field theory being different on the two sides of a codimension-one defect placed at $x_3=0$ and three of the scalar fields of 
${\cal N}=4$ SYM theory carrying vacuum expectation values (vevs) given by the generators of a $k$-dimensional irreducible representation of \su{2} for $x_3>0$.  This setup partly breaks conformal symmetry as well as supersymmetry. Conformal symmetry is reduced from $SO(4,2)$ to $SO(3,2)$ and the supersymmetry is reduced to three-dimensional $\mathcal{N}=4$~\cite{DeWolfe:2001pq,Erdmenger:2002ex}. The presence of the defect implies that operators can acquire non-vanishing
one-point functions of the form~\cite{Cardy:1984bb}
\begin{equation}
\langle {\cal O}_{\Delta}  \rangle (x)= \frac{C}{x_3^{\Delta}},
\end{equation}
with $\Delta$ denoting the conformal dimension, 
and due to the vevs this can happen already at tree level for certain scalar operators. 
 Using the language of integrability, it was possible to express in one compact formula the tree-level one-point functions of all bulk single-trace  scalar operators of the defect CFT~\cite{deLeeuw:2015hxa,Buhl-Mortensen:2015gfd,deLeeuw:2016umh,deLeeuw:2018mkd}.
 Furthermore, by a rather demanding field-theory calculation involving the diagonalization of the highly non-trivial mass matrix  using fuzzy spherical harmonics, it was possible to extend the compact formula for one-point functions to one-loop order in the $SU(2)$ sector of the theory~\cite{Buhl-Mortensen:2016pxs,Buhl-Mortensen:2016jqo,Buhl-Mortensen:2017ind}.
 What is more, the one-loop computation allowed for a comparison with a prediction originating from supergravity \cite{Nagasaki:2012re} and despite the partial breaking of both conformal and supersymmetry a perfect match was 
 found~\cite{Buhl-Mortensen:2016pxs,Buhl-Mortensen:2016jqo}.
 More precisely, the supergravity computation involved taking
the double-scaling limit~\cite{Nagasaki:2011ue}%
\footnote{This double-scaling limit is reminiscent of the Berenstein-Maldacena-Nastase limit \cite{Berenstein:2002jq}, which breaks down at four-loop order \cite{Bern:2006ew,Beisert:2006ez,Cachazo:2006az}. While the present double-scaling limit breaks down for non-protected operators already at one-loop order, it holds for protected operators such as $\tr Z^L$ to at least $(L-1)$-loop order \cite{Buhl-Mortensen:2017ind}.}
\begin{equation}
\lambda\rightarrow \infty,\hspace{0.5cm} k\rightarrow \infty, \hspace{0.5cm} \frac{\lambda}{k^2} \hspace{0.5cm}
\mbox{fixed},
\label{dsl}
\end{equation}
where  $\lambda$ is the 't~Hooft coupling, 
and performing a perturbative expansion in   $\lambda/k^2$. From the result of this computation, a prediction for the ratio of the one-loop and the tree-level value of the one-point function of the chiral primary $\tr Z^L$ in the double-scaling limit could be
inferred~\cite{Buhl-Mortensen:2016pxs}.
 
\begin{figure}
\centering
 \begin{minipage}[b]{0.48\textwidth}
 \def\svgwidth{\textwidth}
 %% Creator: Inkscape inkscape 0.92.2, www.inkscape.org
%% PDF/EPS/PS + LaTeX output extension by Johan Engelen, 2010
%% Accompanies image file 'brane_stack.pdf' (pdf, eps, ps)
%%
%% To include the image in your LaTeX document, write
%%   \input{<filename>.pdf_tex}
%%  instead of
%%   \includegraphics{<filename>.pdf}
%% To scale the image, write
%%   \def\svgwidth{<desired width>}
%%   \input{<filename>.pdf_tex}
%%  instead of
%%   \includegraphics[width=<desired width>]{<filename>.pdf}
%%
%% Images with a different path to the parent latex file can
%% be accessed with the `import' package (which may need to be
%% installed) using
%%   \usepackage{import}
%% in the preamble, and then including the image with
%%   \import{<path to file>}{<filename>.pdf_tex}
%% Alternatively, one can specify
%%   \graphicspath{{<path to file>/}}
%% 
%% For more information, please see info/svg-inkscape on CTAN:
%%   http://tug.ctan.org/tex-archive/info/svg-inkscape
%%
\begingroup%
  \makeatletter%
  \providecommand\color[2][]{%
    \errmessage{(Inkscape) Color is used for the text in Inkscape, but the package 'color.sty' is not loaded}%
    \renewcommand\color[2][]{}%
  }%
  \providecommand\transparent[1]{%
    \errmessage{(Inkscape) Transparency is used (non-zero) for the text in Inkscape, but the package 'transparent.sty' is not loaded}%
    \renewcommand\transparent[1]{}%
  }%
  \providecommand\rotatebox[2]{#2}%
  \newcommand*\fsize{\dimexpr\f@size pt\relax}%
  \newcommand*\lineheight[1]{\fontsize{\fsize}{#1\fsize}\selectfont}%
  \ifx\svgwidth\undefined%
    \setlength{\unitlength}{551.15285609bp}%
    \ifx\svgscale\undefined%
      \relax%
    \else%
      \setlength{\unitlength}{\unitlength * \real{\svgscale}}%
    \fi%
  \else%
    \setlength{\unitlength}{\svgwidth}%
  \fi%
  \global\let\svgwidth\undefined%
  \global\let\svgscale\undefined%
  \makeatother%
  \begin{picture}(1,0.82704315)%
    \lineheight{1}%
    \setlength\tabcolsep{0pt}%
    \put(0,0){\includegraphics[width=\unitlength,page=1]{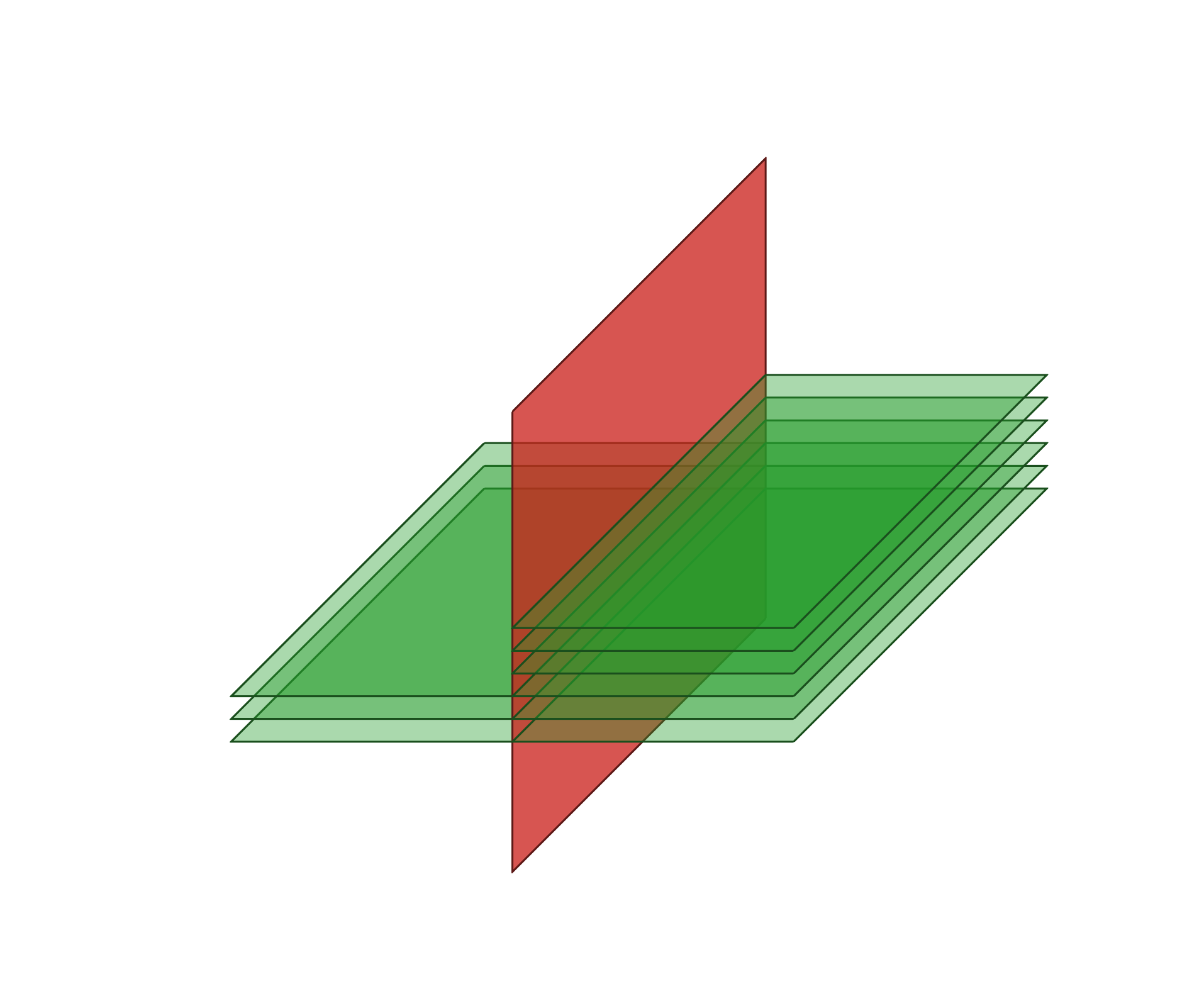}}%
    \put(0.28049988,0.03189463){\color[rgb]{0,0,0}\makebox(0,0)[lt]{\lineheight{0}\smash{\begin{tabular}[t]{l}D7 probe brane\end{tabular}}}}%
    \put(0.77990788,0.27547393){\color[rgb]{0,0,0}\makebox(0,0)[lt]{\lineheight{0}\smash{\begin{tabular}[t]{l}$N$ D3\end{tabular}}}}%
    \put(0.01605426,0.41527255){\color[rgb]{0,0,0}\makebox(0,0)[lt]{\lineheight{0}\smash{\begin{tabular}[t]{l}$N-k_1k_2$ D3\end{tabular}}}}%
    \put(0,0){\includegraphics[width=\unitlength,page=2]{brane_stack.pdf}}%
    \put(0.36410302,0.66495506){\color[rgb]{0,0,0}\makebox(0,0)[lt]{\lineheight{0}\smash{\begin{tabular}[t]{l}$x_3$\end{tabular}}}}%
    \put(0.23913,0.78992863){\color[rgb]{0,0,0}\makebox(0,0)[lt]{\lineheight{0}\smash{\begin{tabular}[t]{l}$x_4, x_5, x_6, x_7, x_8$\end{tabular}}}}%
    \put(0.15953827,0.53637571){\color[rgb]{0,0,0}\makebox(0,0)[lt]{\lineheight{0}\smash{\begin{tabular}[t]{l}$x_0, x_1, x_2$\end{tabular}}}}%
  \end{picture}%
\endgroup%

 \end{minipage}
 \begin{minipage}[b]{0.48\textwidth}
 \def\svgwidth{\textwidth}
 %% Creator: Inkscape inkscape 0.92.2, www.inkscape.org
%% PDF/EPS/PS + LaTeX output extension by Johan Engelen, 2010
%% Accompanies image file 'coordinate_system.pdf' (pdf, eps, ps)
%%
%% To include the image in your LaTeX document, write
%%   \input{<filename>.pdf_tex}
%%  instead of
%%   \includegraphics{<filename>.pdf}
%% To scale the image, write
%%   \def\svgwidth{<desired width>}
%%   \input{<filename>.pdf_tex}
%%  instead of
%%   \includegraphics[width=<desired width>]{<filename>.pdf}
%%
%% Images with a different path to the parent latex file can
%% be accessed with the `import' package (which may need to be
%% installed) using
%%   \usepackage{import}
%% in the preamble, and then including the image with
%%   \import{<path to file>}{<filename>.pdf_tex}
%% Alternatively, one can specify
%%   \graphicspath{{<path to file>/}}
%% 
%% For more information, please see info/svg-inkscape on CTAN:
%%   http://tug.ctan.org/tex-archive/info/svg-inkscape
%%
\begingroup%
  \makeatletter%
  \providecommand\color[2][]{%
    \errmessage{(Inkscape) Color is used for the text in Inkscape, but the package 'color.sty' is not loaded}%
    \renewcommand\color[2][]{}%
  }%
  \providecommand\transparent[1]{%
    \errmessage{(Inkscape) Transparency is used (non-zero) for the text in Inkscape, but the package 'transparent.sty' is not loaded}%
    \renewcommand\transparent[1]{}%
  }%
  \providecommand\rotatebox[2]{#2}%
  \newcommand*\fsize{\dimexpr\f@size pt\relax}%
  \newcommand*\lineheight[1]{\fontsize{\fsize}{#1\fsize}\selectfont}%
  \ifx\svgwidth\undefined%
    \setlength{\unitlength}{507.39454365bp}%
    \ifx\svgscale\undefined%
      \relax%
    \else%
      \setlength{\unitlength}{\unitlength * \real{\svgscale}}%
    \fi%
  \else%
    \setlength{\unitlength}{\svgwidth}%
  \fi%
  \global\let\svgwidth\undefined%
  \global\let\svgscale\undefined%
  \makeatother%
  \begin{picture}(1,0.80123619)%
    \lineheight{1}%
    \setlength\tabcolsep{0pt}%
    \put(0,0){\includegraphics[width=\unitlength,page=1]{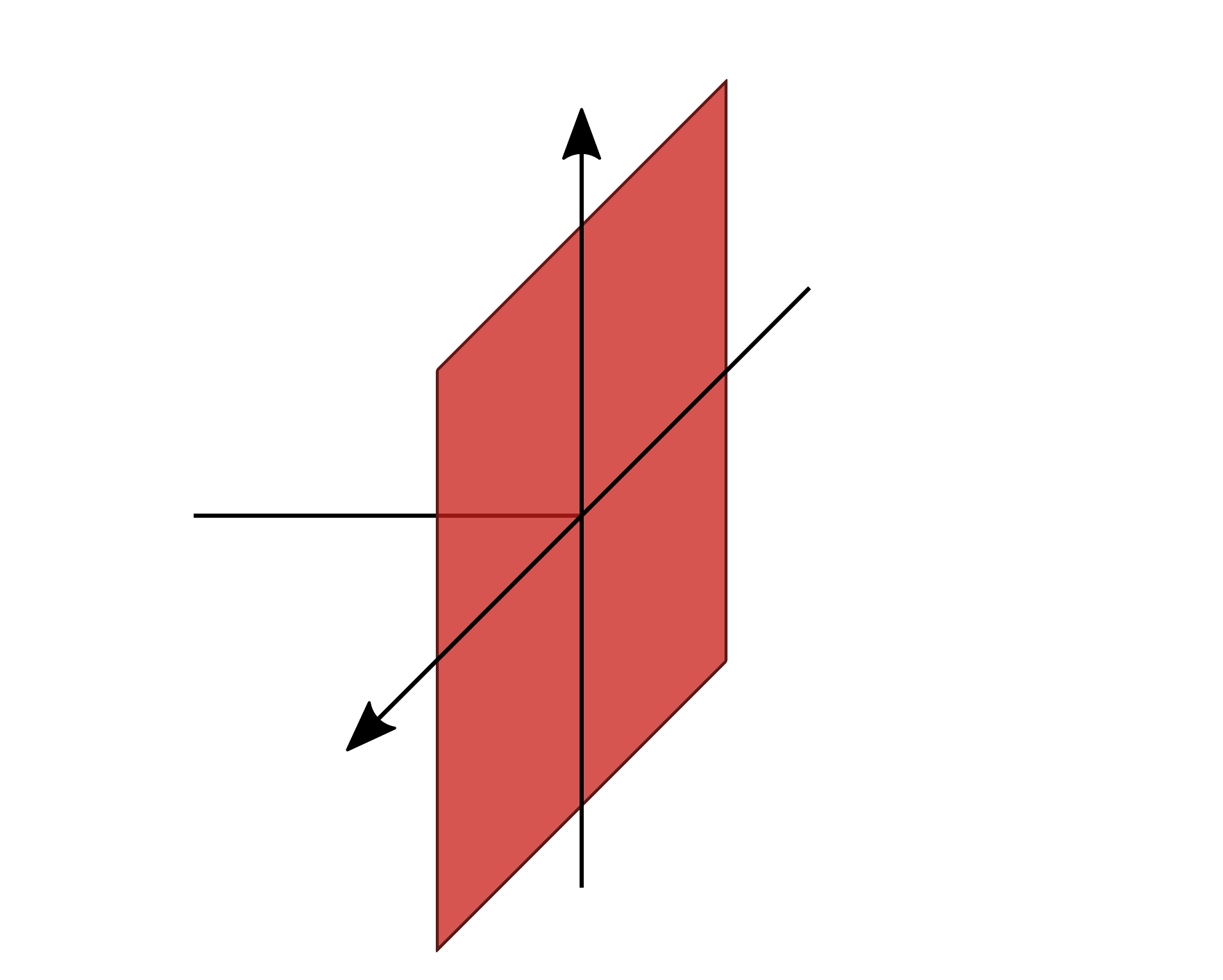}}%
    \put(0.75774808,0.41315383){\color[rgb]{0,0,0}\makebox(0,0)[lt]{\lineheight{0}\smash{\begin{tabular}[t]{l}$x_3$\end{tabular}}}}%
    \put(0.45923792,0.73396793){\color[rgb]{0,0,0}\makebox(0,0)[lt]{\lineheight{0}\smash{\begin{tabular}[t]{l}$x_0$\end{tabular}}}}%
    \put(0.13597535,0.19864059){\color[rgb]{0,0,0}\makebox(0,0)[lt]{\lineheight{0}\smash{\begin{tabular}[t]{l}$x_1, x_2$\end{tabular}}}}%
    \put(0,0){\includegraphics[width=\unitlength,page=2]{coordinate_system.pdf}}%
    \put(0.55185097,0.13859224){\color[rgb]{0,0,0}\makebox(0,0)[lt]{\lineheight{0}\smash{\begin{tabular}[t]{l}broken $U(N)$\end{tabular}}}}%
    \put(0.01743879,0.59337226){\color[rgb]{0,0,0}\makebox(0,0)[lt]{\lineheight{0}\smash{\begin{tabular}[t]{l}$U(N-k_1k_2)$\end{tabular}}}}%
  \end{picture}%
\endgroup%

 \end{minipage}
 \caption{Brane configuration in string theory (left) and the dual field-theory picture (right) with different gauge groups on each side of the defect at $x_3 = 0$. 
 }
 \label{fig:setup}
\end{figure}

A similar prediction can be extracted from a supergravity computation performed in a closely related but completely non-supersymmetric setup, namely that of a D3-D7 probe-brane system~\cite{Kristjansen:2012tn}.
The D3-D7 probe-brane system has two configurations which are of relevance for us, namely one where the geometry of the D7 brane is $AdS_4\times S^2 \times S^2$ and one where the geometry is $AdS_4\times S^4$.  
In both cases, the configuration has to be stabilized by
adding either fluxes $k_1$ and $k_2$ 
on the two $S^2$'s~\cite{Bergman:2010gm} 
or a non-trivial instanton bundle on the $S^4$~\cite{Myers:2008me}.  These flux-stabilized configurations
have interesting applications from the condensed matter perspective giving rise to strongly coupled Dirac fermions in 2+1 dimensions,
 see 
e.g.~\cite{Myers:2008me,Bergman:2010gm,Grignani:2012jh,Kristjansen:2012ny,Kristjansen:2013hma,Hutchinson:2014lda,
Jokela:2013hta,Jokela:2014wsa}.  The former configuration has a dual defect CFT where
all six scalar fields of ${\cal N}=4$ SYM theory are assigned vevs in the form of generators of the $(k_1\times k_2)$-dimensional irreducible representation
of $\su{2}\times \su{2}$ on one side of the defect; see Figure \ref{fig:setup}. 
In the latter case, only five out of the scalar fields are assigned vevs and these transform in an irreducible $SO(5)$ representation.  For both cases, it is possible to introduce a double-scaling parameter and to evaluate the one-point function as an expansion in this parameter~\cite{Kristjansen:2012tn}. Furthermore, in both cases the system is stable if the double-scaling parameter is sufficiently small. 
Reference~\cite{Kristjansen:2012tn} gives the leading order result of this evaluation and the higher orders can be extracted by a straightforward extension of this work. For the $AdS_4\times S^2\times S^2$ symmetric configuration, the double-scaling limit is introduced as follows~\cite{Kristjansen:2012tn}:
\begin{equation}
 \label{eq:double-scaling-limit}
\lambda\rightarrow \infty,\hspace{0.5cm} k_1,k_2\rightarrow \infty, \hspace{0.5cm} 
\frac{\lambda}{(k_1^2+k_2^2)} \hspace{0.5cm}
\mbox{fixed}.
\end{equation}
Keeping also the ratio $k_1 / k_2$ finite and assuming $(k_1 - k_2)$ to be of the same order as $k_1$ and $k_2$, the supergravity prediction for the one-point function of the unique $SO(3)\times SO(3)$-symmetric chiral primary  of
(even) length $L$ reads
\begin{align}
 \label{eq:trZL-string-theory}
 \begin{split}
 \frac{\langle \mathcal{O}_{L} \rangle }{ \langle \mathcal{O}_{L} \rangle_{\mathrm{tree}}}=
 1 + \frac{\lambda}{4 \pi^2 (k_1^2 + k_2^2)} \frac{1}{(L - 1) (k_1^2 + k_2^2)^2} \Bigg(
    4 (k_1 k_2)^2 + (L^3 + 3 L - 2) (k_1^4 + k_2^4)  \\
  +  2 (L - 1) (L + 2) k_1 k_2 (k_1^2 - k_2^2) \cot[(L + 2) \psi_0]
 \Bigg) + \mathcal{O}\Bigg( \frac{\lambda^2}{(k_1^2 + k_2^2)^2} \Bigg)
 ,
 \end{split}
\end{align}
where $\psi_0=\arctan (k_1/k_2)$. Notice that the prediction carries over to any other chiral primary with a non-trivial
projection on an $SO(3)\times SO(3)$-symmetric one, such as e.g.\ $\tr Z^L$. 
For the $AdS_4\times S^4$ configuration, supergravity also gives a prediction for the one-point function, however, with less structure as only one parameter is involved.
In the remainder of this paper, we shall demonstrate how the rather intricate prediction~(\ref{eq:trZL-string-theory}) can be reproduced via a solid field-theory calculation. The major challenge of the computation is the diagonalization of the
mass matrix of the theory, which requires a significant further development of the technique based on fuzzy spherical harmonics 
introduced in~\cite{Buhl-Mortensen:2016pxs,Buhl-Mortensen:2016jqo}.  The challenge is even bigger in the case of the
$SO(5)$-symmetric vevs. Our refined method works for that case as well but with considerably more effort.  We
plan to return to this case in a future publication~\cite{inprogress:2018}. With the present work, we do not only provide a detailed  positive test of AdS/dCFT in a situation where supersymmetry is completely broken; we also set up a perturbative framework which makes possible the evaluation of numerous other quantities in the defect CFT in question. 

Our paper is structured as follows. In Section \ref{sec:mass-matrix}, we diagonalize the highly non-trivial mass matrix that arises due to the vevs. In Section \ref{sec:propgagators}, we determine the resulting propagators of the mass eigenstates, which take the form of $AdS_4$ propagators, and subsequently the propagators of the fields occurring in the action.
Having thus set up the framework for calculating quantum corrections in this defect CFT, we calculate the first quantum correction to the classical solution in Section \ref{sec:one-loop-vevs}, which we find to be non-vanishing.
We proceed to calculate the one-loop correction to the one-point function of general single-trace operators, and in particular to $\tr Z^L$, in Section \ref{sec:one-loop-onepfun}.
In Section \ref{sec:conclusion}, we conclude with an outlook on possible future directions and interesting problems our perturbative framework can be applied to.
Several appendices contain our conventions (Appendix \ref{sec:conventions-general}) as well as details on technical parts of the calculations (Appendices \ref{sec:appendix-colour-flavour-part}--\ref{sec:color-traces-so3-so3}).

\section{Mass matrix}
\label{sec:mass-matrix}

In this section, we diagonalize the mass matrix that arises due to the scalar vevs.
Following the strategy of~\cite{Buhl-Mortensen:2016pxs,Buhl-Mortensen:2016jqo}, we begin by expanding the action around the classical solution in Section~\ref{sec:expansion-action}.
We then proceed to diagonalize the mass matrices for the bosons and fermions in Sections~\ref{sec:boson-mass-matrix} and~\ref{sec:fermion-mass-matrix}, respectively. We summarize the result in Section \ref{sec:summary-spectrum-complicated-bosons}.

\subsection{Expansion of the action}
\label{sec:expansion-action}

The defect CFT we study contains two types of fields: the ones of $\mathcal{N} = 4$ SYM theory transforming in the adjoint of the gauge group and the fundamental fields living on the three-dimensional defect.
However, the fields living on the defect will not contribute to the one-loop one-point functions of bulk%
\footnote{Note that `bulk' refers to four-dimensional Minkowski space without the defect; it should not be confused with the bulk of the dual $AdS_5$.}
operators as explained in~\cite{Buhl-Mortensen:2016jqo}, and we accordingly neglect the corresponding part of the action.
The action for the bulk fields is the one of standard $\mathcal N = 4$ SYM theory in four dimensions,
\begin{align}
 \label{eq:action-n4-sym}
  S_{\mathcal N = 4} = \frac{2}{\gym^2} \int \diff^4 x \, \tr \Bigg(
& - \frac{1}{4} F_{\mu \nu} F^{\mu \nu}
  - \frac{1}{2} D_{\mu} \phi_i D^{\mu} \phi_i
  + \frac{i}{2} \bar \psi \gamma^\mu D_\mu \psi \\
& + \frac{1}{4} [ \phi_i, \phi_j ] [ \phi_i, \phi_j ]
  + \frac{1}{2} \sum_{i = 1}^3 \bar \psi G^i [ \phi_i, \psi]
  + \frac{1}{2} \sum_{i = 4}^6 \bar \psi G^i [ \phi_i, \gamma_5 \psi]
 \Bigg).
\nonumber
\end{align}
We describe in Appendix~\ref{sec:conventions-general} our field-theory conventions, which follow the ones of \cite{Buhl-Mortensen:2016jqo}. In particular, we explicitly give the matrices $G^i$ ($i = 1, \ldots, 6$), which arise in the reduction from ten-\ to four-dimensional SYM theory. 
The $\psi_i$ for $i = 1, \ldots, 4$ are four-dimensional Majorana fermions, and all these fields transform in the adjoint of $U(N)$,
\begin{align}
 D_\mu \phi_i = \partial_\mu \phi_i - i [ A_\mu, \phi_i ], \quad
 D_\mu \psi_i = \partial_\mu \psi_i - i [ A_\mu, \psi_i ].
\end{align}
The classical equations of motion of~\eqref{eq:action-n4-sym} are
\begin{align}
 \label{eq:classical-eom-sym}
 \nabla^2 \pcl_i = \left[ \pcl_j, \left[ \pcl_j, \pcl_i \right] \right], \quad
 i = 1, \ldots, 6,
\end{align}
where we are setting the fermions and gauge fields to zero classically, and are looking for time-independent solutions for the scalars.
A solution to the equations of motion for the six scalar fields with $SO(3) \times SO(3)$ symmetry 
is~\cite{Kristjansen:2012tn}\footnote{The prefactor $\frac{1}{x_3}$ ensures scale invariance of the defect field theory 
and is important for
the dual probe-brane interpretation. 
A set-up where the classical fields were similar but not carrying the $\frac{1}{x_3}$ prefactor was studied in~\cite{Chatzistavrakidis:2009ix}, where in order to stabilize the system extra mass and interaction terms were added to the 
${\cal N}=4$ SYM action.}
\begin{align}
 \label{eq:classical-solution-so3-so3}
 \begin{split}
 \pcl_i(x) &= - \frac{1}{x_3} \left( t_i^{k_1} \otimes \mathds{1}_{k_2} \right) \oplus 0_{N-k_1 k_2} \quad \text{for} \quad i = 1, 2, 3, \\
 \pcl_i(x) &= - \frac{1}{x_3} \left( \mathds{1}_{k_1} \otimes t_{i-3}^{k_2} \right) \oplus 0_{N-k_1 k_2} \quad \text{for} \quad i = 4, 5, 6.
 \end{split}
\end{align}
Here the matrices $t_i^{k_a}$ constitute the $k_a$-dimensional irreducible representation of \su{2}; thus, the solution has $\su{2} \times \su{2}$ symmetry.
In the case $k_1=1$ or $k_2=1$, the vevs \eqref{eq:classical-solution-so3-so3} reduce to the ones in the supersymmetric D3-D5 setup \cite{Buhl-Mortensen:2016jqo}; hence, we will always assume $k_1,k_2\geq2$.
The classical solution \eqref{eq:classical-solution-so3-so3} applies for $x_3 > 0$ and is responsible for breaking the gauge group from $U(N)$ to $U(N - k_1 k_2)$ for $x_3 > 0$.
All other fields vanish classically in this region.
For $x_3 < 0$, all fields have gauge group $U(N - k_1 k_2)$ and the vevs for these fields vanish.

We expand the action around the classical solution as
\begin{align}
 \label{eq:expansion-fields}
 \phi_i(x) = \pcl_i(x) + \pt_i(x).
\end{align}
The gauge fixing is implemented by introducing fermionic ghost fields $c$ and $\bar{c}$ transforming as Lorentz scalars, following \cite{Alday:2009zm,Buhl-Mortensen:2016jqo}.
The terms in the expanded action that are linear in $\pt_i$ vanish by the classical equations of motion.
All fields have a canonically normalized (quadratic) kinetic term,
\begin{align}
 \label{eq:kinetic-term-N4}
 S_{\mathrm{kin}} =
 \frac{2}{\gym^2} \int \mathrm{d}^4 x \; \tr \Bigg(
 \frac{1}{2} A_\mu \partial_\nu \partial^\mu A^\nu +
 \frac{1}{2} \pt_i \partial_\nu \partial^\nu \pt_i +
 \frac{i}{2} \bar{\psi} \gamma^\mu \partial_\mu \psi +
 \bar{c} \partial_\mu \partial^\mu c
 \Bigg).
\end{align}
The mass term for the bosons becomes
\begin{align}
 \label{eq:mass-term-bosons}
 \begin{split}
  S_{\text{m,b}} =
  \frac{2}{\gym^2} \int \mathrm{d}^4 x \; \tr \Bigg(
  & - \frac{1}{2} \pt_j [ \pcl_i, [ \pcl_i, \pt_j ] ]
    - \pt_i [ [ \pcl_i, \pcl_j ] , \pt_j ] \\
  & - \frac{1}{2} A_\mu [ \pcl_i, [\pcl_i, A^\mu ] ]
    + 2 i [ A^\mu, \pt_i ] \partial_\mu \pcl_i
  \Bigg),
 \end{split}
\end{align}
while the mass term for the four Majorana fermions $\psi_i$ and the ghosts $c$ and $\bar{c}$ is
\begin{align}
 \label{eq:mass-term-fermions}
 S_{\text{m,f}} =
 \frac{2}{\gym^2} \int \mathrm{d}^4 x \; \tr \Bigg(
  \frac{1}{2} \sum_{i=1}^3 \bar{\psi} G^i [ \pcl_i, \psi ] +
  \frac{1}{2} \sum_{i=4}^6 \bar{\psi} G^i [ \pcl_i, \gamma_5 \psi ] -
  \sum_{i=1}^6 \bar{c} [ \pcl_i, [ \pcl_i, c ]]
 \Bigg).
\end{align}
The expanded action also contains cubic and quartic interaction vertices between the different fields.
The cubic interactions are given by
\begin{align}
 \label{eq:action-cubic-vertices}
 \begin{aligned}
 S_{\text{cubic}} 
  = \frac{2}{\gym^2} \int \mathrm{d}^4 x \; \tr \Bigg(
    i [ A^\mu, A^\nu ] \partial_\mu A_\nu  +
      [ \pcl_i, \pt_j ] [ \pt_i, \pt_j ]   +
    i [ A^\mu, \pt_i ] \partial_\mu \pt_i  +
      [ A_\mu, \pcl_i ] [A^\mu, \pt_i ] \\ +
    \frac{1}{2} \bar{\psi} \gamma^\mu [ A_\mu, \psi ]  +
    \frac{1}{2} \sum_{i=1}^3 \bar{\psi} G^i [ \pt_i, \psi ] +
    \frac{1}{2} \sum_{i=4}^6 \bar{\psi} G^i [ \pt_i, \gamma_5 \psi ] +
    i (\partial_\mu \bar c) [ A_\mu, c ] -
    \bar{c} [ \pcl_i, [ \pt_i, c ]]
  \Bigg).
 \end{aligned}
\end{align}
The quartic interaction vertices are identical to the quartic vertices present in the action~\eqref{eq:action-n4-sym}.
They do not play a role for the one-loop correction to the one-point functions of bulk operators, 
starting to contribute only at two-loop order \cite{Buhl-Mortensen:2016jqo}.

The mass terms \eqref{eq:mass-term-bosons} and~\eqref{eq:mass-term-fermions} are not diagonal, neither in flavor nor in color, and have to be diagonalized in order to obtain the mass spectrum of the theory and thus the propagators.
Moreover, note that unlike actual mass terms, the terms \eqref{eq:mass-term-bosons} and~\eqref{eq:mass-term-fermions} depend on the inverse distance to the defect via the vevs \eqref{eq:classical-solution-so3-so3}. This dependence can be understood in terms of an effective $AdS_4$ space, as was found in \cite{Nagasaki:2011ue,Buhl-Mortensen:2016jqo} and is discussed in detail in Section \ref{sec:propgagators}.

In the remainder of the paper, we will use Euclidean signature.

\subsection{Boson mass matrix}
\label{sec:boson-mass-matrix}

In this section, we will treat the mass term for the bosons, while the mass term for the fermions will be treated in Section~\ref{sec:fermion-mass-matrix}.

Inserting the classical solution~\eqref{eq:classical-solution-so3-so3} into the mass term~\eqref{eq:mass-term-bosons} for the bosons, the latter can be written as
\begin{align}
 \label{eq:mass-term-bosons-inserted}
 \begin{split}
  S_{\text{m,b}} =
  \frac{2}{\gym^2} \int \mathrm{d}^4 x
  &
  \frac{1}{x_3^2} \tr \Bigg(
  -\frac{1}{2} \sum_{j=1}^6 \pt_j \left[ (L^{(1)})^2 + (L^{(2)})^2 \right] \pt_j
  -\frac{1}{2} A_\mu \left[ (L^{(1)})^2 + (L^{(2)})^2 \right] A^\mu
  \\
  &
  + i \sum_{i,j,k = 1}^3 \epsilon_{ijk} \pt_i L^{(1)}_j \pt_k
  + i \sum_{i,j,k = 1}^3 \epsilon_{ijk} \pt_{i+3} L^{(2)}_j \pt_{k+3}
  \\
  &
  + i \sum_{i=1}^3 \left[ \pt_i L^{(1)}_i A_3 - A_3 L^{(1)}_i \pt_i \right]
  + i \sum_{i=1}^3 \left[ \pt_{i+3} L^{(2)}_i A_3 - A_3 L^{(2)}_i \pt_{i+3} \right]
  \Bigg).
 \end{split}
\end{align}
The operators $L^{(1)}_i$ and $L^{(2)}_i$ for $i = 1, 2, 3$ are defined as the adjoint of the classical solution,
\begin{align}
 L^{(1)}_i
 \equiv \ad \left[ 
  \left( t^{k_1}_i \otimes \mathds{1}_{k_2} \right) 
  \oplus 0_{N - k_1 k_2} 
 \right], \quad
 L^{(2)}_i
 \equiv \ad \left[ 
  \left( \mathds{1}_{k_1} \otimes t^{k_2}_i \right) 
  \oplus 0_{N - k_1 k_2}
 \right],
\end{align}
where as usual $\left( \ad A \right) B \equiv [A, B]$. They satisfy the commutation relations of $\su{2} \times \su{2}$,
\begin{align}
 \left[L^{(1)}_i, L^{(1)}_j\right] = i \epsilon_{ijk} L^{(1)}_k, \quad
 \left[L^{(2)}_i, L^{(2)}_j\right] = i \epsilon_{ijk} L^{(2)}_k, \quad
 \left[L^{(1)}_i, L^{(2)}_j\right] = 0.
\end{align}
Furthermore, we write $(L^{(a)})^2 \equiv \sum_i (L^{(a)}_i)^2$ for the quadratic Casimirs corresponding to the two sectors with $a = 1, 2$.
We will use their eigenvalues $\ell_1 (\ell_1 + 1)$ and $\ell_2 (\ell_2 + 1)$ to label irreducible representations of $\su{2} \times \su{2}$ by $(\ell_1, \ell_2)$.
As in~\cite{Buhl-Mortensen:2016jqo}, we find that we can distinguish two types of bosons:
if their mass term is already diagonal in flavor the fields are called ``easy'' bosons, while the ones for which flavor and color mix are called ``complicated''.

We rewrite \eqref{eq:mass-term-bosons-inserted} as
\begin{align}
 \label{eq:mass-term-bosons-matrix-form}
  S_{\text{m,b}} =
  \frac{2}{\gym^2} \int \mathrm{d}^4 x
  \left( \frac{-1}{2 \, x_3^2} \right)
  \tr \Bigg(
  & E^\dagger \left[ (L^{(1)})^2 + (L^{(2)})^2 \right] E \\
  & + \tilde C^\dagger \left[ 
      (L^{(1)})^2 + (L^{(2)})^2 - 2 \tilde S^{(1)}_i L^{(1)}_i - 2 \tilde S^{(2)}_i L^{(2)}_i 
   \right] \tilde C
  \Bigg),
\nonumber
\end{align}
where we have grouped the fields into vectors of easy and complicated fields $E$ and $\tilde C$ respectively,
\begin{align}
 E =
 \begin{pmatrix}
  A_0 \\ A_1 \\ A_2
 \end{pmatrix},
 \quad
 \tilde C =
 \begin{pmatrix}
  \pt_1 \\ \vdots \\ \pt_6 \\ A_3
 \end{pmatrix}.
\end{align}
The seven-dimensional matrices $\tilde{S}^{(1)}_i$ and $\tilde{S}^{(2)}_i$ act on the flavor index while the operators $L^{(1)}_i$ and $L^{(2)}_i$ act on the color part of the quantum fields.
We see from \eqref{eq:mass-term-bosons-matrix-form} that for the easy fields we only need to diagonalize the operator $(L^{(1)})^2 + (L^{(2)})^2$ in color space.
The mass term for the complicated fields mixes different flavors by means of the matrices $\tilde{S}^{(1)}_i$ and $\tilde{S}^{(2)}_i$ and we will have to diagonalize the color and flavor part simultaneously.
Note that compared to the solution where only three scalar fields get non-trivial $SO(3)$-symmetric vevs
 studied in~\cite{Buhl-Mortensen:2016jqo}, all scalars $\pt_i$ are now complicated bosons and only the three components of the gauge fields $A_0, A_1, A_2$ and the ghost field remain easy.
We will denote the eigenvalues of the matrices inside the trace in~\eqref{eq:mass-term-bosons-matrix-form} by $m^2$.

\subsubsection{Decomposition of the color matrices and easy fields}
\label{sec:decomposition-colour-matrices}

In order to proceed with the diagonalization, we decompose the color part of a generic field $\Phi$ in blocks:
\begin{align}
 \label{eq:field-decomposition}
 \Phi = [\Phi]_{n,n'} {E^n}_{n'} + [\Phi]_{n,a} {E^n}_{a} + [\Phi]_{a,n} {E^a}_{n} + [\Phi]_{a,a'} {E^a}_{a'},
\end{align}
with $n, n' = 1, \ldots, k_1 k_2$ and $a, a' = k_1 k_2 + 1, \ldots, N$.
Here ${E^n}_{n'}$ are $N \times N$ matrices with a single non-vanishing entry, namely a $1$ at position $(n, n')$.
The fields $[ \Phi ]_{n,a}$ and $[ \Phi ]_{a,n}$ will often be referred to as fields in the off-diagonal block.

The fields $[ \Phi ]_{a,a'}$ in the $(N - k_1 k_2) \times (N - k_1 k_2)$ block are massless since 
\begin{align}
 L^{(1)}_i \, {E^a}_{a'} = \left[ \left( t_i^{k_1} \otimes \mathds{1}_{k_2} \right) \oplus 0_{N - k_1 k_2}, E^a_{\phantom{a}a'} \right] = 0,
\end{align}
and similarly for $L^{(2)}_i$.
One can think of this result as the statement that the indices $a$ and $a'$ are singlets under $\su{2} \times \su{2}$.

The matrices ${E^n}_{a}$ and ${E^a}_{n}$ transform in the $(k_1 \times k_2$)-dimensional irreducible representation of $\su{2} \times \su{2}$,
\begin{align}
\begin{split}
 L^{(1)}_i \, {E^n}_{a} &= {E^{n'}}_{a} [t_i^{k_1} \otimes \mathds{1}_{k_2}]_{n',n}, \quad
 L^{(1)}_i \, {E^a}_{n} = - [t_i^{k_1} \otimes \mathds{1}_{k_2}]_{n,n'} {E^{a}}_{n'}, \\
 L^{(2)}_i \, {E^n}_{a} &= {E^{n'}}_{a} [\mathds{1}_{k_1} \otimes t_i^{k_2}]_{n',n}, \quad
 L^{(2)}_i \, {E^a}_{n} = - [\mathds{1}_{k_1} \otimes t_i^{k_2}]_{n,n'} {E^{a}}_{n'}.
\end{split}
\end{align}
Equivalently, each index $n$ transforms in the same representation as $t_i$, namely the one with spins $\ell_1 = \frac{k_1-1}{2}$ and $\ell_2 = \frac{k_2 - 1}{2}$.
It follows that the matrices ${E^n}_{a}$ and ${E^a}_{n}$ already diagonalize the quadratic Casimir operators,
\begin{equation}
 (L^{(1)})^2 \, {E^n}_{a} = \frac{k_1^2-1}{4}{E^{n}}_{a},\qquad (L^{(1)})^2 \, {E^a}_{n}= \frac{k_1^2-1}{4}{E^{a}}_{n},
\end{equation}
and analogously for $(L^{(2)})^2$.
The matrices ${E^n}_{a}$ and ${E^a}_{n}$ transform into each other under Hermitian conjugation, and this behavior carries over to the fields $[ \Phi ]_{n,a}$ and $[ \Phi ]_{a,n}$ in the off-diagonal block:
\begin{align}
 \left( E^{n}_{\phantom{n}a} \right)^\dag = E^{a}_{\phantom{a}n}, \quad
 \left[ \Phi \right]_{n,a}^\dag 
 \equiv \left( \left[ \Phi \right]_{n,a} \right)^\dag 
 = \left[ \Phi \right]_{a,n}.
\end{align}
Moreover, they are orthogonal and normalized in the sense that
\begin{equation}
 \label{eq:states-orthogonality-so3_offdiagonal}
\begin{aligned}
 &\tr\left[ \left( E^{n}_{\phantom{n}a} \right)^\dag E^{n'}_{\phantom{n}a'}\right]=\delta^{n n'}\delta_{a a'},\quad
 \tr \left[\left( E^{a}_{\phantom{a}n} \right)^\dag E^{n'}_{\phantom{n}a'}\right]=0,\\
 &\tr \left[\left( E^{a}_{\phantom{a}n} \right)^\dag E^{a'}_{\phantom{a}n'}\right]=\delta^{a a'}\delta_{n n'},\quad
 \tr \left[\left( E^{n}_{\phantom{n}a} \right)^\dag E^{a'}_{\phantom{a}n'}\right]=0.
\end{aligned}
\end{equation}
For easy fields $[ \Phi ]_{n,a}$ and $[ \Phi ]_{a,n}$, for which $(L^{(1)})^2+(L^{(2)})^2$ is the complete mass term, we thus find the masses 
\begin{align}
\label{eq:easy-masses-off-diagonal}
 \measy^2 \equiv \frac{k_1^2-1}{4}+\frac{k_2^2-1}{4},
\end{align}
which have multiplicity $2 k_1 k_2 (N - k_1 k_2)$.

Finally, the matrices ${E^n}_{n'}$ contain two $n$ indices, and therefore they transform as the product of two $(k_1 \times k_2)$-dimensional irreducible representations of $\su{2} \times \su{2}$.
This product is reducible and decomposes as
\begin{align}
 \label{eq:decomposition-adjoint-su2}
 \left( \frac{k_1 - 1}{2}, \frac{k_2 - 1}{2} \right) \otimes \left( \frac{k_1 - 1}{2}, \frac{k_2 - 1}{2} \right)
 = \bigoplus_{\ell_1 = 0}^{k_1 - 1} \bigoplus_{\ell_2 = 0}^{k_2 - 1} (\ell_1, \ell_2),
\end{align}
where $(\ell_1, \ell_2)$ is the $\su{2} \times \su{2}$ representation with spins $\ell_1$ and $\ell_2$ and dimension $(2 \ell_1 + 1) \times (2 \ell_2 + 1)$.
Note that the fields $[\Phi]_{n,a}$ and $[\Phi]_{a,n}$ in the off-diagonal block have spins $\ell_1 = \tfrac{k_1 - 1}{2}$ and $\ell_2 = \tfrac{k_2 - 1}{2}$, which appears as one of the terms in the decomposition~\eqref{eq:decomposition-adjoint-su2}.
Thus, any results for the masses in the off-diagonal blocks can be obtained from the result in the $k_1 k_2 \times k_1 k_2$ block by the simple replacement rule
\begin{align}
 \label{eq:replacement-off-diagonal-block-so3-so3}
 \ell_1 \rightarrow \frac{k_1 - 1}{2} \quad \text{and} \quad \ell_2 \rightarrow \frac{k_2 - 1}{2}.
\end{align}
This justifies that in the following we will mostly focus on the $k_1 k_2 \times k_1 k_2$ block.

In the case of the field theory where only three of the scalars get non-trivial $SO(3)$-symmetric vevs, dual to the D3-D5 probe-brane setup, the mass term for the easy bosons is $L^2$.
In~\cite{Buhl-Mortensen:2016jqo}, it was found that the diagonalization in the corresponding $k\times k$ block could be solved by expressing the fields in a basis of fuzzy spherical harmonics $\hat{Y}^{m}_{\ell}$ constituting an irreducible spin-$\ell$ representation of \su{2}.
In the present case, the mass term for the easy bosons contains the operator $(L^{(1)})^2 + (L^{(2)})^2$, and since $(L^{(1)})^2$ and $(L^{(2)})^2$ commute with each other, we can diagonalize them simultaneously. 
The eigenstates of $(L^{(1)})^2 + (L^{(2)})^2$ are therefore the tensor products $\hat{Y}^{m_1}_{\ell_1} \otimes \hat{Y}^{m_2}_{\ell_2}$ of two fuzzy spherical harmonics.
We use this basis to express the fields in the $k_1 k_2 \times k_1 k_2$ block as
\begin{align}
\label{eq:fuzzy-expansion}
 \sum_{n,n' = 1}^{k_1 k_2} [\Phi]_{n,n'} {E^n}_{n'} 
 = \sum_{\ell_1 = 0}^{k_1 - 1} \sum_{\ell_2 = 0}^{k_2 - 1} \sum_{m_1 = - \ell_1}^{\ell_1} \sum_{m_2 = - \ell_2}^{\ell_2}
 \Phi_{\ell_1,m_1; \ell_2, m_2} \hat Y_{\ell_1}^{m_1} \otimes \hat Y_{\ell_2}^{m_2}.
\end{align}

The properties of the basis states $\hat{Y}^{m_1}_{\ell_1} \otimes \hat{Y}^{m_2}_{\ell_2}$ follow from the properties of the fuzzy spherical harmonics $\hat Y_\ell^m$, which are reviewed in Appendix~\ref{sec:conventions-su2-and-Ylm}.
An important property is the behavior under Hermitian conjugation, which carries over to the field components $\Phi_{\ell_1, m_1; \ell_2, m_2}$:
\begin{align}
\begin{split}
 \left( \hat{Y}^{m_1}_{\ell_1} \otimes \hat{Y}^{m_2}_{\ell_2} \right)^\dagger = (-1)^{m_1} (-1)^{m_2} \hat{Y}^{-m_1}_{\ell_1} \otimes \hat{Y}^{-m_2}_{\ell_2}, \\
 \left( \Phi_{\ell_1,m_1; \ell_2,m_2} \right)^\dagger = (-1)^{m_1} (-1)^{m_2} \Phi_{\ell_1,-m_1; \ell_2, -m_2}.
\end{split}
\end{align}
The operators $L^{(1)}_i$ and $L^{(2)}_i$ act on the basis states as
\begin{align}
 \label{eq:ladder-operators-action-Ylm}
 \begin{split}
  (L^{(1)})^2   \; \hat{Y}^{m_1}_{\ell_1} \otimes \hat{Y}^{m_2}_{\ell_2}
  &= \ell_1 (\ell_1 + 1) \; \hat{Y}^{m_1}_{\ell_1} \otimes \hat{Y}^{m_2}_{\ell_2}, \\
  L^{(1)}_3     \; \hat{Y}^{m_1}_{\ell_1} \otimes \hat{Y}^{m_2}_{\ell_2}
  &= \sqrt{\ell_1 (\ell_1 + 1)} \langle \ell_1, m_1; 1, 0 | \ell_1, m_1 \rangle \; \hat{Y}^{m_1}_{\ell_1} \otimes \hat{Y}^{m_2}_{\ell_2}, \\
  L^{(1)}_{\pm} \; \hat{Y}^{m_1}_{\ell_1} \otimes \hat{Y}^{m_2}_{\ell_2}
  &= \mp \sqrt{ 2 \ell_1 (\ell_1 + 1) } \langle \ell_1, m_1; 1, \pm 1 | \ell_1, m_1 \pm 1 \rangle \; \hat{Y}^{m_1 \pm 1}_{\ell_1} \otimes \hat{Y}^{m_2}_{\ell_2},
 \end{split}
\end{align}
with the ladder operators $L^{(1)}_{\pm} = L^{(1)}_1 \pm i L^{(1)}_2$ and analogous expressions for $(L^{(2)})^2$, $L^{(2)}_3$ and $L^{(2)}_{\pm}$.
Here and in the following, $\langle \ell, m_{\ell}; s, m_s | j, m_j \rangle$ denotes the \su{2} Clebsch-Gordan coefficient for coupling the two angular momenta $\ell$ and $s$ to the total angular momentum $j$.
For the case $s = 1$ and $j = \ell$ in~\eqref{eq:ladder-operators-action-Ylm}, they are
\begin{align}
 \label{eq:clebsch-gordan-j-l}
 \langle \ell, m; 1, \pm 1 | \ell, m \pm 1 \rangle &= \mp \frac{\sqrt{ \ell (\ell + 1) - m (m \pm 1) }}{\sqrt{2 \ell (\ell + 1)}}, \;
 \langle \ell, m; 1, 0 | \ell, m \rangle = \frac{m}{\sqrt{ \ell (\ell + 1) }}.
\end{align}
Furthermore, the basis states are orthogonal and normalized such that
\begin{align}
 \label{eq:states-orthogonality-so3}
  \tr \left[ \left(\hat{Y}^{m_1'}_{\ell_1'} \otimes \hat{Y}^{m_2'}_{\ell_2'} \right)^\dagger \hat{Y}^{m_1}_{\ell_1} \otimes \hat{Y}^{m_2}_{\ell_2} \right]
  = \delta_{\ell_1', \ell_1} \; \delta_{\ell_2', \ell_2} \; \delta_{m_1, m_1'} \; \delta_{m_2, m_2'}.
\end{align}

Using this basis, we see that the mass eigenvalues of the fields $\Phi_{\ell_1, m_1; \ell_2, m_2}$ are
\begin{align}
\label{eq:easy-masses-diagonal}
& \measy^2 \equiv \ell_1 (\ell_1 + 1) + \ell_2 (\ell_2 + 1),
\end{align}
where we must take all combinations of $\ell_1 = 0, \ldots, k_1 - 1$ and
$\ell_2 = 0, \ldots, k_2 - 1$. 
The multiplicity is the dimension of the corresponding $\su{2} \times \su{2}$ representation, i.e.\ ${(2 \ell_1 + 1) (2 \ell_2 + 1)}$.
As discussed before, the masses of the fields in the $(N-k_1 k_2)\times(N - k_1 k_2)$ block are zero.
Finally, the masses \eqref{eq:easy-masses-off-diagonal} in the $k_1 k_2 \times(N - k_1 k_2)$ and the $(N - k_1 k_2) \times k_1 k_2$ blocks are indeed obtained from \eqref{eq:easy-masses-diagonal} by the replacement rule~\eqref{eq:replacement-off-diagonal-block-so3-so3}.
We summarize the masses of the easy fields in Table~\ref{tab:masses-easy-so3-so3}.

\renewcommand*{\arraystretch}{1.5}
\begin{table}
  \centering
  \begin{tabular}{c c}
    \toprule
    $m^2$ & Multiplicity \\
    \midrule
    $\ell_1 ( \ell_1 + 1 ) + \ell_2 ( \ell_2 + 1 )$  & $(2 \ell_1 + 1) (2 \ell_2 + 1)$  \\
    $( k_1^2 - 1 )/{4} + ( k_2^2 - 1 )/{4}$          & $2 k_1 k_2 (N - k_1 k_2)$        \\
    0                                                & $(N - k_1 k_2) (N - k_1 k_2)$    \\
    \bottomrule
  \end{tabular}
  \caption{Masses for the easy bosons $A_0$, $A_1$ and $A_2$ (as well as the ghosts $c$), including the $k_1 k_2 \times k_1 k_2$, the $k_1 k_2 \times (N - k_1 k_2)$ and the $(N-k_1 k_2) \times (N - k_1 k_2)$ blocks. Here $\ell_1 = 0, \ldots, k_1 - 1$ and $\ell_2 = 0, \ldots, k_2 - 1$.}
  \label{tab:masses-easy-so3-so3}
\end{table}
\renewcommand*{\arraystretch}{1.0}

\subsubsection{Complicated fields}
\label{sec:complicated-fields-so3-so3}

For the complicated fields the decomposition in terms of $\su{2} \times \su{2}$ representations is not sufficient, because we also need to solve the problem of flavor mixing.
Since $(L^{(1)})^2 + (L^{(2)})^2$ commutes with $\tilde{S} \cdot L \equiv \tilde{S}^{(1)}_i L^{(1)}_i + \tilde{S}^{(2)}_i L^{(2)}_i$ we can diagonalize the two terms in~\eqref{eq:mass-term-bosons-matrix-form} simultaneously.
Thus the masses will have the form $\ell_1 (\ell_1 + 1) + \ell_2 (\ell_2 + 1) - 2 \lambda$, where $\lambda$ are the eigenvalues of the mixing matrix $\tilde{S}\cdot L$.

\paragraph{Rewriting the matrices \texorpdfstring{$\tilde S_i$}{S\_i}}
The seven-dimensional matrices $\tilde{S}_i$ are given in block form by
\begin{align}
 \label{eq:smatrices-7by7}
\tilde S_i\equiv \tilde S^{(1)}_i =
 \begin{pmatrix}
  \tilde{T_i} & 0 & \tilde{R}_i \\
  0 & 0 & 0 \\
  \tilde{R}_i^\dagger & 0 & 0
 \end{pmatrix},
 \quad
 \tilde S_{i+3}\equiv\tilde S^{(2)}_{i} =
 \begin{pmatrix}
  0 & 0 & 0 \\
  0 & \tilde{T}_i & \tilde{R}_i \\
  0 & \tilde{R}_i^\dagger & 0
 \end{pmatrix},
 \quad 
 i = 1,2,3.
\end{align}
In the previous equation, $\tilde R_j$ is a $3 \times 1$ matrix that has an $i$ in the $j$-th component and zeros everywhere else, namely $(\tilde R_j)_k = i \, \delta_{jk}$. On the other hand, the three-dimensional matrices $\tilde T_i$ are given by
\begin{align}
 \tilde T_1 =
 \begin{pmatrix}
  0 & 0 & 0  \\
  0 & 0 & -i \\
  0 & i & 0  \\
 \end{pmatrix}, \quad
 \tilde T_2 =
 \begin{pmatrix}
  0 & 0 & i  \\
  0 & 0 & 0  \\
  -i & 0 & 0 \\
 \end{pmatrix}, \quad
 \tilde T_3 =
 \begin{pmatrix}
  0 & -i & 0 \\
  i & 0 & 0 \\
  0 & 0 & 0  \\
 \end{pmatrix}.
\end{align}
These matrices form an irreducible representation of the \su{2} Lie algebra, so they can be brought into the usual form for the spin-one representation
\begin{align}
 \label{eq:t-su2-spin-one}
 T_1 =
 \frac{1}{\sqrt{2}}
 \begin{pmatrix}
  0 & 1 & 0  \\
  1 & 0 & 1 \\
  0 & 1 & 0  \\
 \end{pmatrix}, \quad
 T_2 =
 \frac{1}{\sqrt{2}}
 \begin{pmatrix}
  0 & -i & 0  \\
  i &  0 & -i \\
  0 &  i & 0  \\
 \end{pmatrix}, \quad
 T_3 =
 \begin{pmatrix}
  1 & 0 & 0  \\
  0 & 0 & 0  \\
  0 & 0 & -1 \\
 \end{pmatrix},
\end{align}
using the unitary transformation
\begin{align}
 \label{eq:flavour-transf-bosons-so3}
 U = \frac{1}{\sqrt{2}}
 \begin{pmatrix}
  -1 & 0 & 1 \\
  -i & 0 & -i \\
  0 & \sqrt{2} & 0 \\
 \end{pmatrix}.
\end{align}
Hence, the matrices $\tilde S_i$ can be rewritten as
\begin{align}
 \label{eq:s-7dim-transformed}
 S^{(1)}_i + S^{(2)}_j = V^\dag \left( \tilde{S}^{(1)}_i + \tilde{S}^{(2)}_j \right) V =
  \begin{pmatrix}
  T_i & 0 & R_i \\
  0 & T_j & R_j \\
  R_i^\dagger & R_j^\dagger & 0
 \end{pmatrix},
\end{align}
with
\begin{align}
  T_i = U^\dag \tilde T_i U, \quad
  R_i = U^\dag \tilde R_i, \quad
  V =
  \begin{pmatrix}
   U & 0 & 0 \\
   0 & U & 0 \\
   0 & 0 & 1
  \end{pmatrix}.
\end{align}
The vector of complicated fields has to be transformed accordingly:
\begin{align}
 \label{eq:flavour-transformation-vecC}
 C = V^\dagger \tilde C =
 \begin{pmatrix}
  C^{(1)} \\ C^{(2)} \\ A_3
 \end{pmatrix},
\end{align}
where the three-dimensional vectors $C^{(1)}$ and $C^{(2)}$ are defined by
\begin{align}
 \label{eq:flavour-transformation-vecC-1}
 C^{(1)} \equiv
 \begin{pmatrix}
  C^{(1)}_{+} \\ C^{(1)}_{0} \\ C^{(1)}_{-}
 \end{pmatrix}
 \equiv
 \begin{pmatrix}
  \tfrac{1}{\sqrt{2}} (- \pt_1 + i \pt_2) \\
  \pt_3 \\
  \tfrac{1}{\sqrt{2}} (+ \pt_1 + i \pt_2) \\
 \end{pmatrix}, \quad
 C^{(2)} \equiv
 \begin{pmatrix}
  C^{(2)}_{+} \\ C^{(2)}_{0} \\ C^{(2)}_{-}
 \end{pmatrix}
 \equiv
 \begin{pmatrix}
  \tfrac{1}{\sqrt{2}} (- \pt_4 + i \pt_5) \\
  \pt_6 \\
  \tfrac{1}{\sqrt{2}} (+ \pt_4 + i \pt_5) \\
 \end{pmatrix}.
\end{align}
The subscripts $+, -, 0$ denote the eigenvalues with respect to $T_3$.
One can also check that
\begin{align}
 \label{eq:ri-li-explicit}
 R_i^\dag L^{(1)}_i
 =  i \left(\frac{L^{(1)}_+}{\sqrt{2}}, -L^{(1)}_3, -\frac{L^{(1)}_-}{\sqrt{2}} \right), \quad
 R_i^\dag L^{(2)}_i
 =  i \left(\frac{L^{(2)}_+}{\sqrt{2}}, -L^{(2)}_3, -\frac{L^{(2)}_-}{\sqrt{2}} \right).
\end{align}

After the flavor transformation \eqref{eq:flavour-transf-bosons-so3}, the seven-dimensional matrix that mixes the flavors in the mass term for the complicated bosons is
\begin{align}
 \label{eq:sdotl-full}
 S \cdot L =
 S^{(1)}_i L^{(1)}_i + S^{(2)}_i L^{(2)}_i = 
 \begin{pmatrix}
  T_i L^{(1)}_i & 0 & R_i L^{(1)}_i \\
  0 & T_i L^{(2)}_i & R_i L^{(2)}_i \\
  R_i^\dag L^{(1)}_i & R_i^\dag L^{(2)}_i & 0
 \end{pmatrix}.
\end{align}

In the diagonalization of \eqref{eq:sdotl-full}, we have to distinguish the cases where one $\ell_a$ is $0$ and where both $\ell_a$ are bigger than $0$.%
\footnote{The case where $\ell_1=\ell_2=0$ is trivial as the corresponding fields are massless.}
For simplicity, we begin with the easier case where one $\ell_a$ is $0$.
Note that this formally reduces the diagonalization problem to the one  where only three of the scalar fields get non-trivial
$SO(3)$-symmetric vevs
that was solved in~\cite{Buhl-Mortensen:2016pxs,Buhl-Mortensen:2016jqo}.
We will now present a different solution to this diagonalization problem that has a straightforward generalization to the classical solution with $SO(3) \times SO(3)$ symmetry considered in this paper.
In the following, we also drop all references to $a$.

\paragraph{Diagonalization of \texorpdfstring{$T_i L_i$}{T\_i L\_i}}

After the flavor transformation in the previous section, the four-dimensional matrix $S \cdot L \equiv S_i L_i$ has the form
\begin{align}
 \label{eq:sdotl-full-so3}
 S_i L_i = 
  \begin{pmatrix}
  T_i L_i & R_i L_i \\
  R_i^\dag L_i & 0
  \end{pmatrix}.
\end{align}
It is important to realize that if we find an eigenvector of $T_i L_i$ that is annihilated by $R_i^\dagger L_i$ we can obtain an eigenvector of $S \cdot L$ by padding it with a zero to make it four-dimensional.
We will thus first look for states $\Phi$ such that
\begin{align}
 \label{eq:cond-eigenstates-3x3-annihilated}
 T_i L_i \; \Phi = \lambda_{\Phi} \; \Phi
 \quad \text{and} \quad
 R_i^\dagger L_i \; \Phi = 0.
\end{align}
This does not yield all eigenstates of $S \cdot L$, but we will see that the remaining ones are obtained by diagonalizing a simple $2 \times 2$ matrix.

If we define a total ``angular momentum'' operator $J_i = L_i + T_i$, then 
\begin{align}
 T_i L_i = \frac{1}{2} \left( J^2 - L^2 - T^2 \right) = \frac{1}{2} \left( J^2 - L^2 - 2 \right).
\end{align}
Hence, the diagonalization of the term $T_i L_i$ reduces to the problem of finding a set of common eigenstates for $J^2$, $J_3$ and $L^2$.
This is the well-known problem of addition of angular momentum, which can be solved using Clebsch-Gordan coefficients.
The matrices $T_i$ form the three-dimensional (spin-one) representation of \su{2} and the matrices $L_i$ form the spin-$\ell$ representation.
Thus, the fields $(C_{m_s})_{\ell m}$ in~\eqref{eq:flavour-transformation-vecC-1} have well-defined quantum numbers $\ell$, $m$ and $m_s$ for $L^2$, $L_3$ and $T_3$ respectively.
The fields with total angular momentum $j$, magnetic quantum number $m_j$ and angular momentum $\ell$ are found in terms of Clebsch-Gordan coefficients $\langle \ell, m; s, m_s | j, m_j \rangle$ by
\begin{align}
 \label{eq:eigenstate-j-mj-l-general-so3}
 B_{j, m_j; \ell} &=
 \sum_{m_s = -1}^{+1} \sum_{m = -\ell}^{\ell} \delta_{m + m_s, m_j} \langle \ell, m; 1, m_s | j, m_j \rangle \; (C_{m_s})_{\ell m}.
\end{align}
Here the total angular momentum can in general take the three values $j = \ell, \ell \pm 1$. For the case $\ell = 0$, however, there is only one total angular momentum $j = 1$; this necessitates the aforementioned distinction between $\ell_a=0$ and $\ell_a\neq 0$.
The dependence on $\ell$ will generally be dropped, and we will use the notation $(B_\alpha)_{j,m_j} \equiv B_{j, m_j; \ell = j - \alpha}$. For example, the state $B_+$ has total angular momentum $j = \ell + 1$ and $m_j = - \ell - 1, \ldots, \ell + 1$.
Using this notation and summing explicitly over $m$, \eqref{eq:eigenstate-j-mj-l-general-so3} becomes
\begin{align}
 \label{eq:rel-balpha-calpha-so3}
 (B_\alpha)_{j,m_j} = \sum_{m_s = -1}^{+1} 
 \langle \ell - \alpha, m_j - m_s; 1, m_s | j, m_j \rangle 
 \; (C_{m_s})_{\ell-\alpha, m_j - m_s}.
\end{align}
We can write out the basis states corresponding to~\eqref{eq:eigenstate-j-mj-l-general-so3} in vector form.
Since the $3 \times 3$ matrices $T_i$ are the standard spin-one representation of \su{2}, cf.\ \eqref{eq:t-su2-spin-one}, we have
\begin{align}
 T_3 \, \hat{e}_{m_s} = m_s \, \hat{e}_{m_s}
 \quad
 \text{with}
 \quad
 \hat{e}_{+1} = \begin{pmatrix} 1 \\ 0 \\ 0 \end{pmatrix}, \quad
 \hat{e}_{ 0} = \begin{pmatrix} 0 \\ 1 \\ 0 \end{pmatrix}, \quad
 \hat{e}_{-1} = \begin{pmatrix} 0 \\ 0 \\ 1 \end{pmatrix}.
\end{align}
The basis states that are eigenstates of $J^2$, $J_3$ and $L^2$ can thus be written as
\begin{align}
 \label{eq:eigenstate-j-mj-l-matrix-form}
 \begin{split}
 \hat Y_{j, m_j; \ell}
 & \equiv
 \sum_{m_s = -1}^{+1} \langle \ell, m_j - m_s; 1, m_s | j, m_j \rangle \,
 \hat Y_\ell^{m_j - m_s} \otimes \hat{e}_{m_s} \\
 &=
  \begin{pmatrix}
  {\arraycolsep=1.0pt
  \begin{array}{rl}
  \langle \ell, m_j - 1; 1, +1 | j, m_j \rangle & \hat Y_\ell^{m_j - 1} \\
  \langle \ell, m_j; 1,  0 | j, m_j     \rangle & \hat Y_\ell^{m_j}     \\
  \langle \ell, m_j + 1; 1, -1 | j, m_j \rangle & \hat Y_\ell^{m_j + 1} \\
  \end{array}
  }
  \end{pmatrix}.
 \end{split}
\end{align}
The Clebsch-Gordan coefficients for the case $j = \ell$ were given in~\eqref{eq:clebsch-gordan-j-l}.
For $j = \ell \pm 1$, we have
\begin{align}
 \label{eq:clebsch-gordan-j-lpm1}
 \begin{split}
 \langle \ell, m; 1, \pm 1 | \ell + 1, m \pm 1 \rangle
 &= \frac{ \sqrt{(\ell + 1 \pm m) (\ell + 2 \pm m)} }{ \sqrt{ 2 (\ell + 1) (2 \ell + 1) } }, \\
 \langle \ell, m; 1, 0 | \ell + 1, m \rangle
 &= \frac{ \sqrt{(\ell + 1 - m) (\ell + 1 + m)} }{ \sqrt{ (\ell + 1) (2 \ell + 1) } }, \\
 \langle \ell, m; 1, \pm 1 | \ell - 1, m \pm 1 \rangle
 &= \frac{ \sqrt{(\ell - 1 \mp m) (\ell \mp m)} }{ \sqrt{ 2 \ell (2 \ell + 1) } }, \\
 \langle \ell, m; 1, 0 | \ell - 1, m \rangle
 &= \frac{ \sqrt{(\ell - m) (\ell + m)} }{ \sqrt{ \ell (2 \ell + 1) } }.
 \end{split}
\end{align}
We find three sets of eigenstates for $j = \ell \pm 1$ and $j = \ell$ with eigenvalues
\begin{equation}
 \label{eq:eigenvalue-ell}
\begin{aligned}
 T_i L_i \; \hat Y_{j = \ell + 1, m_j; \ell} 
 &= \ell \; \hat Y_{j = \ell + 1, m_j; \ell}, \\
 T_i L_i \; \hat Y_{j = \ell, m_j; \ell} 
 &= - \hat Y_{j = \ell, m_j; \ell}, \\
 T_i L_i \; \hat Y_{j = \ell - 1, m_j; \ell}
 &= (-\ell - 1) \; \hat Y_{j = \ell - 1, m_j; \ell}.
\end{aligned}
\end{equation}
We will show below that the first and the last states satisfy the second condition in~\eqref{eq:cond-eigenstates-3x3-annihilated}, namely
\begin{align}
 R_i^\dagger L_i \; \hat Y_{ j , m_j ; \, j \pm 1 } = 0.
\end{align}
The fields $B_{\pm}$ can thus be made into eigenstates of $S \cdot L$ by padding with zeros.
The multiplicity of the corresponding eigenvalue is the dimension of the \su{2} representation, i.e.\ $2j + 1=2(\ell\pm1)+1$.

\paragraph{Diagonalization of the remaining \texorpdfstring{$2 \times 2$}{2x2} matrix}

We can expand the complicated scalars in the basis of total angular momentum eigenstates and $A_3$ in the basis of fuzzy spherical harmonics $\hat Y_{\ell, m}$, so that the four-dimensional vector of complicated fields is
\begin{align}
 \label{eq:complicated-fields-expanded-so3}
 C = 
 \begin{pmatrix}
 \sum_{j, m_j, \ell} B_{j, m_j; \, \ell} \hat Y_{j, m_j; \ell}
 \\[1.5ex]
 \sum_{\ell, m} (A_3)_{\ell, m}  \hat Y_{\ell}^m
 \end{pmatrix}.
\end{align}
We know how $T_i L_i$ acts on the basis states $\hat Y_{j, m_j; \ell}$ obtained from the Clebsch-Gordan procedure from~\eqref{eq:eigenvalue-ell}.
Now we will calculate how $R_i^\dagger L_i$, i.e.\ the last row in $S \cdot L$ as given in \eqref{eq:sdotl-full-so3}, acts on $Y_{j, m_j; \ell}$.
Using that the ladder operators act as given in~\eqref{eq:ladder-operators-action-Ylm} together with~\eqref{eq:ri-li-explicit} and the completeness relation of the Clebsch-Gordan coefficients, one obtains
\begin{align}
\nonumber
  R_i^\dagger L_i \, \hat Y_{ j, m_j; \ell }
   &= -i \sqrt{\ell (\ell + 1)} \sum_{m_s}
     \langle \ell, m_j - m_s; 1, m_s | j, m_j \rangle 
     \langle \ell, m_j - m_s; 1, m_s | \ell, m_j \rangle
     \; \hat Y_\ell^{m_j}\\
   &= -i \, \delta_{j,\ell} \sqrt{\ell (\ell + 1)} \; \hat Y_\ell^{m_j}.
  \label{eq:action-RL-state}
\end{align}
This vanishes unless $j = \ell$.
The states $\hat Y_{j, m_j; \ell}$ with $j = \ell \pm 1$ are thus annihilated by $R_i^\dagger L_i$ and can simply be padded with a zero block to give eigenstates of $S \cdot L$ as we have claimed before.
Using~\eqref{eq:eigenvalue-ell} and~\eqref{eq:action-RL-state}, we can find the matrix elements of both $T_i L_i$ and $R_i L_i$: 
\begin{align}
 \label{eq:action-TL-RL-states-so3}
 \begin{aligned}
  \tr \left( \hat Y_{j', m'; \ell'}^\dag \, 
             T_i L_i \, 
             \hat Y_{j, m; \ell} \right)
  & = \mu_{j,\ell} \, \delta_{m,m'} \delta_{\ell,\ell'} 
      \delta_{j,j'}, \\
  \tr \left( (\hat Y_{\ell'}^{m'})^\dag \, 
             R_i^\dag L_i \, \hat Y_{j, m; \ell} \right)
  & = - i \, \delta_{m,m'} \, \delta_{\ell,\ell'} \, \delta_{j,\ell'}
      \sqrt{\ell (\ell + 1)}, \\
  \tr \left( \hat Y_{j', m'; \ell'}^\dag \, R_i L_i \, \hat Y_\ell^m \right)
  & = + i \, \delta_{m,m'} \, \delta_{\ell,\ell'} \, \delta_{\ell,j'}
      \sqrt{\ell (\ell + 1)}.
 \end{aligned}
\end{align}
The matrix elements $\mu_{j,\ell}$ in the first line are $\mu_{\ell + 1, \ell} = \ell$, $\mu_{\ell, \ell} = -1$ and $\mu_{\ell - 1, \ell} = - \ell - 1$, cf.~\eqref{eq:eigenvalue-ell}.
The third line follows naturally from complex conjugation of the second line and $L_i^\dag = L_i$. 

We now insert the vector of complicated fields $C$ given in \eqref{eq:complicated-fields-expanded-so3} into the flavor mixing term in the action, obtaining
\begin{align}
 \label{eq:mixing-a3-b-so3}
 \begin{split}
 \tr \left[ C^\dag S_i L_i \, C \right]
  = \sum_{\ell = 1}^{k-1}  
   \Bigg[&
    \ell     \sum_{m = - \ell -1}^{\ell + 1}
      (B_+)_{\ell + 1 , m}^\dag (B_+)_{\ell + 1 , m} -
    (\ell+1) \sum_{m = - \ell + 1}^{\ell - 1}
      (B_-)_{\ell - 1 , m}^\dag (B_-)_{\ell - 1 , m} \\
 &\! + \sum_{m=-\ell}^\ell 
   \begin{pmatrix} 
     (B_0)_{\ell,m}^\dag & (A_3)_{\ell,m}^\dag
   \end{pmatrix}
   \begin{pmatrix}
      -1 & -i \sqrt{\ell (\ell + 1)} \\
      +i \sqrt{\ell (\ell + 1)} & 0
   \end{pmatrix}
   \begin{pmatrix} 
     (B_0)_{\ell,m} \\ (A_3)_{\ell,m}
   \end{pmatrix} \Bigg].
 \end{split}
\end{align}
The fields $B_{\pm}$ diagonalize the full $4 \times 4$ matrix as we discussed before.
What remains to be diagonalized is the $2 \times 2$ matrix in the last line of the previous equation.
Note in particular that this matrix does not depend on the magnetic quantum number.
The fields that achieve the diagonalization are
\begin{align}
 \begin{split}
 D_+ &= \frac{1}{\sqrt{2 \ell + 1}} \left( -i \sqrt{\ell} B_0 + \sqrt{\ell + 1} A_3 \right), \\
 D_- &= \frac{1}{\sqrt{2 \ell + 1}} \left( i \sqrt{\ell + 1} B_0 + \sqrt{\ell} A_3 \right),
 \end{split}
\end{align}
with eigenvalues $\lambda_+ = \ell$ and $\lambda_- = - \ell - 1$.
Notice from this result that the masses are integer numbers, even though from~\eqref{eq:mixing-a3-b-so3} we could have expected square roots in the spectrum.
This is actually an indication that the spectrum can be obtained in a simpler way, namely only using Clebsch-Gordan coefficients as in~\cite{Buhl-Mortensen:2016jqo}.

This concludes the diagonalization of the $4 \times 4$ sub-block of the seven-dimensional flavor mixing matrix, which is relevant for the case where one $\ell_a$ is $0$.
We summarize the result in Table \ref{tab:summary-spectrum-so3-so3-l0}.
We have effectively rederived the spectrum of the bosons for the classical solution considered in~\cite{Buhl-Mortensen:2016jqo} where only three scalar fields get non-trivial $SO(3)$-symmetric vevs.
Our method is however different and can be extended to the present classical solution with $SO(3) \times SO(3)$ symmetry.
In particular, we will find a natural generalization of the $2 \times 2$ matrix in~\eqref{eq:mixing-a3-b-so3}.

\renewcommand*{\arraystretch}{1.5}
\begin{table}
  \centering
  \begin{tabular}{c c c}
    \toprule
    Mass eigenstate & Mass $m^2$ & Multiplicity \\
    \midrule
    $B_+$ & $\ell_1 (\ell_1 - 1)$       & $2 \ell_1 + 3$ \\
    $B_-$ & $(\ell_1 + 1) (\ell_1 + 2)$ & $2 \ell_1 - 1$ \\
    $D_+$ & $\ell_1 (\ell_1 - 1)$       & $2 \ell_1 + 1$ \\
    $D_-$ & $(\ell_1 + 1) (\ell_1 + 2)$ & $2 \ell_1 + 1$ \\
    \bottomrule
  \end{tabular}
 \caption{Masses and eigenstates of the complicated bosons in the $k_1 k_2 \times k_1 k_2$ block for the case $\ell_2 = 0$ and $\ell_1 = 1, \ldots, k_1 - 1$. 
 The case $\ell_1 = 0$  and $\ell_2 = 1, \ldots, k_1 - 1$ is obtained by relabeling. In the case $\ell_1=\ell_2=0$, the masses vanish, while the case $\ell_1\neq 0$ and $\ell_2\neq0$ is shown in Table \ref{tab:summary-spectrum-so3-so3}.}
 \label{tab:summary-spectrum-so3-so3-l0}
\end{table}
\renewcommand*{\arraystretch}{1.0}

\paragraph{Full mixing matrix}

Let us now diagonalize the full seven-dimensional matrix \eqref{eq:sdotl-full} in the case where $\ell_1\neq0$ and $\ell_2\neq0$.
Following the steps discussed for the $4 \times 4$ sub-block relevant for the case where one $\ell_a=0$, we define fields $B^{(1)}$ and $B^{(2)}$ with total angular momentum in each sector.
As before, they are given in terms of Clebsch-Gordan coefficients by
\begin{align}
 \label{eq:eigenstate-j-mj-l-general-so3-so3}
 (B^{(1)})_{j_1, m_1, \ell_1; \ell_2, m_2} &=
 \sum_{m_s = -1}^{+1} \langle \ell_1, m_1 - m_s; 1, m_s | j_1, m_1 \rangle \;
 (C^{(1)}_{m_s})_{\ell_1, m_1; \ell_2, m_2}, \\
 (B^{(2)})_{\ell_1, m_1; j_2, m_2, \ell_2} &=
 \sum_{m_s = -1}^{+1} \langle \ell_2, m_2 - m_s; 1, m_s | j_2, m_2 \rangle \;
 (C^{(2)}_{m_s})_{\ell_1, m_1; \ell_2, m_2}.
\end{align}
We can also write out the corresponding basis states explicitly:
\begin{align}
 {(\hat Y^{(1)})}_{j_1, m_1, \ell_1; \ell_2, m_2} \equiv 
 \hat Y_{ j_1, m_1; \ell_1 } \otimes \hat Y_{\ell_2}^{m_2}, \quad
 {(\hat Y^{(2)})}_{\ell_1, m_1; j_2, m_2, \ell_2} \equiv 
 \hat Y_{\ell_1}^{m_1} \otimes \hat Y_{j_2, m_2; \ell_2}.
\end{align}
Now using the natural generalization of the matrix elements in~\eqref{eq:action-TL-RL-states-so3}, one can see that the four fields $B_\pm^{(1)}$ and $B_\pm^{(2)}$ diagonalize the full $7 \times 7$ matrix \eqref{eq:sdotl-full}. It remains to diagonalize a $3 \times 3$ matrix, which is a simple generalization of~\eqref{eq:mixing-a3-b-so3}:
\begin{align}
 \label{eq:mixing-a3-b}
  \begin{pmatrix}
    (B_0^{(1)})^\dagger & (B_0^{(2)})^\dagger & (A_3)^\dagger
  \end{pmatrix}
  \begin{pmatrix}
    -1 & 0 & - i \sqrt{\ell_1 (\ell_1 + 1)} \\
    0 & -1 & - i \sqrt{\ell_2 (\ell_2 + 1)} \\
    + i \sqrt{\ell_1 (\ell_1 + 1)} & + i \sqrt{\ell_2 (\ell_2 + 1)} & 0
  \end{pmatrix}
  \begin{pmatrix}
    B_0^{(1)} \\ B_0^{(2)} \\ A_3
  \end{pmatrix}.
\end{align}
Here we have dropped the quantum numbers from the fields to unclutter the notation.
This matrix has eigenvalues
\begin{align}
 \label{eq:eigenvalues-3x3}
 \lambda_0 = -1, \quad
 \lambda_\pm = -\frac{1}{2} \pm \sqrt{\ell_1(\ell_1+1) + \ell_2(\ell_2+1) + \tfrac{1}{4}},
\end{align}
and the corresponding diagonal fields are 
\begin{align}
 \label{eq:diagonal-fields-3x3}
  \begin{split}
  D_0 &= \frac{1}{\sqrt{N_0}} \left( 
   - \sqrt{\ell_2 (\ell_2 + 1)} B_0^{(1)}
   + \sqrt{\ell_1 (\ell_1 + 1)} B_0^{(2)}
  \right), \\
  D_\pm &= \frac{1}{\sqrt{N_\pm}} \left( 
    i \sqrt{\ell_1 (\ell_1 + 1)} B_0^{(1)}
  + i \sqrt{\ell_2 (\ell_2 + 1)} B_0^{(2)}
  + \lambda_{\mp} A_3
  \right),
  \end{split}
\end{align}
with
\begin{align}
 \label{eq:normalization-constants-product-case}
  \begin{aligned}
  N_{\pm} 
  & = \lambda_{\mp} (\lambda_{\mp} - \lambda_{\pm}) \\
  & = \frac{1}{2} \left( 1 + 4 \ell_1 (\ell_1 + 1) + 4 \ell_2 (\ell_2 + 1) \pm
  \sqrt{ 1 + 4 \ell_1 (\ell_1 + 1) + 4 \ell_2 (\ell_2 + 1) } \right), \\
  N_{0} 
  & = - \lambda_+ \lambda_- = \ell_1 (\ell_1 + 1) + \ell_2 (\ell_2 + 1). 
  \end{aligned}
\end{align}
Since $\lambda_\pm$ contains a square root, it is clear that it is impossible to obtain the spectrum of masses using only a Clebsch-Gordan decomposition, but a more general procedure like the one we have presented is required.

\renewcommand*{\arraystretch}{1.5}
\begin{table}
  \centering
  \begin{tabular}{c r l c}
    \toprule
    Mass eigenstate & \multicolumn{2}{c}{Mass $m^2$} & Multiplicity \\
    \midrule
    $ B_{+}^{(1)} $ &
    $m^{2}_{{(1)},+} =$ & $\ell_1 (\ell_1 - 1) + \ell_2 (\ell_2 + 1)$ &
    $(2 \ell_1 + 3) (2 \ell_2 + 1)$ \\
    $ B_{-}^{(1)} $ &
    $m^{2}_{{(1)},-} =$ & $ (\ell_1 + 1)(\ell_1 + 2) + \ell_2 (\ell_2 + 1)$ &
    $(2 \ell_1 - 1) (2 \ell_2 + 1)$ \\
    $ B_{+}^{(2)} $ &
    $m^{2}_{{(2)},+} =$ & $ \ell_1 (\ell_1 + 1) + \ell_2 (\ell_2 - 1)$ &
    $(2 \ell_1 + 1) (2 \ell_2 + 3)$ \\
    $ B_{-}^{(2)} $ &
    $m^{2}_{{(2)},-} =$ & $ \ell_1 (\ell_1 + 1) + (\ell_2 + 1)(\ell_2 + 2)$ &
    $(2 \ell_1 + 1) (2 \ell_2 - 1)$ \\
    $ D_0 $ &
    $m^{2}_{0} =$ & $ \ell_1 (\ell_1 + 1) + \ell_2 (\ell_2 + 1) + 2$ &
    $(2 \ell_1 + 1) (2 \ell_2 + 1)$ \\
    $ D_+ $ &
    $m^{2}_{+} =$ & $ \ell_1 (\ell_1 + 1) + \ell_2 (\ell_2 + 1) - 2 \lambda_+$ &
    $(2 \ell_1 + 1) (2 \ell_2 + 1)$ \\
    $ D_- $ &
    $m^{2}_{-} =$ & $ \ell_1 (\ell_1 + 1) + \ell_2 (\ell_2 + 1) - 2 \lambda_-$ &
    $(2 \ell_1 + 1) (2 \ell_2 + 1)$ \\
    \bottomrule
  \end{tabular}
  \caption{Masses and eigenstates of the complicated bosons in the $k_1 k_2 \times k_1 k_2$ block in the $SO(3) \times SO(3)$-symmetric case. One must consider all combinations of $\ell_1 = 1, \ldots, k_1 - 1$ and $\ell_2 = 1, \ldots, k_2 - 1$. The masses for the fields in the off-diagonal blocks are obtained by the replacements $\ell_1 \rightarrow \tfrac{k_1 - 1}{2}$ and $\ell_2 \rightarrow \tfrac{k_2 - 1}{2}$, while the corresponding multiplicities are obtained by the same replacement followed by a multiplication with $2(N-k_1k_2)$.}
  \label{tab:summary-spectrum-so3-so3}
\end{table}
\renewcommand*{\arraystretch}{1.0}

\subsection{Fermion mass matrix}
\label{sec:fermion-mass-matrix}

Inserting the classical solution~\eqref{eq:classical-solution-so3-so3} into the mass term for the Majorana fermions~\eqref{eq:mass-term-fermions}, we find 
\begin{align}
 \label{eq:mass-term-fermions-inserted}
 S_{\text{m,f}} =
 \frac{2}{\gym^2} \int \mathrm{d}^4 x \left( \frac{-1}{2 \, x_3} \right) \tr \Bigg(
    \sum_{i = 1}^{3} \bar{\psi}_{j} (G^{(1)}_i)_{jk} L^{(1)}_i \psi_k +
    \sum_{i = 1}^{3} \bar{\psi}_{j} (G^{(2)}_i)_{jk} L^{(2)}_{i} (\gamma_5 \psi_k)
 \Bigg),
\end{align}
where $G^{(1)}_i\equiv G_i$ and $G^{(2)}_i\equiv G_{i+3}$ for $i=1,2,3$.
Since $[G^{(1)}_i, G^{(2)}_j] = 0$ and $[L^{(1)}_i, L^{(2)}_j] = 0$, we can diagonalize both terms in~\eqref{eq:mass-term-fermions-inserted} simultaneously.
We give the form of the matrices $G^{(1)}_i$ and $G^{(2)}_i$ in Appendix~\ref{sec:conventions-general} using the same conventions as~\cite{Buhl-Mortensen:2016jqo}.
From~\cite{Buhl-Mortensen:2016jqo}, we also know that the matrices $G^{(1)}_i$ can be transformed into block-diagonal form with
\begin{align}
 U = \frac{1}{\sqrt{2}}
 \begin{pmatrix}
   0 & -i & -1 &  0 \\
   0 &  1 &  i &  0 \\
   -1 &  0 &  0 &  i \\
   i & 0 & 0 & -1
 \end{pmatrix}
 \quad \Rightarrow \quad
 U^\dagger G^{(1)}_i U = -
 \begin{pmatrix}
   \sigma_i & 0 \\
   0 & \sigma_i
 \end{pmatrix}
 = - \mathds{1}_2 \otimes \sigma_i.
\end{align}
Here $\sigma_i$ are the usual Pauli matrices.
Acting with $U$ on the remaining matrices $G^{(2)}_i$ gives
\begin{align}
  U^\dagger G^{(2)}_i U
  = i \, \sigma_i \otimes \mathds{1}_2.
\end{align}
The extra factor of $i$ is consistent with the fact that the matrices $G^{(2)}_i$ are anti-Hermitian and it is also required to make the term with $\gamma_5$ in~\eqref{eq:mass-term-fermions-inserted} Hermitian.
On the fermions, the transformation $U$ yields
\begin{align}
 \label{eq:vec-fermions-transformed}
 U^\dagger
 \begin{pmatrix}
  \psi_1 \\ \psi_2 \\ \psi_3 \\ \psi_4
 \end{pmatrix}
 =
 \frac{1}{\sqrt{2}}
 \begin{pmatrix}
  -\psi_3 -i \psi_4 \\
  i \psi_1 + \psi_2 \\
  -\psi_1 - i \psi_2 \\
  -i \psi_3 - \psi_4
 \end{pmatrix}
 =
 \begin{pmatrix}
  C_{++} \\ C_{-+} \\ C_{+-} \\ C_{--}
 \end{pmatrix}
 \equiv C_F.
\end{align}
Here the subscripts on $C_{m_{s_1},m_{s_2}}$ indicate that the field has spin $\tfrac{1}{2}$ and magnetic quantum number $m_{s_1}$ with respect to $\frac{1}{2}\mathds{1}_2 \otimes \sigma_3$, and spin $\tfrac{1}{2}$ and magnetic quantum number $m_{s_2}$ with respect to $\frac{1}{2} \sigma_3 \otimes \mathds{1}_2$. The fields also have orbital angular momentum $\ell_a$ and magnetic quantum number $m_a$ with respect to $L^{{(a)}}$ for $a = 1, 2$. This problem is closely related to the one studied in~\cite{Buhl-Mortensen:2016jqo}, with the difference that here we have two copies of the spin-orbit coupling problem. 

To diagonalize the mass matrix, we define the total angular momentum operators
\begin{align}
 J^{(1)}_i = L^{(1)}_i + \frac{1}{2} \mathds{1}_2 \otimes \sigma_i, \quad
 J^{(2)}_i = L^{(2)}_i + \frac{1}{2} \sigma_i \otimes \mathds{1}_2,
\end{align}
so the terms inside the trace in~\eqref{eq:mass-term-fermions-inserted} take the form
\begin{align}
 - \bar{C}_F \left[
    (J^{(1)})^2 - (L^{(1)})^2 - \frac{1}{2} \left(\frac{1}{2} + 1 \right) 
 \right] C_F
 + \bar{C}_F \left[
    (J^{(2)})^2 - (L^{(2)})^2 - \frac{1}{2} \left(\frac{1}{2} + 1 \right) 
 \right] (i \gamma_5 ) \, C_F.
\end{align}
The notation $\bar{C}_F$ means the following:
transpose the four-dimensional vector of fermions $C_F$ as given in~\eqref{eq:vec-fermions-transformed} and take the Dirac conjugate $\bar{\psi} \equiv \psi^\dagger \gamma^0$ of each fermion inside of it.
The explicit formula for the diagonal fields in terms of the Clebsch-Gordan coefficients is given by
\begin{align}
 B^{j_1 j_2}_{\ell_1, m_{j_1}; \ell_2, m_{j_2}}
 &=
 \sum_{\substack{m_{s_1}, m_1 \\ m_{s_2}, m_2}}
 \langle \ell_1, m_1; \tfrac{1}{2}, m_{s_1} | j_1, m_{j_1} \rangle
 \langle \ell_2, m_2; \tfrac{1}{2}, m_{s_2} | j_2, m_{j_2} \rangle
 (C_{m_{s_1}, m_{s_2}})_{\ell_1, m_1; \ell_2, m_2},
\end{align}
where the total angular momentum is $j_a = \ell_a \pm \tfrac{1}{2}$. 
In total, there are four combinations from combining $j_1 = \ell_1 \pm \tfrac{1}{2}$ with $j_2 = \ell_2 \pm \tfrac{1}{2}$ in all possible ways, each with a multiplicity of $(2 j_1 + 1) (2 j_2 + 1)$.
The eigenvalues of each term in~\eqref{eq:mass-term-fermions-inserted} are
\begin{align}
 j (j + 1) - \ell (\ell + 1) - \frac{1}{2} \left(\frac{1}{2} + 1 \right)
 =
 \begin{cases}
 \ell &\quad \text{for} \quad j = \ell + \tfrac{1}{2}, \\
 -\ell - 1 & \quad \text{for} \quad j = \ell - \tfrac{1}{2}.
 \end{cases}
\end{align}

After the diagonalization, the quadratic part of the action for the fermions takes the schematic form
\begin{align}
  \label{eq:fermion-action-after-diagonalization}
  S = \frac{2}{\gym^2} \int \mathrm{d}^4 x \sum_\alpha
  \tr \left[ \;
    \frac{i}{2} \bar{B}_\alpha \gamma^\mu \partial_\mu B_\alpha -
    \frac{1}{2 x_3} \bar{B}_\alpha 
    \left( c_\alpha + i \, d_\alpha \gamma_5 \right) B_\alpha \;
  \right].
\end{align}
Here the index $\alpha$ is running over all the diagonal fields $B$.
We will now use a chiral rotation to rewrite this action in a form where the mass term is positive and does not contain the $i\gamma_5$ part. 
Following the procedure described in~\cite{Burgess:2007zi}, one finds that the required transformation is
\begin{align}
 \label{eq:chiral_rot_psi}
 B_{\alpha}
 =   \cos \left( \tfrac{\theta}{2} \right) B_{\alpha}'
 - i \sin \left( \tfrac{\theta}{2} \right) \gamma_5 B_{\alpha}',
 \qquad \theta \equiv \arg ( c + i d ).
\end{align}
Notice that this transformation preserves the Majorana property, namely the fields $B_{\alpha}'$ are also Majorana fermions.
Using this transformation, one can check that the resulting action has the form
\begin{align}
  \label{eq:fermion-action-after-chiral-rotation}
  S = \frac{2}{\gym^2} \int \mathrm{d}^4 x \sum_\alpha
  \tr \left[ \;
    \frac{i}{2} \bar{B}_{\alpha}' \gamma^\mu \partial_\mu B_{\alpha}' -
    \frac{m_\alpha}{2 x_3} \bar{B}_{\alpha}' B_{\alpha}' \;
  \right],
\end{align}
with $m_{\alpha} = \left| c_{\alpha} + i d_{\alpha} \right| = \sqrt{c_{\alpha}^2 + d_{\alpha}^2}$.
We list the values of $c_{\alpha}$, $d_{\alpha}$ and $m_{\alpha}$ along with their multiplicities in Table~\ref{tab:fermion-masses}.

\renewcommand*{\arraystretch}{1.5}
\begin{table}
  \centering
  \begin{tabular}[h]{c c c r l c}
    \toprule
    Mass eigenstate & $c$ & $d$ & \multicolumn{2}{c}{Mass $m = | c + i d |$} & Multiplicity \\
    \midrule
    $B^{\ell_1 + \tfrac{1}{2}, \ell_2 + \tfrac{1}{2}}$ &
    $-\ell_1$ & $\ell_2$ & $m_{++} =$ & $\sqrt{\ell_1^2 + \ell_2^2}$ &
    $(\ell_1 + 1) (\ell_2 + 1)$ \\
    $B^{\ell_1 + \tfrac{1}{2}, \ell_2 - \tfrac{1}{2}}$ &
    $-\ell_1$ & $-\ell_2 - 1$ & $m_{+-} =$ & $\sqrt{\ell_1^2 + (\ell_2 + 1)^2}$ &
    $(\ell_1 + 1) \ell_2$ \\
    $B^{\ell_1 - \tfrac{1}{2}, \ell_2 + \tfrac{1}{2}}$ &
    $\ell_1 + 1$ & $\ell_2$ & $m_{-+} =$ & $\sqrt{(\ell_1 + 1)^2 + \ell_2^2}$ &
    $\ell_1 (\ell_2 + 1)$ \\
    $B^{\ell_1 - \tfrac{1}{2}, \ell_2 - \tfrac{1}{2}}$ &
    $\ell_1 + 1$ & $-\ell_2 - 1$ & $m_{--} =$ & $\sqrt{(\ell_1 + 1)^2 + (\ell_2 + 1)^2}$ &
    $\ell_1 \ell_2$ \\
    \bottomrule
  \end{tabular}
  \caption{Eigenvalues and eigenstates of the fermions in the $SO(3) \times SO(3)$-symmetric case in the $k_1 k_2 \times k_1 k_2$ block. One must consider all combinations of $\ell_1 = 0, \ldots, k_1 - 1$ and $\ell_2 = 0, \ldots, k_2 - 1$. For the definition of $c$ and $d$, see \eqref{eq:fermion-action-after-diagonalization}. The values for $c$, $d$ and $m$ for the fields in the off-diagonal blocks are obtained by the replacements $\ell_1 \rightarrow \tfrac{k_1 - 1}{2}$ and $\ell_2 \rightarrow \tfrac{k_2 - 1}{2}$, while the corresponding multiplicities are obtained by the same replacement followed by a multiplication with $2(N-k_1k_2)$.}
  \label{tab:fermion-masses}
\end{table}
\renewcommand*{\arraystretch}{1.0}

\subsection{Summary of the spectrum}
\label{sec:summary-spectrum-complicated-bosons}

We have now derived the spectrum for the defect CFT with $SO(3) \times SO(3)$-symmetric vevs.
For the easy bosons (and the ghosts), we had to diagonalize the operator $(L^{(1)})^2 + (L^{(2)})^2$ which was achieved by expanding the fields in the $k_1 k_2 \times k_1 k_2$ block in fuzzy spherical harmonics.
The fields in the off-diagonal blocks were already eigenstates of this operator.
We list the masses and multiplicities of the easy bosons in Table~\ref{tab:masses-easy-so3-so3}.

For the complicated bosons, the mass term reads
\begin{align}
 (L^{(1)})^2 + (L^{(2)})^2 - 2 S \cdot L,
\end{align}
where the term $S \cdot L$ is responsible for mixing fields of different flavor.
Knowing that $(L^{(1)})^2 + (L^{(2)})^2$ is diagonalized by an expansion in fuzzy spherical harmonics, we have subsequently obtained the eigenstates of $S \cdot L$ in two steps.
Since we were coupling the spin-$\ell$ with the spin-one representation of \su{2}, we had to distinguish between the case where either $\ell_1$ or $\ell_2$ were zero and the case where both $\ell_a$ were non-zero.
The case $\ell_a = 0$ formally reduced the diagonalization problem to the one solved in~\cite{Buhl-Mortensen:2016jqo}, which we solved using a slightly different approach that was also applicable to the second case where both $\ell_1 \neq 0$ and $\ell_2 \neq 0$.
For this case, we first diagonalized the $3 \times 3$ blocks $T_i L^{(1)}_i$ and $T_i L^{(2)}_i$ using angular momentum coupling.
The eigenstates with $j_1 = \ell_1 \pm 1$ and $j_2 = \ell_2 \pm 1$ could trivially be padded with zeros to give eigenstates of the full matrix and their eigenvalues are given in \eqref{eq:eigenvalue-ell}.
For the remaining eigenstates, we had to diagonalize the $3 \times 3$ matrix in~\eqref{eq:mixing-a3-b} and found $D_\pm$ and $D_0$ in~\eqref{eq:diagonal-fields-3x3} with eigenvalues $\lambda_\pm$ and $\lambda_0$ in~\eqref{eq:eigenvalues-3x3}.
Adding the contribution from $(L^{(1)})^2+(L^{(2)})^2$, we obtain the masses shown in Table~\ref{tab:summary-spectrum-so3-so3-l0} for the case where one of the $\ell_a$ is zero and in Table~\ref{tab:summary-spectrum-so3-so3} for the general case where $\ell_1 \neq 0$ and $\ell_2 \neq 0$.
Note that we are only listing the masses and multiplicities for the fields $[\Phi]_{n,n'}$ in the $k_1 k_2 \times k_1 k_2$ block here.
To obtain the masses and multiplicities of the fields in the off-diagonal block, we use the replacement rule~\eqref{eq:replacement-off-diagonal-block-so3-so3}.
The multiplicity also receives an extra factor of $2 (N - k_1 k_2)$ from the size of the two blocks.
Additionally there are $(N - k_1 k_2) \times (N - k_1 k_2)$ massless fields $[\Phi]_{a,a'}$.

Finally, we found that the spectrum of the fermions could be obtained by simply employing the angular momentum techniques from~\cite{Buhl-Mortensen:2016jqo} for each sector.
The only additional step was the chiral rotation which allowed us to trade the term with $i \gamma_5$ in the action for a standard mass term.
The fermion spectrum is shown in Table~\ref{tab:fermion-masses}.

Let us compare the spectrum for the defect CFT with $SO(3) \times SO(3)$-symmetric vevs dual to the D3-D7 probe-brane system derived here to the one for the defect CFT  dual to the D3-D5 probe-brane system, where only three scalar fields get non-trivial $SO(3)$-symmetric vevs, derived in~\cite{Buhl-Mortensen:2016jqo}.
In the D3-D5 system, the spectrum can be derived using Clebsch-Gordan coefficients only, i.e.\ it is not necessary to employ the two-step process that we used to rederive it here.
In the D3-D7 system however, Clebsch-Gordan coefficients are not sufficient as can be seen from the appearance of square roots in the mass eigenvalues.
Furthermore, in the D3-D5 system, supersymmetry was visible in the spectrum.
Defining $\nu = \sqrt{m^2 + \tfrac{1}{4}}$ for the bosons and comparing it with the mass $|m_f|$ of the fermions, one could see that the steps between these parameters were half-integers.
This could be attributed to supersymmetry in $AdS_{4}$, where the conformal dimensions are given by $\Delta = \tfrac{3}{2} + \nu$ for the bosons and $\Delta = \tfrac{3}{2} + |m_f|$ for the fermions.
The conformal dimensions within one supermultiplet however differ by $\tfrac{1}{2}$ which implies the observed relation between $\nu$ and $|m_f|$.
In the present case, we can only relate three of the masses that appear in the spectrum of the bosons; namely, we find the relation
\begin{align}
 \label{eq:relation-nus}
 \nu_- = \sqrt{m_-^2 + \tfrac{1}{4}} = \nu_{\mathrm{easy}} + 1, \quad
 \nu_+ = \sqrt{m_+^2 + \tfrac{1}{4}} = \nu_{\mathrm{easy}} - 1.
\end{align}
This is consistent with the fact that supersymmetry is broken in the D3-D7 system.

\section{Propagators}
\label{sec:propgagators}

In this section, we take into account the effect that the $x_3$-dependence of the `masses' has on the propagators of the scalars (Subsection~\ref{sec:scalar-propagators}) and the fermions (Subsection~\ref{sec:fermionic-propagators}), following~\cite{Buhl-Mortensen:2016jqo}. We then derive the propagators of the flavor eigenstates that occur in the action in terms of the propagators of the mass eigenstates.
Thus, this section provides the framework for doing perturbative calculations in this defect CFT.

\subsection{Scalar propagators}
\label{sec:scalar-propagators}

The propagator for a generic scalar field with mass term $\tfrac{m^2}{x_3^2}$ is the solution to
\begin{align}
  \left( - \partial_\mu \partial^\mu + \frac{m^2}{x_3^2} \right) K^{m^2}(x,y) 
  = \frac{\gym^2}{2} \, \delta(x-y).
\end{align}
As noted in~\cite{Nagasaki:2011ue}, the propagator of a scalar with mass $\tfrac{m^2}{x_3^2}$ in $(d + 1)$-dimensional Minkowski space is related to the propagator of a scalar with constant mass $\tilde m^2$ in $AdS_{d+1}$.
The relation is explicitly given by
\begin{align}
 \label{eq:spacetime-propagator-flat-ads-relation}
 K^{m^2}(x, y) = \frac{\gym^2}{2} (x_3 y_3)^{-\tfrac{d - 1}{2}} K_{AdS}^{\tilde{m}^2}(x,y),
 \quad
 \tilde{m}^2 = m^2 - \frac{d^2 - 1}{4}.
\end{align}
In our case, $d$ is the dimension of the defect, i.e.\ $d=3$.
Using that $\tilde{m}^2 = \Delta (\Delta - d)$ in $AdS_{d+1}$, we find that the scaling dimension $\Delta$ is
\begin{align}
 \Delta = \frac{d}{2} + \nu, \quad
 \nu \equiv \sqrt{m^2 + \tfrac{1}{4}}.
\end{align}
A closed expression for the scalar propagator in $AdS_{d+1}$ using Euclidean signature can be found e.g.\ in~\cite{Ammon:2015wua}:
\begin{align}
 \label{eq:spacetime-propagator-ads}
 K_{AdS}^{\Delta}(x, y)
 =
 \frac{\Gamma(\Delta) \, \xi(x, y)^{\Delta}}
      {2^\Delta (2\Delta - d) \pi^{d/2} \Gamma(\Delta - \tfrac{d}{2}) } \,
 {}_2F_1\left( \tfrac{\Delta}{2}, \tfrac{\Delta + 1}{2}; \Delta - \tfrac{d}{2} + 1; 
               \xi^2(x, y) \right)
\end{align}
with
\begin{align}
 \xi(x, y) = \frac{2 x_3 y_3}{x_3^2 + y_3^2 + (x_0 - y_0)^2 + (x_1 - y_1)^2 + (x_2 - y_2)^2}.
\end{align}
For the Feynman-diagram calculation, we will require the propagator evaluated at $x = y$.
In this case, the propagator diverges (in the UV) and needs to be regularized. Our regularization of choice is dimensional regularization (or rather dimensional reduction, as we discuss below). Moreover, we want to keep the codimension of the defect at $1$, such that its dimension becomes $d=3-2\epsilon$.
The expression \eqref{eq:spacetime-propagator-ads} cannot be used in this case.
Instead,
\begin{align}
 \label{eq:spacetime-propagator-regularized}
 \begin{split}
 K^{\nu}(x, x)
 &= \frac{\gym^2}{2} \frac{1}{16 \pi^2 x_3^2} \bigg[ m^2 \bigg(-\frac{1}{\epsilon} - \log(4 \pi) + \gamma_{\mathrm{E}} - 2 \log(x_3) + 2 \Psi( \nu + \tfrac{1}{2} ) - 1 \bigg) - 1 \bigg],
 \end{split}
\end{align}
which is derived from an integral representation of  \eqref{eq:spacetime-propagator-ads}, see \cite{Buhl-Mortensen:2016jqo}. 
Above, $\gamma_{\mathrm{E}}$ denotes the Euler-Mascheroni constant and $\Psi$ denotes the digamma function.

\subsection{Fermionic propagators}
\label{sec:fermionic-propagators}

After the chiral rotation, the action for the Majorana fermions takes the form
\begin{align}
\label{eq:action-after-chiral-rot}
  S = \frac{2}{\gym^2} \int \mathrm{d}^4 x \, \tr \left[ \;
     \frac{i}{2} \bar \psi' \gamma^\mu \partial_\mu \psi'
   - \frac{m}{2 x_3} \bar \psi' \psi'
  \right],
\end{align}
where the mass $m > 0$, cf.\ \eqref{eq:fermion-action-after-chiral-rotation}.
The fermionic propagator is the solution to
\begin{align}
 \left( -i \gamma^\mu \partial_\mu + \frac{m}{x_3} \right) K_F^m(x,y)
 = \frac{\gym^2}{2} \delta(x-y).
\end{align}
These propagators were derived in~\cite{Buhl-Mortensen:2016jqo,Kawano:1999au},
\begin{align}
 \label{eq:exact-fermionic-propagator}
 K_F^{m}(x,y) 
 = 
 \left[ i \gamma^\mu \partial_\mu + \frac{m}{x_3} \right]
 \left[ 
    K^{\nu = m - \frac{1}{2}}(x,y) \mathcal P_- +
    K^{\nu = m + \frac{1}{2}}(x,y) \mathcal P_+
 \right], 
\end{align}
with $\mathcal P_\pm = \frac{1}{2} ( 1 \pm i \gamma^3 )$ and $K^\nu(x,y)$ being the bosonic propagator.

The fermionic propagator will later be required in the calculation of the one-loop correction to the classical solution (Section~\ref{sec:one-loop-vevs}), where fermions can circulate in a loop.
As all spinor indices have to be contracted in this case, we will be interested in the spinor trace of the propagator.
Using \eqref{eq:spacetime-propagator-regularized}, one can show that the trace of the fermionic propagator, regularized for $x = y$, is \cite{Buhl-Mortensen:2016jqo}
\begin{align}
 \label{eq:fermion-prop-regularized}
 \tr K_F^{m}(x,x)
= \frac{\gym^2}{8 \pi^2 x_3^3}
 \Bigg[ & 
   m^3 + m^2 - 3m - 1 \\
 &  + m(m^2 - 1) \left( -\frac{1}{\epsilon} - \log(4 \pi) + \gamma_{\mathrm{E}} - 2 \log(x_3) + 2 \Psi(m) - 2 \right)
 \Bigg].
\nonumber
\end{align}

It will later be convenient to have an expression for the propagators between the fermion fields before the chiral rotation.
Before the chiral rotation, the action takes the form~\eqref{eq:fermion-action-after-diagonalization},
\begin{align}
  S = \frac{2}{\gym^2} \int \mathrm{d}^4 x \, \tr \left[ \;
     \frac{i}{2} \bar \psi \gamma^\mu \partial_\mu \psi
   - \frac{1}{2 x_3} \bar \psi (c + i d \gamma_5) \psi
  \right].
\end{align}
Here $\psi$ could be any of the fields $B_{\alpha}$, either in the $k_1 k_2 \times k_1 k_2 $, the $(N-k_1 k_2 )\times k_1 k_2 $ or the $k_1 k_2 \times(N-k_1 k_2 )$ block.
Since the mass $m$ is related to the parameters $c$ and $d$ by $m = \left| c + i d \right|$, the propagators between the original fields $\psi$ and chirally rotated fields $\psi'$ are
\begin{align}
 \label{eq:propagator-fermions-old-paper}
 \left\langle \psi(x) \bar \psi(y) \right\rangle 
  = \tilde{K}_F^{c, d}(x,y), \qquad
 \left\langle \psi'(x) \bar \psi'(y) \right\rangle = K_F^{m = |c + i d|}(x,y).
\end{align}
Using the transformation \eqref{eq:chiral_rot_psi}, one can see that the relation between them is
\begin{align}
 \label{eq:propagator-fermions-old-paper-new}
 \tilde{K}_F^{c, d}
  = \cos^2 \left( \tfrac{\theta}{2} \right) K_F^{|c + i d|}
  - \sin^2 \left( \tfrac{\theta}{2} \right) \gamma_5 K_F^{|c + i d|} \gamma_5
  - \sin   \left( \tfrac{\theta}{2} \right) 
    \cos   \left( \tfrac{\theta}{2} \right) \{ \gamma_5, K_F^{|c + i d|} \},
\end{align}
where $\theta \equiv \arg ( c + i d )$.
We will always be interested in the trace of this propagator, possibly multiplied by $i\gamma_5$. Using the explicit form of the fermionic propagator~\eqref{eq:exact-fermionic-propagator} and trigonometric identities, we find
\begin{align}
 \label{eq:trace-chiral-rotation}
 \tr \tilde{K}_F^{c, d}
 =  \frac{c}{|m|} \tr K_F^{m = |c + i d|}, \qquad
 \tr \left( i \gamma_5 \tilde{K}_F^{c, d} \right)
 =  \frac{d}{|m|} \tr K_F^{m = |c + i d|}.
\end{align}

\subsection{Color and flavor part of the propagators}
\label{sec:propagators-colour-flavour}

In Sections~\ref{sec:boson-mass-matrix} and~\ref{sec:fermion-mass-matrix} we have found the mass eigenstates of the theory, and the propagators between them can be obtained as described in Sections~\ref{sec:scalar-propagators} and~\ref{sec:fermionic-propagators}. 
However, it will prove convenient to also derive the propagators between the fields that originally appeared in the action of $\mathcal{N} = 4$ SYM theory, namely the six scalars, the gauge field, the Majorana fermions and the ghosts. 
The reason is that it would be extremely cumbersome to rewrite the interaction vertices~\eqref{eq:action-cubic-vertices} in terms of the diagonal fields.
Note that we are still giving the propagators for the color components $[\Phi]_{n,a}$ and $[\Phi]_{a,n}$ defined in \eqref{eq:field-decomposition} as well as $\Phi_{\ell_1,m_1; \ell_2, m_2}$ defined in \eqref{eq:fuzzy-expansion}, which partially diagonalize the color part of the mixing problem.%
\footnote{Recall that the massless fields $[\Phi]_{a,a'}$ have ordinary propagators.  The massless fields from the $k_1k_2\times k_1k_2$ block can only propagate for $x_3>0$ and appropriate boundary conditions have to be imposed at the defect for these fields. In the D3-D5 case, supersymmetry puts constraints on the possible choices of boundary 
conditions, cf.\ \cite{Gaiotto:2008sa,deLeeuw:2017dkd},
 but in the present 
case we have no such guidelines. The choice of boundary conditions for these fields, however, will not affect the results 
in the large-$N$ limit.}

To find these propagators, we express the original fields in terms of the diagonal fields. 
For example, for the bosons we have to undo the three steps of the diagonalization: the flavor transformation~\eqref{eq:flavour-transformation-vecC}, the Clebsch-Gordan procedure~\eqref{eq:eigenstate-j-mj-l-general-so3-so3} and the diagonalization of the final $3 \times 3$ matrix~\eqref{eq:diagonal-fields-3x3}.
The details of this calculation are shown in Appendix~\ref{sec:appendix-colour-flavour-part}.

The mass term of the complicated bosons is diagonalized in terms of the fields $B_{\pm}^{(1)}$, $B_{\pm}^{(2)}$, $D_0$ and $D_{\pm}$.
Thus the propagators between these fields are simply the scalar propagators $K^{m^2}(x, y)$ from Section~\ref{sec:scalar-propagators} with the corresponding mass eigenvalue from Table~\ref{tab:summary-spectrum-so3-so3}.
The eigenvalues $\lambda_\pm$ and normalization constants $N_\pm$ and $N_0$ were given in~\eqref{eq:eigenvalues-3x3} and~\eqref{eq:normalization-constants-product-case}, but we repeat them here for convenience:
\begin{align}
 \lambda_{\pm} = -\frac{1}{2} \pm \sqrt{\ell_1 (\ell_1 + 1) + \ell_2 (\ell_2 + 1) + \tfrac{1}{4}}, \;
 N_\pm = \lambda_{\mp} \left( \lambda_{\mp} - \lambda_{\pm} \right), \;
 N_0 = - \lambda_+ \lambda_-.
\end{align}
For the matrix elements of the \su{2} generators $t_i$, we use the shorthand notation
\begin{align}
 [t^{(\ell_1)}_i]_{m_1,m_1'} \equiv [t^{2 \ell_1 + 1}_i]_{\ell_1 - m_1 + 1, \ell_1 - m_1' + 1}, \quad
 [t^{(\ell_2)}_i]_{m_2,m_2'} \equiv [t^{2 \ell_2 + 1}_i]_{\ell_2 - m_2 + 1, \ell_2 - m_2' + 1}.
\end{align}
Explicit expressions for the generators $t_i$ are given in Appendix~\ref{sec:conventions-su2-and-Ylm}.
The propagators involving easy fields are diagonal in flavor, and we find
\begin{align}
 \label{eq:propagator-easy-bosons}
 \begin{split}
 \langle (A_0)_{\ell_1 m_1; \ell_2 m_2} (A_0)^\dagger_{\ell_1' m_1'; \ell_2' m_2'} \rangle
 &= 
   \delta_{\ell_1 \ell_1'} \delta_{\ell_2 \ell_2'} 
   \delta_{m_1,m_1'} \delta_{m_2,m_2'}
   \underbrace{ K^{m^2 = \ell_1(\ell_1 + 1) + \ell_2(\ell_2 + 1)} }_{\equiv K^{\mathrm{easy}}},
 \end{split}
\end{align}
where one could replace $A_0$ with any of the other easy fields $A_1$, $A_2$ or $c$.
For the propagators involving $A_3$ and scalars of different sectors, we find
\begin{align}
  \label{eq:propagator-phis-different-sectors}
  \langle (\pt^{(1)}_i)_{\ell_1 m_1; \ell_2 m_2} (\pt^{(2)}_j)^\dagger_{\ell_1' m_1'; \ell_2' m_2'} \rangle
  &= \delta_{\ell_1 \ell_1'} \delta_{\ell_2 \ell_2'} 
  [t^{(\ell_1)}_i]_{m_1,m_1'} [t^{(\ell_2)}_j]_{m_2,m_2'}
  \underbrace{ \left( \frac{K^{m^2_-}}{N_-} + \frac{K^{m^2_+}}{N_+} - \frac{K^{m_0^2}}{N_0} \right) }_{\equiv K^{\phi}_{\mathrm{opp}}},
  \\
  \label{eq:propagator-phi-a}
  \langle (\pt^{(1)}_i)_{\ell_1 m_1; \ell_2 m_2} (A_3)^\dagger_{\ell_1' m_1'; \ell_2' m_2'} \rangle
  &= - \langle (A_3)_{\ell_1 m_1; \ell_2 m_2} (\pt^{(1)}_i)^\dagger_{\ell_1' m_1'; \ell_2' m_2'} \rangle \\
  &= -i \delta_{\ell_1 \ell_1'} \delta_{\ell_2 \ell_2'} [t^{(\ell_1)}_i]_{m_1 m_1'} \delta_{m_2 m_2'}
   \underbrace{ \left(\frac{\lambda_+}{N_-} K^{m^2_-} + \frac{\lambda_-}{N_+} K^{m^2_+} \right) }_{\equiv K^{\phi,A}},
\nonumber
  \\
  \label{eq:propagators-a-a}
  \langle (A_3)_{\ell_1 m_1; \ell_2 m_2} (A_3)^\dagger_{\ell_1' m_1'; \ell_2' m_2'} \rangle &=
  \delta_{\ell_1 \ell_1'} \delta_{\ell_2 \ell_2'} \delta_{m_1 m_1'} \delta_{m_2 m_2'}
  \underbrace{ \left( \frac{\lambda_+^2}{N_-} K^{m^2_-} + \frac{\lambda_-^2}{N_+} K^{m^2_+} \right) }_{\equiv K^{A,A}},
\end{align}
with $\pt^{(1)}_i \equiv \pt_i$ and $\pt^{(2)}_i \equiv \pt_{i+3}$.
For the propagator between scalars from the same sector, we find
\begin{align}
 \label{eq:propagator-phis-same-sector}
    &\langle (\pt^{(1)}_i)_{\ell_1 m_1; \ell_2 m_2} (\pt^{(1)}_j)^\dagger_{\ell_1' m_1'; \ell_2' m_2'} \rangle
    = \delta_{\ell_1 \ell_1'} \delta_{\ell_2 \ell_2'} \delta_{m_2 m_2'}
     \\ \nonumber
      & \bigg[
      \delta_{ij} \delta_{m_1 m_1'}
      \underbrace{
      \bigg(
        \frac{\ell_1 + 1}{2 \ell_1 + 1} K^{m^2_{{(1)},+}} + \frac{\ell_1}{2 \ell_1 + 1} K^{m^2_{{(1)},-}}
      \bigg)
      }_{\equiv K^{\phi,(1)}_{\mathrm{sing}}}
      - i \epsilon_{ijk} [t^{(\ell_1)}_k]_{m_1, m_1'}
      \underbrace{
      \bigg( \frac{K^{m^2_{{(1)},+}}}{2 \ell_1 + 1} - \frac{K^{m^2_{{(1)},-}}}{2 \ell_1 + 1} \bigg)
      }_{\equiv K^{\phi,(1)}_{\mathrm{anti}}} \\ \nonumber
      &\hphantom{\bigg[} 
      - [t^{(\ell_1)}_i t^{(\ell_1)}_j]_{m_1, m_1'}
      \underbrace{
      \bigg(
        \frac{K^{m^2_{{(1)},+}}}{(2 \ell_1 + 1)(\ell_1 + 1)} + \frac{K^{m^2_{{(1)},-}}}{(2 \ell_1 + 1) \ell_1}
        - \frac{\ell_2 (\ell_2 + 1)}{\ell_1 (\ell_1 + 1)} \frac{K^{m^2_0}}{N_0} - \frac{K^{m^2_-}}{N_-} - \frac{K^{m^2_+}}{N_+}
      \bigg)
      }_{\equiv K^{\phi,(1)}_{\mathrm{sym}}}
    \bigg].
\end{align}
From \eqref{eq:propagator-phi-a} and~\eqref{eq:propagator-phis-same-sector}, the propagators for the other sector are obtained by a simple relabeling, e.g.
\begin{align}
 \label{eq:prop-relabeling}
 \langle (\pt_i^{(2)})_{\ell_1 m_1; \ell_2 m_2} (\pt_j^{(2)})^\dagger_{\ell_1' m_1'; \ell_2' m_2'} \rangle =
 \langle (\pt_i^{(1)})_{\ell_2 m_2; \ell_1 m_1} (\pt_j^{(1)})^\dagger_{\ell_2' m_2'; \ell_1' m_1'} \rangle,
\end{align}
where the (implicit) dependence of the masses on $\ell_1$ and $\ell_2$ must be taken into account as well.
In the following, we will often use the combination of spacetime propagators $K^{\mathrm{easy}}$, $K^{\phi}_{\mathrm{opp}}$, $K^{\phi,A}$, $K^{A,A}$, $K^{\phi,(a)}_{\mathrm{sing}}$, $K^{\phi,(a)}_{\mathrm{anti}}$ and $K^{\phi,(a)}_{\mathrm{sym}}$ defined in \eqref{eq:propagator-easy-bosons}--\eqref{eq:propagator-phis-same-sector}.%
\footnote{%
The cases where either $\ell_1 = 0$ or $\ell_2 = 0$ required special treatment in the diagonalization of the boson mass matrix, see the discussion in Section~\ref{sec:complicated-fields-so3-so3}.
In these cases, the spectrum reduces to the one in Table~\ref{tab:summary-spectrum-so3-so3-l0}, which was originally found in~\cite{Buhl-Mortensen:2016jqo}.
While the boson masses in Table~\ref{tab:summary-spectrum-so3-so3} do not have the correct limit for $\ell_1 = 0$ or $\ell_2 = 0$, the propagators presented in this section indeed reduce to the ones found in~\cite{Buhl-Mortensen:2016jqo}.
}

Before the chiral rotation, the quadratic part of the action for the fermions is diagonalized by the fields $B^{\ell_1 + \tfrac{1}{2}, \ell_2 + \tfrac{1}{2}}$, $B^{\ell_1 + \tfrac{1}{2}, \ell_2 - \tfrac{1}{2}}$, $B^{\ell_1 - \tfrac{1}{2}, \ell_2 + \tfrac{1}{2}}$ and $B^{\ell_1 - \tfrac{1}{2}, \ell_2 - \tfrac{1}{2}}$.
Written in terms of these fields, the action still contains $\gamma_5$. Therefore, the propagators between them are of the form $\tilde{K}^{c,d}_F$ in~\eqref{eq:propagator-fermions-old-paper-new}, where the eigenvalues $c$ and $d$ are given in Table~\ref{tab:fermion-masses}.
In the calculations in this paper, the propagators always appear inside a spinor trace, possibly multiplied by $\gamma_5$, and they can be transformed to the propagators $K^m_F$ by means of \eqref{eq:trace-chiral-rotation} which relates them to the propagators after the chiral rotation.
Undoing the diagonalization of the fermion mass matrix, we find
\begin{align}
 \label{eq:propagator-fermions}
 \begin{split}
  &\langle (\psi_i)_{\ell_1 m_1; \ell_2 m_2} (\overline{\psi_j})_{\ell_1' m_1'; \ell_2' m_2'} \rangle
 = \frac{\delta_{\ell_1 \ell_1'} \delta_{\ell_2 \ell_2'}}{(2 \ell_1 + 1) (2 \ell_2 + 1)}\\
 &\,\,\bigg\{+
   \delta_{ij} \delta_{m_1 m_1'} \delta_{m_2 m_2'}
   \Big[
     \ell_1 \ell_2 \; \tilde{K}_F^{\ell_1 + 1, -\ell_2 - 1}  
 +
     \ell_1 (\ell_2 + 1) \; \tilde{K}_F^{\ell_1 + 1, \ell_2} \\
     &\hphantom{\,\,\bigg\{+
   \delta_{ij} \delta_{m_1 m_1'} \delta_{m_2 m_2'}
   \Big[}
     + (\ell_1 + 1) \ell_2 \; \tilde{K}_F^{-\ell_1, - \ell_2 - 1} +
     (\ell_1 + 1) (\ell_2 + 1) \; \tilde{K}_F^{-\ell_1, \ell_2}
   \Big] \\
   & \quad
   - [G^{(1)}_n]_{ij} [t^{(\ell_1)}_n]_{m_1 m_1'} \delta_{m_2 m_2'}
   \Big[
     (\ell_2 + 1) \left(  \tilde{K}_F^{-\ell_1, \ell_2} -  \tilde{K}_{F}^{\ell_1 + 1, \ell_2} \right) \\
    &\qquad\hphantom{ \,  \quad
   - [G^{(1)}_n]_{ij} [t^{(\ell_1)}_n]_{m_1 m_1'} \delta_{m_2 m_2'}
   \Big[} +
     \ell_2 \left(  \tilde{K}_F^{-\ell_1, -\ell_2 - 1} -  \tilde{K}_F^{\ell_1 + 1, - \ell_2 - 1} \right)
   \Big] \\
   & \quad
   - i [G^{(2)}_n]_{ij} [t^{(\ell_2)}_n]_{m_2 m_2'} \delta_{m_1 m_1'}
   \Big[
     (\ell_1 + 1) \left(  \tilde{K}_F^{-\ell_1, \ell_2} -  \tilde{K}_{F}^{-\ell_1, -\ell_2 - 1} \right) \\
     &\qquad\hphantom{ \quad
   - i [G^{(2)}_n]_{ij} [t^{(\ell_2)}_n]_{m_2 m_2'} \delta_{m_1 m_1'}
   \Big[}+
     \ell_1 \left(  \tilde{K}_F^{\ell_1 + 1, \ell_2} -  \tilde{K}_F^{\ell_1 + 1, - \ell_2 - 1} \right)
   \Big] \\
   & \quad
   +i [G^{(1)}_{n_1} G^{(2)}_{n_2}]_{ij} 
      [t^{(\ell_1)}_{n_1}]_{m_1 m_1'} [t^{(\ell_2)}_{n_2}]_{m_2 m_2'}
   \Big[
      \tilde{K}_F^{\ell_1 + 1, -\ell_2 - 1} -  \tilde{K}_F^{\ell_1 + 1, \ell_2} -  \tilde{K}_F^{-\ell_1, -\ell_2 - 1} +  \tilde{K}_F^{-\ell_1, \ell_2}
   \Big]
 \bigg\}.
 \end{split}
\end{align}

The propagators given so far are valid for fields in the $k_1 k_2 \times k_1 k_2$ block, not 
the fields in the $(N - k_1 k_2) \times k_1 k_2$ and $k_1 k_2 \times (N - k_1 k_2)$ blocks.
As we argued in Section~\ref{sec:decomposition-colour-matrices}, we can simply replace
\begin{align}
 \label{eq:replacement-rule-propagators}
 \ell_1 \rightarrow \frac{k_1 - 1}{2}
 \quad \text{and} \quad
 \ell_2 \rightarrow \frac{k_2 - 1}{2}
\end{align}
everywhere to obtain the masses for the fields in the off-diagonal blocks.
For the fields themselves, we replace $(\Phi)_{\ell_1 m_1; \ell_2 m_2} \rightarrow [\Phi]_{n,a}$.
To obtain the corresponding mass eigenstates, we have to replace the matrices $\hat{Y}^{m_1}_{\ell_1} \otimes \hat{Y}^{m_2}_{\ell_2}$ by ${E^n}_{a}$, resulting in a replacements of the orthonormality condition \eqref{eq:states-orthogonality-so3} with \eqref{eq:states-orthogonality-so3_offdiagonal} and similar changes in the non-diagonal matrix part.
We find for the propagators between the easy fields,
\begin{align}
 \label{eq:propagators-offdiagonal-easy-fields}
 \langle [A_0]_{n,a} [A_0]_{n',a'}^\dagger \rangle = \delta_{a,a'} \delta_{n,n'} K^{\mathrm{easy}},
\end{align}
where as above $A_0$ could be any of the easy fields $A_0$, $A_1$, $A_2$ and $c$.
For the remaining propagators, we find
\begin{align} \label{eq:propagators-offdiagonal-1}
 \langle [\pt^{(1)}_i]_{n,a} [\pt^{(2)}_j]_{n',a'}^\dagger \rangle
 &=
 \delta_{a,a'} [t^{k_1}_i \otimes t^{k_2}_j]_{n,n'} K^{\phi}_{\mathrm{opp}}, \\
 \label{eq:propagators-offdiagonal-2}
 \langle [\pt^{(1)}_i]_{n,a} [A_3]_{n',a'}^\dagger \rangle
 &=
 - \langle [A_3]_{n,a} [\pt^{(1)}_i]^\dagger_{n',a'} \rangle
 =
 -i \delta_{a,a'} [t^{k_1}_{i} \otimes \mathds{1}_{k_2}]_{n,n'} K^{\phi,A}, \\
 \label{eq:propagators-offdiagonal-3}
 \langle [A_3]_{n,a} [A_3]_{n',a'}^\dagger \rangle
 &=
 \delta_{a,a'} \delta_{n,n'} K^{A,A}
\end{align}
and
\begin{align}
\begin{split}
 \label{eq:propagators-offdiagonal-4}
 \langle [\pt^{(1)}_i]_{n,a} [\pt^{(1)}_j]_{n',a'}^\dagger \rangle
 =
 \delta_{a,a'} \Big[
 &
 \delta_{ij} \delta_{n,n'} K^{\phi,(1)}_{\mathrm{sing}}
 - i \epsilon_{ijk} [t^{k_1}_k \otimes \mathds{1}_{k_2}]_{n,n'} K^{\phi,(1)}_{\mathrm{anti}} \\
 &
 - [t^{k_1}_i t^{k_1}_j \otimes \mathds{1}_{k_2}]_{n,n'} K^{\phi,(1)}_{\mathrm{sym}}
 \Big].
\end{split}
\end{align}
As above, we can simply obtain the expressions for the scalars from the other sectors from~\eqref{eq:propagators-offdiagonal-2} and~\eqref{eq:propagators-offdiagonal-4}, e.g.
\begin{align}
 \langle [\pt^{(2)}_i]_{n,a} [A_3]_{n',a'}^\dagger \rangle
 &=
 -i \delta_{a,a'} [\mathds{1}_{k_1} \otimes t^{k_2}_{i}]_{n,n'} K^{\phi,A}.
\end{align}
Note that it is understood that the replacement rule~\eqref{eq:replacement-rule-propagators} is applied everywhere, 
in particular also in $K^{\mathrm{easy}}$,  $K^{\phi}_{\mathrm{opp}}$, $K^{\phi,A}$, $K^{A,A}$, $K^{\phi,(1)}_{\mathrm{sing}}$, $K^{\phi,(1)}_{\mathrm{anti}}$, and $K^{\phi,(1)}_{\mathrm{sym}}$ defined in \eqref{eq:propagator-easy-bosons}--\eqref{eq:propagator-phis-same-sector}.
No new complications arise for the fermions in the off-diagonal block and it is straightforward to obtain the propagators between them from~\eqref{eq:propagator-fermions}.

\section{One-loop corrections to the classical solution}
\label{sec:one-loop-vevs}

With the propagators at hand, we are now able to study many different quantities perturbatively.
In this section, we start by calculating the first quantum correction to the classical solution, i.e.\ to the vevs of the scalars.
While it is not observable itself, it occurs as a part of the calculation of many observables, including the one-loop corrections to one-point functions of scalar single-trace operators considered in the subsequent section.
We find that the first quantum correction to the scalar vevs is non-vanishing, unlike in the D3-D5 system, where the vevs of the scalars were not corrected at one-loop order~\cite{Buhl-Mortensen:2016jqo}.

The one-loop vacuum expectation value of the scalars is~\cite{Buhl-Mortensen:2016jqo}
\begin{align}
 \label{eq:one-loop-corr-vevs}
 % First contraction
 \contraction{
 \langle \phi_i \rangle_{\text{1-loop}}(x)
 = 
  }{\pt_i}{(x)
  \int \text{d}^4 y 
  \sum_{\Phi_1, \Phi_2, \Phi_3} 
  V_3(}{\Phi}
 % Second contraction
 \contraction{
 \langle \phi_i \rangle_{\text{1-loop}}(x)
 = 
  \pt_i(x) 
  \int \text{d}^4 y 
  \sum_{\Phi_1, \Phi_2, \Phi_3} 
  V_3(\Phi_1(y),}{\Phi}{_2(y), }{\Phi}
 \langle \phi_i \rangle_{\text{1-loop}}(x)
  = 
   \pt_i(x) 
   \int \text{d}^4 y 
   \sum_{\Phi_1, \Phi_2, \Phi_3} 
   V_3(\Phi_1(y), \Phi_2(y), \Phi_3(y))
.
\end{align}
Here, the sum of all the contractions of cubic interactions occurs where one of the fields, which we call $\Phi_1$, remains uncontracted. The field $\Phi_1$ is then contracted with $\pt_i$ and the position of the interaction is integrated over to obtain $\langle \phi_i \rangle_{\text{1-loop}}$.%
\footnote{The only conceivable contribution of the defect fields at one-loop order is through a cubic defect vertex $V_3$. However, the defect fields $\Phi_2$ and $\Phi_3$ are massless in this case, resulting in a massless tadpole integral that vanishes due to conformal symmetry.}

The calculation of \eqref{eq:one-loop-corr-vevs} requires the evaluation of propagators at the same spacetime points, i.e.\ $K^\nu(y, y)$ and $\tr K_F^m(y, y)$.
This introduces divergences which we regularize using dimensional regularization, cf.\ \eqref{eq:spacetime-propagator-regularized} and~\eqref{eq:fermion-prop-regularized}.
Dimensional regularization in $4 - 2 \epsilon$ dimensions changes the number of components of the gauge field to $n_A = 4 - 2 \epsilon$ while keeping the number of scalars and fermions fixed.
This breaks supersymmetry and is therefore not a convenient regularization scheme for standard $\mathcal{N} = 4$ theory;
for instance, non-renormalization theorems due to supersymmetry are only applicable if supersymmetry is preserved by the regulator.
Usually, supersymmetry can be restored in dimensional reduction by introducing additional $2 \epsilon$ scalars in the action~\cite{Siegel:1979wq,Capper:1979ns}, which has been successfully applied in $\mathcal{N} = 4$ theory (see e.g.~\cite{Erickson:2000af,Nandan:2014oga} and references therein).%
\footnote{For sufficiently high loop orders, dimensional reduction is known to become inconsistent though~\cite{Siegel:1980qs,Avdeev:1981vf,Avdeev:1982np,Avdeev:1982xy}.}
In the defect theory, the regularization procedure must be chosen in a way that is compatible with the theory without the defect, i.e.\ with $\mathcal{N} = 4$ SYM theory.
The reason is that the entire UV behavior of the theory with defect is governed by the theory without the defect.
One can see this by considering the scalar propagator~\eqref{eq:spacetime-propagator-flat-ads-relation} in the limit $x \rightarrow y$, where it reduces to the propagator for a scalar in $\mathcal{N} = 4$ SYM theory in four dimensions.
In the following, we will therefore work in a version of dimensional reduction where we introduce $2 \epsilon$ scalars behaving as the easy components of the gauge fields.
We also note that dimensional reduction has been applied successfully in~\cite{Buhl-Mortensen:2016jqo} for the D3-D5 system, where it was crucial for the one-loop correction to the vevs to vanish.

We will work in the planar limit, where $N \to \infty$ and $\gym \to 0$, such that the 't~Hooft coupling $\lambda = N \gym^2$ remains fixed.
The computation of $\langle \phi_i \rangle_{\text{1-loop}}$ is technically involved, so we present it in detail in Appendix~\ref{sec:one-loop-vevs-appendix}, while here we will focus on the results.
We find that the one-loop correction to the scalar vevs is
\begin{align}
 \label{eq:correction-vevs}
 \begin{split}
 \big\langle \phi^{(a)}_i \big\rangle (x) 
 &=\big\langle \phi^{(a)}_i \big\rangle_{\mathrm{tree}} (x) +
 \big\langle \phi^{(a)}_i \big\rangle_{\mathrm{1-loop}} (x) + \mathcal{O}(\lambda^2)\\
 &= \left(1+\frac{\lambda}{16 \pi^2} W^{(a)}(k_1, k_2)+\mathcal{O}(\lambda^2) \right)\big\langle \phi^{(a)}_i \big\rangle_{\mathrm{tree}} (x)
, \end{split}
\end{align}
for $a=1,2$.
This result is valid for arbitrary $k_1,k_2\geq 2$, and the functions $W^{(1)}(k_1,k_2)$ and $W^{(2)}(k_1, k_2)$ are
\begin{align}
\begin{alignedat}{3}
 W^{(1)}(k_1, k_2)
 \label{eq:effective-vertex-W1}
 &=
  - \frac{1}{2} \left( 3 \measy^2 - 4 + \frac{16}{k_1^2 + k_2^2 - 2} \right) \Psi \big(\nu _{\text{easy}} + \tfrac{1}{2}\big)
  \\ &\phaneq
  - \frac{\big(k_1-2\big) \big(k_1+3\big) }{2 k_1 \big(k_1-1\big)}m_{{(1)},-}^2 \Psi \big(\nu_{{(1)},-} + \tfrac{1}{2}\big)
  - \frac{\big(k_2-2\big) }{2 k_2}m_{{(2)},-}^2 \Psi \big(\nu_{{(2)},-}+\tfrac{1}{2}\big)
 \\ &\phaneq
  - \frac{\big(k_1+2\big) \big(k_1-3\big) }{2 k_1 \big(k_1+1\big)}m_{{(1)},+}^2 \Psi \big(\nu_{{(1)},+}+\tfrac{1}{2}\big)
  - \frac{\big(k_2+2\big) }{2 k_2}m_{{(2)},+}^2 \Psi \big(\nu_{{(2)},+}+\tfrac{1}{2}\big)
 \\ &\phaneq
  -
   \Bigg( \frac{1}{2} + \frac{4}{k_1^2 + k_2^2 - 2} \frac{\big(k_2^2 - 1\big) }{\big(k_1^2-1\big) } \Bigg) m_0^2 \Psi \big(\nu_0+\tfrac{1}{2}\big)
  + \frac{1}{2} - \frac{8}{k_1^2 + k_2^2 - 2}
 \\ &\phaneq
  + \frac{\big(k_1+1\big) \big(k_2-1\big)}{k_1 k_2} \big(m_{--}^2-1\big) \Bigg( \Psi \big(m_{--}\big) + \frac{1}{2 m_{--}} \Bigg)
 \\ &\phaneq
  + \frac{\big(k_1+1\big) \big(k_2+1\big)}{k_1 k_2} \big(m_{-+}^2-1\big) \Bigg( \Psi \big(m_{-+}\big) + \frac{1}{2 m_{-+}} \Bigg)
 \\ &\phaneq
  + \frac{\big(k_1-1\big) \big(k_2-1\big)}{k_1 k_2} \big(m_{+-}^2-1\big) \Bigg( \Psi \big(m_{+-}\big) + \frac{1}{2 m_{+-}} \Bigg)
 \\ &\phaneq
  + \frac{\big(k_1-1\big) \big(k_2+1\big)}{k_1 k_2} \big(m_{++}^2-1\big) \Bigg( \Psi \big(m_{++}\big) + \frac{1}{2 m_{++}} \Bigg)
\end{alignedat}
\end{align}
and
\begin{align}
 \label{eq:effective-vertex-W2}
 W^{(2)}(k_1, k_2) = W^{(1)}(k_2, k_1). 
\end{align}
The masses and $\nu = \sqrt{m^2 + \tfrac{1}{4}}$ are functions of $k_1$ and $k_2$ that are explicitly given in Tables~\ref{tab:masses-easy-so3-so3}, \ref{tab:summary-spectrum-so3-so3} and \ref{tab:fermion-masses}, where in the latter two the replacement $\ell_a\to\frac{k_a-1}{2}$ is understood.
While we have suppressed this dependence in \eqref{eq:effective-vertex-W1}, it is understood to be taken into account in \eqref{eq:effective-vertex-W2}.
Note that we have used~\eqref{eq:relation-nus} to write $\Psi(\nu_- + \tfrac{1}{2})$ and $\Psi(\nu_+ + \tfrac{1}{2})$ in terms of $\Psi(\nu_{\mathrm{easy}} + \tfrac{1}{2})$.

On top of the planar limit, we can employ the double-scaling limit introduced in~\eqref{eq:double-scaling-limit}. We find
\begin{align}
 \label{eq:correction-vevs-dslimit}
 \begin{split}
 \big\langle \phi^{(1)}_i \big\rangle_{\mathrm{1-loop}} (x) 
  \simeq 
  - \frac{\lambda}{4\pi^2(k_1^2+k_2^2)} 
    \frac{2 k_2^4}{(k_1^2 + k_2^2)^2}
    \big\langle \phi_i^{(1)} \big\rangle_{\mathrm{tree}}, \\
 \big\langle \phi^{(2)}_i \big\rangle_{\mathrm{1-loop}} (x) 
 \simeq 
  - \frac{\lambda}{4\pi^2(k_1^2+k_2^2)} 
    \frac{2 k_1^4}{(k_1^2 + k_2^2)^2}
    \big\langle \phi_i^{(2)} \big\rangle_{\mathrm{tree}},
 \end{split}
\end{align}
where $\simeq$ signifies that we are only keeping the leading powers in $k_1$ and $k_2$. 
Notice that the expansion yields a result that has the desired expansion in the double-scaling parameter $\tfrac{\lambda}{ (k_1^2 + k_2^2)}$.

Finally, let us note that the one-loop corrections to the vevs of all other fields are vanishing.

\section{One-loop corrections to single-trace operators}
\label{sec:one-loop-onepfun}

In this section, we consider planar one-point functions of gauge-invariant bulk operators of the defect CFT.
We start with general single-trace operators (Subsection~\ref{sec:general-one-point-functions}) following~\cite{Buhl-Mortensen:2016jqo} and then specialize to the 1/2-BPS operator $\tr Z^L$ (Subsection \ref{sec:one-loop-trZL}).
In particular, we consider operators with well-defined scaling dimensions $\Delta$, normalized such that in the theory without the defect the two-point functions are%
\footnote{%
The latter requirement is necessary for the one-point functions to be observable. In general, only $\langle\mathcal{O}\rangle/||\mathcal{O}||$ is observable, where the norm $||\mathcal{O}||$ is given by the two-point function far away from the defect.}
\begin{align}
 \langle \mathcal{O}_{a}(x) \mathcal{O}_{b}(y) \rangle 
 = \frac{\delta_{a b}}{|x - y|^{2 \Delta_a}}.
\end{align}
On the grounds of conformal symmetry, we know that the one-loop one-point function of these operator in the defect CFT will be of the form
\begin{align}
 \langle \mathcal{O}_{\Delta(\lambda)} (x) \rangle
 = \frac{c}{x_3^{\Delta_0 + \gamma}} 
 = \frac{c}{x_3^{\Delta_0}} \Big( 1 + \gamma \log x_3 + \ldots \Big),
\end{align}
where $\Delta_0$ is the bare and $\gamma$ the anomalous conformal dimension of the operator.

\subsection{General single-trace operators}
\label{sec:general-one-point-functions}

We will consider a general single-trace operator built out of the scalars,
\begin{align}
 \label{eq:def-general-onepfun}
 \mathcal{O}(x) = 
 \mathcal{O}^{i_1 i_2 \ldots i_L} 
 \tr ( \phi_{i_1} \phi_{i_2} \ldots \phi_{i_L} ) (x),
\end{align}
which is required to have a well-defined scaling dimension. At leading order, this requires the operator $\mathcal{O}$ to be an eigenstate of the one-loop dilatation operator and hence the wave function $\mathcal{O}^{i_1 i_2 \ldots i_L}$ to be a solution of the one-loop Bethe ansatz \cite{Minahan:2002ve}.

We can evaluate the one-point function of this operator at tree level by inserting the classical solution~\eqref{eq:classical-solution-so3-so3} for the fields $\phi_i$:
\begin{align}
 \label{eq:def-tree-onepfun}
 \langle \mathcal{O} \rangle_{\text{tree}}(x) =
 \mathcal{O}^{i_1 i_2 \ldots i_L} 
 \tr ( \pcl_{i_1} \pcl_{i_2} \ldots \pcl_{i_L} ) (x). 
\end{align}
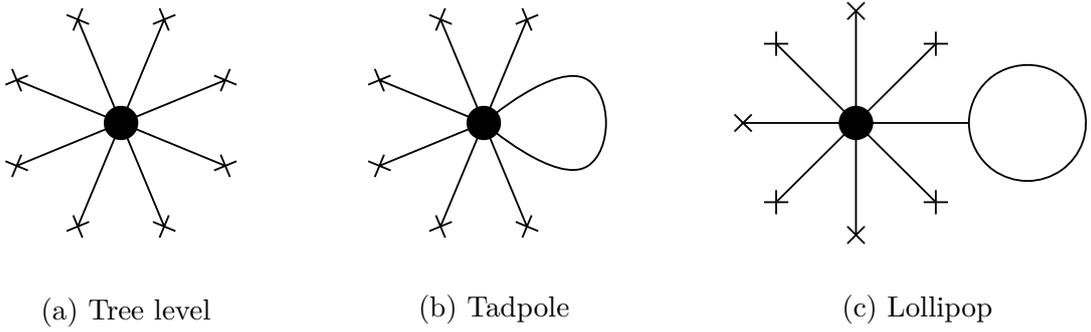
\begin{figure}[!tbp]
 \centering
 \begin{minipage}[b]{0.99\textwidth}
 \def\svgwidth{\textwidth}
 %% Creator: Inkscape inkscape 0.92.2, www.inkscape.org
%% PDF/EPS/PS + LaTeX output extension by Johan Engelen, 2010
%% Accompanies image file 'feynman_diagrams.pdf' (pdf, eps, ps)
%%
%% To include the image in your LaTeX document, write
%%   \input{<filename>.pdf_tex}
%%  instead of
%%   \includegraphics{<filename>.pdf}
%% To scale the image, write
%%   \def\svgwidth{<desired width>}
%%   \input{<filename>.pdf_tex}
%%  instead of
%%   \includegraphics[width=<desired width>]{<filename>.pdf}
%%
%% Images with a different path to the parent latex file can
%% be accessed with the `import' package (which may need to be
%% installed) using
%%   \usepackage{import}
%% in the preamble, and then including the image with
%%   \import{<path to file>}{<filename>.pdf_tex}
%% Alternatively, one can specify
%%   \graphicspath{{<path to file>/}}
%% 
%% For more information, please see info/svg-inkscape on CTAN:
%%   http://tug.ctan.org/tex-archive/info/svg-inkscape
%%
\begingroup%
  \makeatletter%
  \providecommand\color[2][]{%
    \errmessage{(Inkscape) Color is used for the text in Inkscape, but the package 'color.sty' is not loaded}%
    \renewcommand\color[2][]{}%
  }%
  \providecommand\transparent[1]{%
    \errmessage{(Inkscape) Transparency is used (non-zero) for the text in Inkscape, but the package 'transparent.sty' is not loaded}%
    \renewcommand\transparent[1]{}%
  }%
  \providecommand\rotatebox[2]{#2}%
  \newcommand*\fsize{\dimexpr\f@size pt\relax}%
  \newcommand*\lineheight[1]{\fontsize{\fsize}{#1\fsize}\selectfont}%
  \ifx\svgwidth\undefined%
    \setlength{\unitlength}{461.56181525bp}%
    \ifx\svgscale\undefined%
      \relax%
    \else%
      \setlength{\unitlength}{\unitlength * \real{\svgscale}}%
    \fi%
  \else%
    \setlength{\unitlength}{\svgwidth}%
  \fi%
  \global\let\svgwidth\undefined%
  \global\let\svgscale\undefined%
  \makeatother%
  \begin{picture}(1,0.33306682)%
    \lineheight{1}%
    \setlength\tabcolsep{0pt}%
    \put(0,0){\includegraphics[width=\unitlength,page=1]{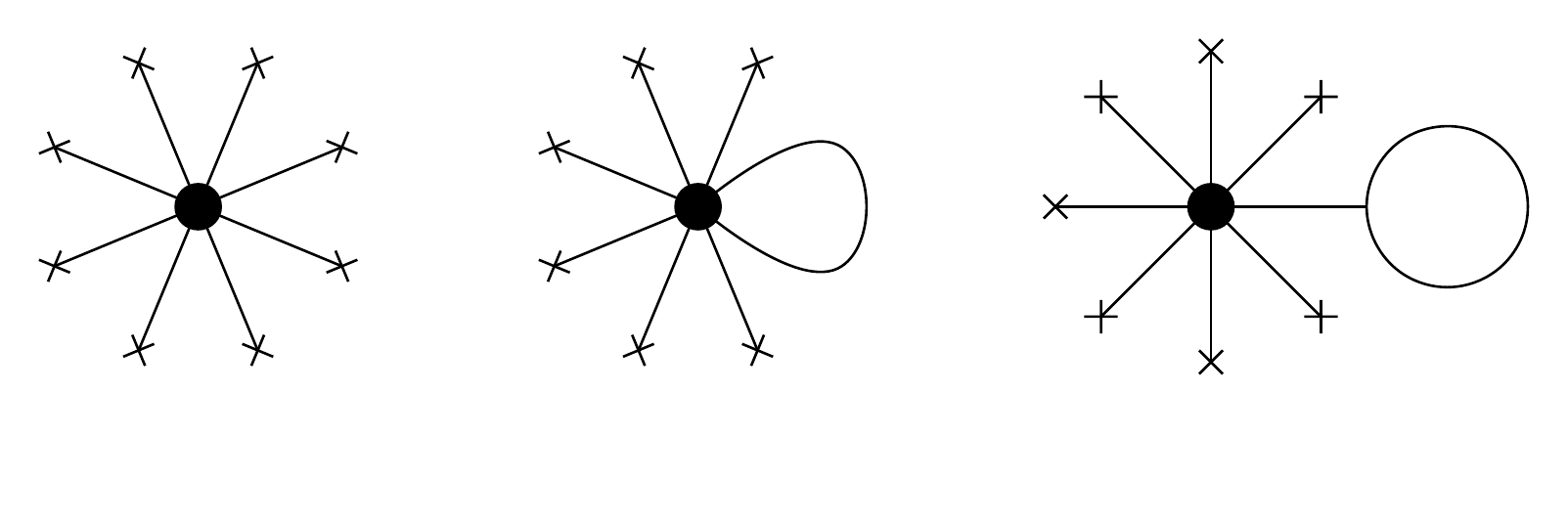}}%
    \put(0.05616155,0.02742199){\color[rgb]{0,0,0}\makebox(0,0)[lt]{\lineheight{1.25}\smash{\begin{tabular}[t]{l}(a) Tree level\end{tabular}}}}%
    \put(0.38783832,0.0295219){\color[rgb]{0,0,0}\makebox(0,0)[lt]{\lineheight{1.25}\smash{\begin{tabular}[t]{l}(b) Tadpole\end{tabular}}}}%
    \put(0.76020815,0.0295219){\color[rgb]{0,0,0}\makebox(0,0)[lt]{\lineheight{1.25}\smash{\begin{tabular}[t]{l}(c) Lollipop\end{tabular}}}}%
  \end{picture}%
\endgroup%

 \end{minipage}
 \caption{Diagrams that contribute at tree level (a) and one-loop order (b)-(c) to a single-trace operator such as $\langle \tr Z^L \rangle_{L=8}$ (in the planar limit).
 The black dot denotes the operator and the crosses signify the insertion of the classical solution.
 }
 \label{fig:feynman-diagrams-one-loop-order}
\end{figure}
At one-loop level, there are two diagrams that contribute to the one-point function, see Figure~\ref{fig:feynman-diagrams-one-loop-order}. Following \cite{Buhl-Mortensen:2016pxs,Buhl-Mortensen:2016jqo}, we will call them lollipop and tadpole diagram.

The lollipop diagram is one-particle reducible and describes the one-loop correction to the classical solution. 
Its contribution is obtained by considering all fields $\phi_i$ at their classical value $\pcl_i$, except for the one at position $i_j$, which is replaced by its one-loop correction. We then sum for all possible values of $j = 1, \ldots L$,
\begin{align}
 \label{eq:def-lol-onepfun}
 \langle \mathcal{O} \rangle_{\text{lol}} (x) = 
 \mathcal{O}^{i_1 i_2 \ldots i_L} 
 \sum_{j = 1}^L
 \tr ( \pcl_{i_1} 
       \ldots \langle \phi_{i_j} \rangle_{\text{1-loop}} \ldots 
       \pcl_{i_L} ) (x).
\end{align}
For a particular $\mathcal O$, this diagram can be evaluated using the correction to the vevs~\eqref{eq:correction-vevs} which we have calculated in the previous section. 

The tadpole diagram is obtained by expanding the fields around the classical solution as $\phi_i = \pcl_i + \pt_i$, and keeping only the quadratic terms in the quantum part $\pt_i$.
The two quantum fields in a particular term of this sum must be Wick contracted, and one obtains
\begin{align}
 \label{eq:def-tadpole-onepfun}
 \begin{aligned}
    \langle \mathcal{O} \rangle_{\text{tad}} (x) 
  & = 
    \sum_{j_1, j_2 = 1}^L
    \mathcal{O}^{i_1 \ldots i_{j_1} \ldots i_{j_2} \ldots i_L} 
    \tr ( \pcl_{i_1} \ldots 
    \contraction{}{\pt}{_{i_{j_1}} \ldots }{\pt}
    \pt_{i_{j_1}} \ldots \pt_{i_{j_2}}
    \ldots \pcl_{i_L} ) (x) \\
  & =
    \sum_{j = 1}^L
    \mathcal{O}^{i_1 \ldots i_{j} i_{j+1} \ldots i_L} 
    \tr ( \pcl_{i_1} \ldots 
          E^n_{\phantom{n}a} E^a_{\phantom{a}n'} 
          \ldots \pcl_{i_L} )
    \langle [\pt_{i_j}]_{n,a} [\pt_{i_{j+1}}]_{a,n'} \rangle.
 \end{aligned}
\end{align}
In the second line, we have used that in the large-$N$ limit only contractions from neighboring fields contribute. 
Moreover, propagators between fields in the off-diagonal block scale like $N-k_1 k_2 \simeq N$, whereas propagators from the $k_1 k_2 \times k_1 k_2$ block would scale like $k_1 k_2 \ll N$, so we are only keeping the former.
One can a priori calculate this diagram for any particular operator $\mathcal{O}$ by using the propagators in~\eqref{eq:propagators-offdiagonal-1} and~\eqref{eq:propagators-offdiagonal-4}.

The one-point function of a general operator $\mathcal{O}$ can receive two additional corrections at one-loop order.
If the contribution from the tadpole diagram in~\eqref{eq:def-tadpole-onepfun} is UV-divergent, the divergence has to be canceled by the renormalization constant $\mathcal{Z} = 1 + \mathcal{Z}_{\text{1-loop}} + \mathcal{O}(\lambda^2)$.
At one-loop order, the corresponding correction to $\langle \mathcal{O} \rangle$ is
\begin{align}
 \label{eq:def-renormalization-onepfun}
 \langle \mathcal{O} \rangle_{\text{1-loop},\mathcal{Z}}(x)
 =
 \langle \mathcal{Z}_{\text{1-loop}} \mathcal{O} \rangle_{\text{tree}}(x).
\end{align}
The second additional correction to $\langle \mathcal{O} \rangle$ arises from the first quantum correction to the wave function $\mathcal{O}^{i_1 i_2 \ldots i_L}$ of the operator.
Since we are considering operators with well-defined conformal dimension at one-loop level, $\mathcal{O}^{i_1 i_2 \ldots i_L}$ is already a one-loop eigenstate found by diagonalizing the one-loop dilatation operator.
The first quantum correction therefore comes from the two-loop eigenstate $\mathcal{O}^{i_1 i_2 \ldots i_L}_{\text{2-loop}}$,
\begin{align}
 \label{eq:def-twoloopeigenstate-onepfun}
 \langle \mathcal{O} \rangle_{\text{1-loop},\mathcal{O}}(x)
 =
 \mathcal{O}^{i_1 i_2 \ldots i_L}_{\text{2-loop}} \tr (\pcl_{i_1} \pcl_{i_2} \cdots \pcl_{i_L})(x).
\end{align}

Thus, the one-loop one-point function of a generic single-trace operator is
\begin{align}
 \langle \mathcal{O} \rangle_{\text{1-loop}}(x)
 =
 \langle \mathcal{O} \rangle_{\text{lol}}(x) +
 \langle \mathcal{O} \rangle_{\text{tad}}(x) +
 \langle \mathcal{O} \rangle_{\text{1-loop},\mathcal{Z}}(x) +
 \langle \mathcal{O} \rangle_{\text{1-loop},\mathcal{O}}(x).
\end{align}
Finally, we note that the planar one-point function of a multi-trace operator is given by the product of the one-point functions of its single-trace factors.

\subsection{One-loop one-point function of \texorpdfstring{$\tr Z^L$}{tr Z**L}}
\label{sec:one-loop-trZL}

We will now particularize the results from the previous subsection for the 1/2-BPS operator $\mathcal O = \tr Z^L$, where ${Z = \phi_3 + i \phi_6}$.
The tree-level one-point function of $\tr Z^L$ is obtained by replacing all fields by their classical value:
\begin{align}
 \label{eq:trZL-tree-result}
 \langle \, \tr Z^L \, \rangle_{\text{tree}}
 = \tr \left[ (Z^{\mathrm{cl}})^L \right]
 \simeq
   \frac{(-i)^L (k_1^2 + k_2^2 )^{\frac{L}{2}+1} 
         \sin \left[ (L+2) \psi_0 \right]}
        {2^{L} x_3^L (L + 1)(L + 2)}.
\end{align}
This and other color traces have been collected in Appendix~\ref{sec:color-traces-so3-so3}.
In the above equation, we have defined the angle $\psi_0 = \arctan (k_1 / k_2)$.
Moreover, the symbol $\simeq$ is used here and in what follows to indicate that we are only keeping the leading-order term in the limit where $k_1$ and $k_2$ are large.
The result vanishes unless $L$ is even, so this will be implicitly assumed in the following discussion.

Now we proceed to study the one-point function of $\tr Z^L$ beyond tree level.
Since the operator $\tr Z^L$ is 1/2-BPS, in the theory without the defect it is protected from quantum corrections; therefore,
$\langle \mathcal{O} \rangle_{\text{1-loop},\mathcal{Z}}(x) =0$ and $ \langle \mathcal{O} \rangle_{\text{1-loop},\mathcal{O}}(x) = 0$.
However, for the latter statement to be true, we must use a renormalization scheme that preserves the supersymmetry of the theory without the defect, and therefore it is required that we use dimensional reduction in our calculation.
We conclude that if we use dimensional reduction, only the lollipop and tadpole diagrams contribute at one-loop order,
\begin{align}
 \label{eq:lol-tad-trZL}
 \langle \, \tr Z^L \, \rangle_{\text{lol}}
 = L \, \tr \left[ (Z^{\mathrm{cl}})^{L-1} 
              \langle Z \rangle_{\text{1-loop}} \right], \quad
 \langle \, \tr Z^L \, \rangle_{\text{tad}}
 = L \, \tr \big[ (Z^{\mathrm{cl}})^{L-2}
              \contraction{}{Z}{}{Z} Z Z \big].
\end{align}
In the remainder of this section, we will evaluate these two diagrams.

To calculate the lollipop diagram, we use~\eqref{eq:lol-tad-trZL} and the one-loop correction to the vevs~\eqref{eq:correction-vevs-dslimit}:
\begin{align}
 \label{eq:trZL-lollipop-result}
 \begin{split}
 \langle \, \tr Z^L \, \rangle_{\text{lol}}
 & \simeq
 \frac{\lambda L}{2 \pi^2 x_3 (k_1^2 + k_2^2)^3} \Big( 
   k_2^4 \, \tr \left[ (Z^{\mathrm{cl}})^{L-1} \, t^{k_1}_3 \otimes \mathds{1}_{k_2} \right] +
   i \,
   k_1^4 \, \tr \left[ (Z^{\mathrm{cl}})^{L-1} \, \mathds{1}_{k_1} \otimes t^{k_2}_3 \right]
 \Big) \\
 & \simeq
 \frac{\lambda (-i)^L (k_1^2 + k_2^2)^{\frac{L}{2} - 3}}
      {2^{L+1}\pi^2(L+1)(L+2) x_3^L}
 \Big(
   (k_2^2 - k_1^2)\left( k_1^4 + k_2^4 + (k_1 k_2)^2 (L + 2) \right) \sin( L \psi_0 )
 \\ & \hspace{7.4cm}
   - k_1 k_2 (k_1^4 + k_2^4) L \cos( L \psi_0 )
 \Big).
 \end{split}
\end{align}
In the second line, we have used~\eqref{eq:color-traces-one-t} in Appendix~\ref{sec:color-traces-so3-so3} to compute the color traces. 

Finally, the contribution from the tadpole diagram~\eqref{eq:lol-tad-trZL} is
\begin{align}
\label{eq:tadpole-more-compact}
\begin{split}
 \langle \tr Z^L \, \rangle_{\mathrm{tad}}
  =
 N L \Bigg(&
   \tr \left[ (Z^{\mathrm{cl}})^{L-2} \mathds{1}_{k_1} \otimes \left( t_3^{k_2} \right)^2 \right]
   K^{\phi,(2)}_{\mathrm{sym}}
 - \tr \left[ (Z^{\mathrm{cl}})^{L-2} \left( t_3^{k_1} \right)^2 \otimes \mathds{1}_{k_2} \right]
   K^{\phi,(1)}_{\mathrm{sym}} \\
 &
 + \tr \left[ (Z^{\mathrm{cl}})^{L-2} \right]
   \left( K^{\phi,(1)}_{\mathrm{sing}} - K^{\phi,(2)}_{\mathrm{sing}} \right)
 + 2 i \, 
   \tr \left[ (Z^{\mathrm{cl}})^{L-2} t_3^{k_1} \otimes t_3^{k_2} \right]
   K^{\phi}_{\mathrm{opp}} 
 \Bigg),
\end{split}
\end{align}
where we have used the propagators~\eqref{eq:propagator-phis-different-sectors} and~\eqref{eq:propagator-phis-same-sector}.
We can expand this expression in the limit where $k_1$ and $k_2$ are large, which combined with the color traces in Appendix~\ref{sec:color-traces-so3-so3} gives
\begin{align}
 \label{eq:trZL-tadpole-result}
 \langle \, \tr Z^L \rangle_{\text{tad}}
 \simeq \frac{\lambda L (-i)^L (k_1^2 + k_2^2)^{\frac{L}{2} - 1}}
        {2^{L+2} \pi^2 (L - 1) (L + 2) x_3^L}
 \Big[
   2 k_1 k_2 \cos(L \psi_0) - (k_1^2 - k_2^2) \sin(L \psi_0)
 \Big].
\end{align}
Notice that the tadpole diagram does not depend on the regulator $\epsilon$ from dimensional regularization.
In fact, even though~\eqref{eq:trZL-tadpole-result} is applicable only in the double-scaling limit, the regulator drops from the tadpole diagram even for finite $k_1$ and $k_2$.
This is an important consistency check;
since $\tr Z^L$ is a 1/2-BPS operator, it should not be renormalized, so we should not find any UV-divergences and the terms proportional to $\tfrac{1}{\epsilon}$ should cancel.

We can combine the tree-level result~\eqref{eq:trZL-tree-result}, the lollipop diagram~\eqref{eq:trZL-lollipop-result} and the tadpole diagram~\eqref{eq:trZL-tadpole-result} to obtain
\begin{align}
 \label{eq:trZL-ratio-field-theory}
  \frac{\langle \tr Z^L \rangle}{\langle \tr Z^L \rangle_{\text{tree}}} 
  = 1 & + \frac{\lambda}{4 \pi^2 (L-1) \left( k_1^2 + k_2^2 \right)^3} \Bigg(
    4 (k_1 k_2)^2 + (L^2 + 3L -2) \left( k_1^4 + k_2^4 \right) \\
    & 
    + 2 (L-1) (L+2) k_1 k_2 \left( k_1^2 - k_2^2 \right) \cot[(L+2)\psi_0]
  \Bigg)
  + \mathcal{O} \left( \frac{\lambda^2}{(k_1^2 + k_2^2)^2} \right).
\nonumber
\end{align}
Note that the result has indeed an expansion in the parameter $\tfrac{\lambda}{(k_1^2 + k_2^2)}$ as suggested by the string-theory dual of the defect CFT.
Moreover, the result \eqref{eq:trZL-ratio-field-theory} precisely agrees with the supergravity prediction \eqref{eq:trZL-string-theory} quoted in the introduction!\footnote{%
To be precise, the supergravity prediction is for the unique $SO(3)\times SO(3)$-symmetric chiral primary operator built from $L$ scalar fields \cite{Kristjansen:2012tn}; while this operator is not equal to $\tr Z^L$, $\tr Z^L$ has a non-vanishing projection on it (induced by the norm from the two-point function far away from the defect), such that the ratio of the one-point function and the tree-level one-point function of both operators coincide.}

\section{Outlook}
\label{sec:conclusion}
While the main result of the present paper is a highly non-trivial positive test of AdS/dCFT for a configuration where supersymmetry is completely broken, an important accompanying achievement is the establishment of a perturbative framework for the $SO(3) \times SO(3)$-symmetric defect CFT involved. A crucial step of this achievement was of course the determination of the exact mass spectrum of the theory using fuzzy spherical harmonics, but an equally essential step was the rewriting of the resulting propagators of the theory in terms of
generators of $\su{2} \times \su{2}$. Worth stressing is also the recognition that dimensional reduction constitutes an appropriate regularization scheme being compatible with the supersymmetry of the underlying bulk CFT which governs the UV behavior of the defect CFT.
We have used our perturbative framework to calculate the one-loop correction to the classical solution in the planar limit and obtained an explicit result for the one-point function of $\tr Z^L$ in the double-scaling limit; in the future, it would be interesting to go to finite $N$ (following \cite{Buhl-Mortensen:2016jqo,Guo_2017}), to obtain explicit results at finite $k_1$ and $k_2$ for $\tr Z^L$ and to go to higher loop orders.
With the perturbative framework in place, the scene is also set for the calculation of quantum corrections to other quantities of interest in the defect CFT, such as other types of correlation functions or Wilson loops. 
In the case of the simpler D3-D5 probe-brane setup, the calculation of a simple Wilson line to one-loop order~\cite{deLeeuw:2016vgp} confirmed the prediction of  a classical string-theory calculation~\cite{Nagasaki:2011ue} consisting of evaluating the area of a minimal surface in the double-scaling limit~(\ref{dsl}).
The circular Wilson loop of the D3-D5 defect CFT was analyzed in~\cite{Aguilera-Damia:2016bqv} and the case of two anti-parallel Wilson lines was considered in a search for a Gross-Ooguri  transition in~\cite{Preti:2017fhw}. 
Finally, the calculation of two-point functions of the defect CFT allowed for data mining in ${\cal N}=4$ SYM theory by means of the boundary conformal bootstrap equations~\cite{deLeeuw:2017dkd}. A special class of two-point functions was considered
in~\cite{Widen:2017uwh}.
 
 In the case of the defect CFT based on the D3-D5 probe-brane setup, where only three of the scalars get non-trivial $SO(3)$-symmetric vevs,
 the one-point function problem showed very strong signs of integrability. Hence, it was possible to express the tree-level one-point function of any scalar operator in a closed formula valid for any value of the representation label $k$~\cite{deLeeuw:2018mkd}.  The formula could be extended
 to one-loop order in the $SU(2)$ sub-sector and a conjecture for an all-loop asymptotic formula for this sub-sector was put forward as well~\cite{Buhl-Mortensen:2017ind}, which extends the match with the supergravity prediction \cite{Nagasaki:2012re} for $\langle \tr Z^L\rangle$ in the double-scaling limit to all loop-orders smaller than $L$.
 The calculation of a
 tree-level one-point function can be formulated as the evaluation of the overlap between a Bethe state describing the operator in question and a so-called matrix product state~\cite{deLeeuw:2015hxa},
 and the apparent integrability of the 
 one-point function problem in the D3-D5 probe-brane set-up was suggested to be a consequence of the matrix product state being annihilated by all the odd charges of the integrable 
 spin chain underlying the spectrum of ${\cal N}=4$ SYM theory~\cite{Piroli:2017sei}. One can explicitly check that the matrix product state of relevance for the computation of one-point functions of the $SO(3)\times SO(3)$-symmetric defect CFT is {\it not}
annihilated by the odd charges of the ${\cal N}=4$ SYM spin chain~\cite{inprogress:2019}. In accordance with this, it has only been possible to derive results
for tree-level one-point functions of non-protected operators on a case by case basis~\cite{inprogress:2019}.  

On the other hand, one can prove that the matrix product state of relevance for the computation of the one-point functions of the earlier mentioned 
$SO(5)$-symmetric defect CFT based on the non-supersymmetric D3-D7 probe-brane system with probe geometry $AdS_4\times S^4$ is indeed annihilated by the odd charges of the ${\cal N}=4$ SYM spin chain~\cite{deLeeuw:2018mkd}.  Although
only a few exact tree-level results and in particular no closed formula exist so far~\cite{deLeeuw:2016ofj},  this observation  indicates that setting up
the perturbative program for the $SO(5)$-symmetric defect CFT could potentially be very rewarding.  We have already taken the
first step in this direction by explicitly determining the mass spectrum of the theory via a further generalization of the method
of fuzzy spherical harmonics~\cite{inprogress:2018}, and we hope to be able to report on the completion of the program in the near future.

\section*{Acknowledgments}

We thank M.\ de Leeuw, A.\ C.\ Ipsen, G.\ Semenoff, K.\ E.\ Vardinghus and K.\ Zarembo for useful discussions. 
C.K.\ and M.W.\ were  supported  in  part  by  FNU  through grant number  DFF-4002-00037. M.W.\ in addition was supported in part by an ERC Starting Grant \mbox{(No.\ 757978)} and a grant from the Villum Fonden.

\appendix

\section{Conventions}
\label{sec:conventions-general}

In this appendix, we summarize our conventions for field-theory calculations (Appendix~\ref{sec:other-conventions}) and fuzzy spherical harmonics (Appendix~\ref{sec:conventions-su2-and-Ylm}).

\subsection{Field-theory conventions}
\label{sec:other-conventions}

Throughout the paper, we choose the metric of Minkowski space to have mostly positive signature, i.e.\ $\eta^{\mu \nu} = \mathrm{diag}(-1, +1, \ldots, +1)$.
We will work in (3+1) dimensions, and we will denote by $d = 3$ the dimension of the codimension-one defect.
For the fermionic fields, we take the four-dimensional $\gamma$-matrices to be
\begin{align}
 \gamma^\mu =
 \begin{pmatrix}
  0 & \sigma^\mu \\
  \bar{\sigma}^\mu & 0
 \end{pmatrix}, \quad
 \gamma^5 = i \gamma^0 \gamma^1 \gamma^2 \gamma^3 =
 \begin{pmatrix}
  -\mathds{1}_2 & 0 \\
  0 & \mathds{1}_2
 \end{pmatrix},
\end{align}
with $\sigma^\mu = (\mathds{1}_2, \sigma^i)$, $\bar{\sigma}^\mu = (\mathds{1}_2, -\sigma^i)$ and $\{ \gamma^\mu, \gamma^\nu \} = -2 \eta^{\mu \nu}$.

For the four-dimensional matrices $G^i$ that appear in the reduction of the spinors in ten dimensions to four dimensions, we use the same conventions as in~\cite{Buhl-Mortensen:2016jqo}:
\begin{align}
\label{eq:def-dimred-g-matrices}
\begin{alignedat}{3}
 & G^1 \equiv G^{(1)}_1 = i
 \begin{pmatrix}
 0 & -\sigma_3 \\ \sigma_3 & 0
 \end{pmatrix},
 \;
 && G^2 \equiv G^{(1)}_2 = i
 \begin{pmatrix}
  0 & \sigma_1 \\ -\sigma_1 & 0
 \end{pmatrix},
 \;
 && G^3 \equiv G^{(1)}_3 =
 \begin{pmatrix}
  \sigma_2 & 0 \\ 0 & \sigma_2
 \end{pmatrix},
 \\
 & G^4 \equiv G^{(2)}_1 = i
 \begin{pmatrix}
  0 & -\sigma_2 \\ -\sigma_2 & 0
 \end{pmatrix},
 \;
 && G^5 \equiv G^{(2)}_2 =
 \begin{pmatrix}
  0 & -\mathds{1}_2 \\ \mathds{1}_2 & 0
 \end{pmatrix},
 \;
 && G^6 \equiv G^{(2)}_3 = i
 \begin{pmatrix}
  \sigma_2 & 0 \\ 0 & -\sigma_2
 \end{pmatrix}.
\end{alignedat}
\end{align}
The matrices in the first line are Hermitian, $(G^{(1)}_i)^\dagger = G^{(1)}_i$, while those in the second line are anti-Hermitian, $(G^{(2)}_i)^\dagger = -G^{(2)}_i$.
Their (anti-)commutation relations are
\begin{equation}
\begin{aligned}
   \left\{ G^{(1)}_i, G^{(1)}_j \right\} &= + 2 \delta_{ij}, \quad&&
\left[  G^{(1)}_i, G^{(1)}_j \right] 
    = - 2 i \epsilon_{ijk} G^{(1)}_k, \\ \left\{ G^{(2)}_i, G^{(2)}_j \right\} &= - 2 \delta_{ij},   
 && \left[  G^{(2)}_i, G^{(2)}_j \right] = - 2 \epsilon_{ijk} G^{(2)}_k.
\end{aligned}
\end{equation}
The two sets commute, $\left[ G^{(1)}_i, G^{(2)}_j \right] = 0$.

\subsection{Lie algebra \texorpdfstring{\su{2}}{su(2)} and fuzzy spherical harmonics}
\label{sec:conventions-su2-and-Ylm}

For the vevs with $SO(3) \times SO(3)$ symmetry, we will need explicit expressions for the generators $t_i$ of the corresponding Lie algebra as well as for the fuzzy spherical harmonics $\hat{Y}^m_{\ell}$ that serve as a basis for the fields in color space.
Those are given here using the same conventions as~\cite{Buhl-Mortensen:2016jqo}.

The basis matrices ${E^i}_j$ are defined to have a 1 at position $(i, j)$, i.e.\ $[{E^i}_j]_{m,n} = \delta_{i,m} \delta_{j,n}$.
We use the same form of the $k$-dimensional matrices $t_i$ of \su{2} that was used in~\cite{deLeeuw:2015hxa}, namely
\begin{align}
 t_+ = \sum_{n=1}^{k-1} c_{k,n} {E^n}_{n+1}, \quad
 t_- = \sum_{n=1}^{k-1} c_{k,n} {E^{n+1}}_n, \quad
 t_3 = \sum_{n=1}^{k} d_{k,n} {E^n}_n,
\end{align}
with the coefficients
\begin{align}
 c_{k,n} = \sqrt{ n (k - n) }, \quad
 d_{k,n} = \frac{1}{2} (k - 2n + 1).
\end{align}
Defining also $t_1 = \tfrac{1}{2} (t_+ + t_-)$ and $t_2 = \tfrac{1}{2i} (t_+ - t_-)$, these matrices satisfy the commutation relations of \su{2},
\begin{align}
 \left[ t_i, t_j \right] = i \epsilon_{ijk} t_k.
\end{align}

The $k$-dimensional matrices $t_i$ can be used to construct \su{2} representations $\hat{Y}^m_{\ell}$ of spin $\ell$, for $\ell = 0, 1, \ldots, k - 1$, cf.~\cite{deWit:1988wri,hoppePhdThesis:1982}.
The $k \times k$ matrices $\hat{Y}^m_\ell$ are essentially given by a symmetric and traceless polynomial of degree $\ell$ in the generators $t_i$,
\begin{align}
 \label{eq:definition-Ylm}
 \hat{Y}^m_{\ell} =
 2^{\ell} \sqrt{\frac{(k - \ell - 1)!}{(k + \ell)!}} \left( \frac{k^2 - 1}{4} \right)^{\ell/2}
 \sum_{i_1, \ldots, i_\ell} f^{\ell m}_{i_1, \ldots, i_\ell} \hat{x}_{i_1} \cdots \hat{x}_{i_{\ell}},
 \quad \ell = 1, \ldots, k - 1,
\end{align}
where the \su{2} generators have been rescaled to
\begin{align}
 \hat{x}_i = \sqrt{\frac{4}{k^2 - 1}} t_i
 \quad \Rightarrow \quad
 \sum_{i} \hat{x}_i \hat{x}_i = \mathds{1}_k,
\end{align}
and the coefficients $f^{\ell m}_{i_1, \ldots, i_\ell}$ implement the symmetry and tracelessness conditions.
Note that the last equation defines the fuzzy two-sphere with coordinates $\hat{x}_i$ and that the construction~\eqref{eq:definition-Ylm} stems from the observation that on a normal two-sphere a basis of functions can be constructed as a homogeneous polynomial in the Cartesian coordinates $x_i$, $i = 1, 2, 3$.
These functions are the well-known spherical harmonics $Y^m_{\ell}$.

We now give some properties of $\hat{Y}^m_{\ell}$ that are important for our purposes.
With the normalization as above, they satisfy
\begin{align}
 \left( \hat{Y}^m_{\ell} \right)^\dagger = (-1)^m \hat{Y}^{-m}_{\ell}
 \quad \text{and} \quad
 \tr \left[ \left( \hat{Y}^m_{\ell} \right)^\dagger \hat{Y}^{m'}_{\ell'} \right] = \delta_{\ell \ell'} \delta_{m m'}.
\end{align}
We also make use of the relation between the generators $t_i$ and $\hat{Y}^m_{\ell}$ for $\ell = 1$, namely
\begin{align}
 \label{eq:relation-t-Ylm-1}
 t_1 = c \left( \hat{Y}^{-1}_1 - \hat{Y}^1_1 \right), \quad
 t_2 = i c \left( \hat{Y}^{-1}_1 + \hat{Y}^1_1 \right), \quad
 t_3 = c \sqrt{2} \hat{Y}^0_1
\end{align}
with
\begin{align}
 \label{eq:relation-t-Ylm-2}
 c = \frac{(-1)^{k+1}}{2} \sqrt{\frac{k (k^2 - 1)}{6}}.
\end{align}

\section{Color and flavor part of the propagators}
\label{sec:appendix-colour-flavour-part}

In this appendix, we derive the propagators between the fields that originally appeared in the action of $\mathcal{N} = 4$ SYM theory.
We focus on the propagators involving the six scalars and the gauge field; the propagators involving the Majorana fermions can be obtained in a similar way.
To obtain the propagators, we will express the original fields in terms of the fields in which the mass term of the action becomes diagonal.
For example, for the complicated bosons with $\ell_1,\ell_2\neq0$, we have to undo the three steps of the diagonalization: the flavor transformation~\eqref{eq:flavour-transformation-vecC}, the Clebsch-Gordan procedure~\eqref{eq:eigenstate-j-mj-l-general-so3-so3} and the diagonalization of the final $3 \times 3$ matrix~\eqref{eq:diagonal-fields-3x3}.

After the flavor transformation, $S \cdot L$ is in the form~\eqref{eq:sdotl-full} and the transformed vector of complicated fields is
\begin{align}
 V^\dagger C =
 \begin{pmatrix}
  C^{(1)} \\ C^{(2)} \\ A_3
 \end{pmatrix},
\end{align}
where $C^{(1)}$ and $C^{(2)}$ were given in~\eqref{eq:flavour-transformation-vecC-1}.
In the $3 \times 3$ blocks $T_i L^{(1)}_i$ and $T_i L^{(2)}_i$, we diagonalize using Clebsch-Gordan coefficients and obtain the eigenstates $(B^{(1)})_{j_1,m_1, \ell_1; \ell_2, m_2}$ and $(B^{(2)})_{\ell_1, m_1; j_2, m_2, \ell_2}$.
The relation to the fields $C^{(a)}_\pm$ and $C^{(a)}_0$ with $a=1,2$ is
\begin{align}
  (C^{(a)}_\pm)_{\ell m} = \sum_{j} \langle \ell, m; 1, \pm 1 | j, m \pm 1 \rangle (B^{(a)})_{j, m \pm 1; \ell}, \;
  (C^{(a)}_0)_{\ell m} = \sum_{j} \langle \ell, m; 1, 0 | j, m  \rangle (B^{(a)})_{j, m; \ell}.
\end{align}
For $j_1 = \ell_1 \pm 1$ and $j_2 = \ell_2 \pm 1$, these fields diagonalize $S \cdot L$ and it only remains to diagonalize the $3 \times 3$ matrix in~\eqref{eq:mixing-a3-b}.
The fields $D_\pm$ and $D_0$ in which the mass term is diagonal were given in~\eqref{eq:diagonal-fields-3x3}.
Inverting this relation, we find
\begin{align}
  \label{eq:nondiagonal-fields-so3-so3}
  \begin{split}
  B_{0}^{(1)} &= 
   - \sqrt{\ell_2 (\ell_2 + 1)} \, \frac{D_0}{\sqrt{N_0}}
   -i \sqrt{\ell_1 (\ell_1 + 1)} \left( 
     \frac{D_+}{\sqrt{N_+}} + \frac{D_-}{\sqrt{N_-}}
   \right), \\
  B_{0}^{(2)} &= 
   + \sqrt{\ell_1 (\ell_1 + 1)}\frac{D_0}{\sqrt{N_0}} 
   -i \sqrt{\ell_2 (\ell_2 + 1)} \left( 
     \frac{D_+}{\sqrt{N_+}} + \frac{D_-}{\sqrt{N_-}}
  \right), \\
  A_3 &= 
      \frac{\lambda_-}{\sqrt{N_+}} D_+ 
    + \frac{\lambda_+}{\sqrt{N_-}} D_-.
  \end{split}
\end{align}

We begin with the propagators between scalars from different sectors and those involving $A_3$ using the notation described in Section~\ref{sec:propagators-colour-flavour}.
They contain at most one \su{2} Clebsch-Gordan coefficient from each sector, which we can express as the matrix element of an \su{2} generator $t_i$.
In particular, we do not yet encounter products of \su{2} generators unlike in the propagators for scalars from the same sector.
For convenience, we define
\begin{align}
 [r^\ell_s]_{m,m'} \equiv
 \sqrt{\ell(\ell + 1)} \langle \ell, m ; 1, s | \ell, m + s \rangle \delta_{m',m+s}, 
\end{align}
for $s = -1, 0, 1$.
One can check that $r_\pm = \mp t_\mp / \sqrt{2}$, $r_0 = t_3$, $r_s^\dag = r_{-s}$ and finally
\begin{align}
 [(r^\ell_s)^\dag]_{m,m'} =
 \sqrt{\ell(\ell + 1)} \langle \ell, m-s ; 1, s | \ell, m \rangle \delta_{m',m-s}.
\end{align}
Using this notation, it will be easier to keep track of factors $\pm 1/\sqrt{2}$.
The propagators involving $A_3$ are
\begin{align}
  \begin{split}
  \langle (C^{(1)}_{s})_{\ell_1 m_1; \ell_2 m_2} (A_3)^\dagger_{\ell_1' m_1'; \ell_2' m_2'} \rangle 
   & = - i \delta_{\ell_1 \ell_1'} \delta_{\ell_2 \ell_2'} \delta_{m_2 m_2'} [r^{\ell_1}_s]_{m_1,m_1'}
   \left( \frac{\lambda_+}{N_-} K^{m^2_-} + \frac{\lambda_-}{N_+} K^{m^2_+} \right), \\
   \langle (A_3)_{\ell_1 m_1; \ell_2 m_2} (C^{(1)}_{s})^\dagger_{\ell_1' m_1'; \ell_2' m_2'} \rangle 
   & = i \delta_{\ell_1 \ell_1'} \delta_{\ell_2 \ell_2'} \delta_{m_2 m_2'} [(r^{\ell_1}_s)^\dag]_{m_1,m_1'} 
   \left( \frac{\lambda_+}{N_-} K^{m^2_-} + \frac{\lambda_-}{N_+} K^{m^2_+} \right), \\
  \langle (A_3)_{\ell_1 m_1; \ell_2 m_2} (A_3)^\dagger_{\ell_1' m_1'; \ell_2' m_2'} \rangle &=
  \delta_{\ell_1 \ell_1'} \delta_{\ell_2 \ell_2'} \delta_{m_1 m_1'} \delta_{m_2 m_2'}
  \left( \frac{\lambda_+^2}{N_-} K^{m^2_-} + \frac{\lambda_-^2}{N_+} K^{m^2_+} \right).
  \end{split}
\end{align}
To obtain the same propagators for $C^{(2)}_{s}$, we simply relabel as in~\eqref{eq:prop-relabeling}.
For the propagators that mix the two blocks, we need
\begin{align}
 \begin{split}
  \langle (B_{0}^{(1)})_{\ell_1 m_1; \ell_2 m_2} (B_{0}^{(2)})^\dagger_{\ell_1' m_1'; \ell_2' m_2'} \rangle
  &=
  \delta_{\ell_1 \ell_1'} \delta_{\ell_2 \ell_2'} \delta_{m_1 m_1'} \delta_{m_2 m_2'} \\
  &\quad \times
  \sqrt{\ell_1 (\ell_1 + 1)} \sqrt{\ell_2 (\ell_2 + 1)} 
  \left( \frac{K^{m^2_-}}{N_-} + \frac{K^{m^2_+}}{N_+} - \frac{K^{m^2_0}}{N_0} \right),
 \end{split}
\end{align}
and we obtain
\begin{equation}
\begin{aligned}
  \langle (C^{(1)}_{s})_{\ell_1 m_1; \ell_2 m_2} (C^{(2)}_{s'})^\dagger_{\ell_1' m_1'; \ell_2' m_2'} \rangle
  &= \delta_{\ell_1 \ell_1'} \delta_{\ell_2 \ell_2'} 
  [r^{\ell_1}_s]_{m_1,m_1'} [(r^{\ell_2}_{s'})^\dag]_{m_2,m_2'} \\
&\phaneq\times  \left( \frac{K^{m^2_-}}{N_-} + \frac{K^{m^2_+}}{N_+} - \frac{K^{m^2_0}}{N_0} \right).
\end{aligned}
 \end{equation}
Converting to the fields $\phi_i$ is a matter of undoing the flavor transformation,
\begin{align}
  \langle (\pt^{(1)}_i)_{\ell_1 m_1; \ell_2 m_2} (\pt^{(2)}_j)^\dagger_{\ell_1' m_1'; \ell_2' m_2'} \rangle
  &= \delta_{\ell_1 \ell_1'} \delta_{\ell_2 \ell_2'} 
  [t^{(\ell_1)}_i]_{m_1,m_1'} [t^{(\ell_2)}_j]_{m_2,m_2'} 
  \left( \frac{K^{m^2_-}}{N_-} + \frac{K^{m^2_+}}{N_+} - \frac{K^{m^2_0}}{N_0} \right), \\
  \begin{split}
  \langle (\pt^{(1)}_i)_{\ell_1 m_1; \ell_2 m_2} (A_3)^\dagger_{\ell_1' m_1'; \ell_2' m_2'} \rangle
  &= - \langle (A_3)_{\ell_1 m_1; \ell_2 m_2} (\pt^{(1)}_i)^\dagger_{\ell_1' m_1'; \ell_2' m_2'} \rangle \\
  &= -i \delta_{\ell_1 \ell_1'} \delta_{\ell_2 \ell_2'} [t^{(\ell_1)}_i]_{m_1 m_1'} \delta_{m_2 m_2'}
  \left(\frac{\lambda_+}{N_-} K^{m^2_-} + \frac{\lambda_-}{N_+} K^{m^2_+} \right),
  \end{split}
\end{align}
with $\pt^{(1)}_i \equiv \pt_i$ and $\pt^{(2)}_i \equiv \pt_{i+3}$.
We obtain the analogue of the last equation for the second sector by relabeling as in~\eqref{eq:prop-relabeling}.

As anticipated, the propagators between scalars from the same sector contain products of Clebsch-Gordan coefficients and are therefore more involved.
For simplicity let us focus on one sector, say the first one for concreteness.
We define the combination $K_{(1)}^0$ as
\begin{align}
 \label{eq:propagators-K0(1)-def}
 \begin{alignedat}{1}
  \langle (B_{0}^{(1)})_{\ell_1 m_1; \ell_2 m_2} (B_{0}^{(1)})^\dagger_{\ell_1' m_1'; \ell_2' m_2'} \rangle
  &=
  \delta_{\ell_1 \ell_1'} \delta_{\ell_2 \ell_2'} \delta_{m_1 m_1'} \delta_{m_2 m_2'} \\
  & \quad \times
  \underbrace{\Big[
  \frac{\ell_2 (\ell_2 + 1)}{N_0} K^{m_0^2} + \ell_1 (\ell_1 + 1) \left( \frac{K^{m_-^2}}{N_-} + \frac{K^{m_+^2}}{N_+} \right)
  \Big]}_{\equiv K_{(1)}^0}.
 \end{alignedat}
\end{align}
The propagators with $C^{(1)}_{0}$ are
\begin{align}
  \begin{split}
    &\langle (C^{(1)}_{\pm})_{\ell_1 m_1; \ell_2 m_2} (C^{(1)}_{0})_{\ell_1' m_1'; \ell_2' m_2'}^\dagger \rangle
    = \delta_{\ell_1 \ell_1'}\delta_{\ell_2 \ell_2'} \frac{[t^{(\ell_1)}_\mp]_{m_1, m_1'}}{\sqrt{2}}\delta_{m_2 m_2'} \\
    & \; \times
    \Bigg(
      - \frac{\ell_1 \mp m_1 - 1}{(2 \ell_1 + 1) \ell_1} K^{m^2_{{(1)},-}}
      + \frac{\ell_1 \pm m_1 + 2}{(2 \ell_1 + 1) (\ell_1 + 1)} K^{m^2_{{(1)},+}}
      + \frac{\mp m_1 - 1}{\ell_1 (\ell_1 + 1)} K_{(1)}^0
      \Bigg), \\
    &\langle (C^{(1)}_{0})_{\ell_1 m_1; \ell_2 m_2} (C^{(1)}_{0})_{\ell_1' m_1'; \ell_2' m_2'}^\dagger \rangle
    = \delta_{\ell_1 \ell_1'} \delta_{\ell_2 \ell_2'}\delta_{m_1 m_1'} \delta_{m_2 m_2'}\\
    & \; \times
    \Bigg(
      \frac{(\ell_1 - m_1 + 1) (\ell_1 + m_1 + 1)}{(2 \ell_1 + 1) (\ell_1 + 1)} K^{m^2_{{(1)},+}}
      + \frac{(\ell_1 - m_1) (\ell_1 + m_1)}{(2 \ell_1 + 1) \ell_1} K^{m^2_{{(1)},-}}
      + \frac{m_1^2}{\ell_1 (\ell_1 + 1)} K_{(1)}^0
      \Bigg).
  \end{split}
\end{align}
The propagators between $C^{(1)}_{\pm}$ are
\begin{align}
  \begin{split}
  &\langle (C^{(1)}_{\pm})_{\ell_1 m_1; \ell_2 m_2} (C^{(1)}_{\pm})_{\ell_1' m_1'; \ell_2' m_2'} \rangle
  = \frac{1}{2} \delta_{\ell_1 \ell_1'}\delta_{\ell_2 \ell_2'}\delta_{m_2 m_2'}
  \Bigg[
    \frac{[t^{(\ell_1)}_\mp t^{(\ell_1)}_\pm]_{m_1, m_1'}}{\ell_1 (\ell_1 + 1)} K_{(1)}^0 \\
    & \qquad
    + \delta_{m_1 m_1'} \Bigg(
    \frac{(\ell_1 \mp m_1) (\ell_1 \mp m_1 - 1)}{(2 \ell_1 + 1) \ell_1} K^{m^2_{{(1)},-}} +
    \frac{(\ell_1 \pm m_1 + 1) (\ell_1 \pm m_1 + 2)}{(2 \ell_1 + 1) (\ell_1 + 1)} K^{m^2_{{(1)},+}}
    \Bigg)
    \Bigg], \\
  &\langle (C^{(1)}_{\pm})_{\ell_1 m_1; \ell_2 m_2} (C^{(1)}_{\mp})_{\ell_1' m_1'; \ell_2' m_2'} \rangle
  = \frac{1}{2} \delta_{\ell_1 \ell_1'}\delta_{\ell_2 \ell_2'} [t^{(\ell_1)}_\mp t^{(\ell_1)}_\mp]_{m_1, m_1'}\delta_{m_2 m_2'} \\
  & \qquad \times
  \left(
    \frac{K^{m^2_{{(1)},-}}}{(2 \ell_1 + 1) \ell_1} -
    \frac{K^{m^2_{{(1)},0}}}{\ell_1 (\ell_1 + 1)} + 
    \frac{K^{m^2_{{(1)},+}}}{(2 \ell_1 + 1) (\ell_1 +1)}
    \right).
  \end{split}
\end{align}
Undoing the flavor transformation and inserting $K^0_{(1)}$ from~\eqref{eq:propagators-K0(1)-def}, we find that the propagator between two scalars from the same sector is
\begin{align}
  \begin{split}
    &\langle (\pt^{(1)}_i)_{\ell_1 m_1; \ell_2 m_2} (\pt^{(1)}_j)^\dagger_{\ell_1' m_1'; \ell_2' m_2'} \rangle
    = \delta_{\ell_1 \ell_1'} \delta_{\ell_2 \ell_2'} \delta_{m_2 m_2'}
     \\
      & \bigg[
      \delta_{ij} \delta_{m_1 m_1'}
      \bigg(
        \frac{\ell_1 + 1}{2 \ell_1 + 1} K^{m^2_{{(1)},+}} + \frac{\ell_1}{2 \ell_1 + 1} K^{m^2_{{(1)},-}}
      \bigg)
      - i \epsilon_{ijk} [t^{(\ell_1)}_k]_{m_1, m_1'}
      \bigg( \frac{K^{m^2_{{(1)},+}}}{2 \ell_1 + 1} - \frac{K^{m^2_{{(1)},-}}}{2 \ell_1 + 1} \bigg)
      \\
      &\hphantom{\bigg[} 
      - [t^{(\ell_1)}_i t^{(\ell_1)}_j]_{m_1, m_1'}
      \bigg(
        \frac{K^{m^2_{{(1)},+}}}{(2 \ell_1 + 1)(\ell_1 + 1)} + \frac{K^{m^2_{{(1)},-}}}{(2 \ell_1 + 1) \ell_1}
        - \frac{\ell_2 (\ell_2 + 1)}{\ell_1 (\ell_1 + 1)} \frac{K^{m^2_0}}{N_0} - \frac{K^{m^2_-}}{N_-} - \frac{K^{m^2_+}}{N_+}
      \bigg)
    \bigg],
  \end{split}
\end{align}
with an analogous expression for the other sector obtained by relabeling as in~\eqref{eq:prop-relabeling}.
We note that the terms with $\delta_{ij}$ and $\epsilon_{ijk}$ are the same as in~\cite{Buhl-Mortensen:2016jqo} and that the last one would vanish in the setup of that reference.

\section{One-loop correction to the scalar vacuum expectation values}
\label{sec:one-loop-vevs-appendix}

In this appendix, we present in detail the calculation of the correction to the scalar vevs summarized in Section~\ref{sec:one-loop-vevs}.
We split the calculation in three parts: we obtain the effective vertex $V_{\text{eff}}$ in Section~\ref{sec:effective-vertex}, the contraction of the vertex with the external field is computed in Section~\ref{sec:stick-lollipop} and finally the remaining spacetime integral is performed in Section~\ref{sec:spacetime-integral}.

\subsection{Calculation of the effective vertex}
\label{sec:effective-vertex}

To compute the one-loop correction to the vevs of the scalars, we will need to know the effective one-particle vertex defined by
\begin{align}
 % Contraction
 \contraction{
 V_{\text{eff}}(y)
 \equiv
 \sum_{\Phi_1, \Phi_2, \Phi_3} 
 V_3(\Phi_1(y), }{\Phi}{_2(y), }{\Phi}
 % Equation
 V_{\text{eff}}(y)
 \equiv
 \sum_{\Phi_1, \Phi_2, \Phi_3} 
 V_3(\Phi_1(y), \Phi_2(y), \Phi_3(y)),
\end{align}
where the sum is carried over all inequivalent contractions of cubic vertices in~\eqref{eq:action-cubic-vertices}.
We will start by calculating all the contractions assuming the limit $N \to \infty$, but keeping $k_1$ and $k_2$ finite.
We will continue to use equal signs in equations where the large-$N$ limit has been used.
Then we will collect all contributions, and show that the regulator $\epsilon$ drops out.

The calculation of the contractions proceeds identically to~\cite{Buhl-Mortensen:2016jqo}, but the propagators are different in the two setups. In this section, capital Latin indices $I,J,K$ will run from 1 to 6, whereas lowercase Latin indices $i, j, k$ will run from 1 to 3. We will perform dimensional reduction at the end of the calculation, so in the intermediate results we will explicitly keep the dependence on the number of fields of each species. 
All contractions come with a factor $\frac{2}{\gym^2}$, which we will include at the end when we add all the contributions.

Since we are working in the large-$N$ limit, all propagators will involve only fields in the off-diagonal block. When we write a general propagator $K^{\cdots}$, it will be the one defined in Section~\ref{sec:propagators-colour-flavour}, but with the replacement $\ell_i \to (k_i - 1) / 2$ implicitly understood.

\paragraph{Simple contractions}

All the contractions in this paragraph can be immediately obtained from~\cite{Buhl-Mortensen:2016jqo} by adapting the notation. The ghost contractions are
\begin{align}
 \label{eq:lolli_ghost_1}
 \contraction{V_{\mathrm{G}} \equiv - \tr \Big(}{\bar c}{[ \pcl_I, [\tilde \phi_I,}{c}
 V_{\mathrm{G}} \equiv - \tr \left( \bar c [ \pcl_I, [\tilde \phi_I,c]] \right)
 & = - n_c \frac{2N}{y_3} K^{\mathrm{easy}} \tr \left(\pt_I t_I\right), \\
 \label{eq:lolli_ghost_2}
 \contraction{i (\partial_\mu \tr \big( }{\bar c}{) [ A^\mu, }{c}
 \tr \left( i (\partial_\mu \bar c) [ A^\mu, c]] \right)
 & = 0.
\end{align}
All the contributions from the vertex that couples three gauge fields vanish due to the symmetry of the propagator,
\begin{align}
 \label{eq:lolli_v1}
 % Contraction 1
 \contraction{
 \tr \big( i[A^\mu, }{A^\nu}{] \partial_\mu }{A_\nu \big)}
 % Contraction 2
 \contraction{
    \tr \big( i[A^\mu, A^\nu] \partial_\mu A_\nu \big)
 = \tr \big( i[}{A^\mu}{, A^\nu] \partial_\mu}{A_\nu \big)}
 % Contraction 3
 \contraction{
    \tr \big( i[A^\mu, A^\nu] \partial_\mu A_\nu \big)
 = \tr \big( i[A^\mu, A^\nu] \partial_\mu A_\nu
 = \tr \big( i[}{A^\mu}{, }{A^\nu}
 % Equation
   \tr \big( i[A^\mu, A^\nu] \partial_\mu A_\nu \big)
 = \tr \big( i[A^\mu, A^\nu] \partial_\mu A_\nu \big)
 = \tr \big( i[A^\mu, A^\nu] \partial_\mu A_\nu \big)
 = 0.
\end{align}
Finally, we consider the vertex $\tr \big(i [A^\mu, \pt_I] \partial_\mu \pt_I\big)$. The first two contractions give
\begin{align}
 \label{eq:lolli_v4_1}
 \contraction{\tr \big( i [A^\mu, }{\tilde \phi_I}{] \partial_\mu}{\tilde \phi_I}
 \tr \big( i [A^\mu, \tilde \phi_I] \partial_\mu \tilde \phi_I \big)
 = 0,
\end{align}
and
\begin{align}
 \label{eq:lolli_v4_2}
 \contraction{V_1 \equiv \tr \big( i [}{A^\mu}{,}{\tilde \phi_I}
 V_1 \equiv \tr \big( i [A^\mu, \tilde \phi_I] \partial_\mu \tilde \phi_I \big)
 = + 2 N \left( \partial_3 K^{A,\phi} \right) \tr \left( \pt_I \, t_I \right).
\end{align}
Note that in the last equation we have carried out an integration by parts to move the derivative from the field to the propagator.
This is allowed because the effective vertex will always be contracted with a scalar $\pt_i$ and then integrated, as in~\eqref{eq:one-loop-corr-vevs}.
For the last contraction, note that we can use~(D.21) from~\cite{Buhl-Mortensen:2016jqo}, because as in that case, we have $K^{\phi,A} \propto K^{\nu - 1} - K^{\nu + 1}$. Thus, we find
\begin{align}
 \label{eq:lolli_v4_3}
 \contraction{V_2 \equiv \tr \big( i [}{A^\mu}{,\tilde \phi_I] \partial_\mu}{\tilde \phi_I}
 V_2 \equiv \tr \big( i [A^\mu, \tilde \phi_I] \partial_\mu \tilde \phi_I \big)
 = + N \left( \partial_3 K^{\phi,A} \right) \tr \left( \pt_I t_I \right).
\end{align}

\paragraph{Interaction of three scalars}
We can rewrite the interaction vertex involving three scalars as
\begin{align}
 \label{eq:lolli_v2_rewrite}
 \tr \big( [\pcl_I,\tilde \phi_J][\tilde \phi_I, \tilde \phi_J] \big)
 = \tr \big( \pt_I [ \pt_J, [\pcl_I, \pt_J]] \big).
\end{align}
There are three inequivalent contractions:
\begin{align}
\label{eq:lolli_v2_1}
  \contraction{V_3 \equiv \tr \big( \pt_I [ }{\pt_J}{, [\pcl_I, }{\pt_J}
  V_3 \equiv \tr \big( \pt_I [ \pt_J, [\pcl_I, \pt_J]] \big)
  &= \frac{2N}{y_3} \left(
     n_{\phi,(1)} K^{\phi,(1)}_{\mathrm{sing}}
   - \frac{k_1^2 - 1}{4} K^{\phi,(1)}_{\mathrm{sym}}
  \right) \tr \left( \pt_I t_I \right) + (1 \leftrightarrow 2),\\
\label{eq:lolli_v2_2}
  \contraction{V_4 \equiv \tr \big( }{\pt_I}{[ \pt_J, [\pcl_I, }{\pt_J}
  V_4 \equiv \tr \big( \pt_I [ \pt_J, [\pcl_I, \pt_J]] \big)
  &= \frac{2N}{y_3} \Bigg[ 
   - K^{\phi,(1)}_{\mathrm{sing}} 
    - \frac{n_{\phi,(1)}-1}{2}
      \left( K^{\phi,(1)}_{\mathrm{anti}} + K^{\phi,(1)}_{\mathrm{sym}} \right) \\
  &\qquad + \frac{k_1^2 - 1}{4} K^{\phi,(1)}_{\mathrm{sym}}
    - \frac{k_2^2 - 1}{4} K^{\phi}_{\mathrm{opp}}
  \Bigg] 
  \tr \left( \pt_i^{(1)} t_i^{(1)} \right) + (1 \leftrightarrow 2),
  \nonumber
\intertext{and}
\label{eq:lolli_v2_3}
  \contraction{V_5 \equiv \tr \big(}{\pt_I}{[}{\pt_J}
  V_5 \equiv \tr \big( \pt_I [ \pt_J, [\pcl_I, \pt_J]] \big)
  &= -\frac{N}{y_3} (n_{\phi,(1)} - 1)
    \left( 2 K^{\phi,(1)}_{\mathrm{anti}} + K^{\phi,(1)}_{\mathrm{sym}} \right)
    \tr \left( \pt_i^{(1)} t_i^{(1)} \right) + (1 \leftrightarrow 2).
\end{align}

\paragraph{Interaction of one scalar with two gauge fields}
Next we rewrite the interaction between one scalar and two gauge fields as
\begin{align}
\label{eq:lolli_v3_rewrite}
 \tr \big( [A^\mu, \pcl_I][A_\mu, \pt_I] \big)
 = \tr \big( \pt_I [ A^\mu, [\pcl_I, A_\mu]] \big).
\end{align}
For $\mu = 0, 1, 2 \equiv i$, there is only one possible contraction:
\begin{align}
 \label{eq:lolli_v3_1}
 \contraction{V_6 \equiv \tr \big( \pt_I [}{A^i}{, [\pcl_I, }{A_i}
 V_6 \equiv \tr \big( \pt_I [ A^i, [\pcl_I, A_i]] \big)
 = n_{A,\mathrm{easy}} \frac{2N}{y_3} K^{\mathrm{easy}} \tr \left(\pt_I t_I\right).
\end{align}
In this contraction the chosen regularization procedure becomes relevant, because in ${d = 3 - 2 \epsilon}$ space dimensions $n_{\mathrm{A,easy}} = 3 - 2 \epsilon$.
We are working in dimensional reduction~\cite{Siegel:1979wq,Capper:1979ns} and should therefore add $2 \epsilon$ scalars to the action that behave exactly as the easy components of the gauge field.
Thus, we should also consider the contraction
\begin{align}
 \label{eq:lolli_v3_1.1}
 \contraction{V_7 \equiv \tr \big( \pt_I [}{A^{2 \epsilon}}{, [\pcl_I, }{A_{2 \epsilon}}
 V_7 \equiv \tr \big( \pt_I [ A^{2 \epsilon}, [\pcl_I, A_{2 \epsilon}]] \big)
 = 2 \epsilon \frac{2N}{y_3} K^{\mathrm{easy}} \tr \left(\pt_I t_I\right).
\end{align}
Adding the previous two equations, we find $n_{\mathrm{A,easy}} + 2 \epsilon = 3$ as a prefactor.
Since $n_{\mathrm{A,easy}}$ only appears in this vertex, we can say that in dimensional reduction $n_{\mathrm{A,easy}} = 3$ exactly.

For $\mu = 3$, there are three possible contractions. The first one gives
\begin{align}
 \label{eq:lolli_v3_2}
 \contraction{V_8 \equiv \tr \big( \pt_I [}{A^3}{, [\pcl_I, }{A_3}
 V_8 \equiv \tr \big( \pt_I [ A^3, [\pcl_I, A_3]] \big)
 = \frac{2N}{y_3} K^{A,A} \tr \left( \pt_I t_I \right),
\end{align}
while the other two do not contribute to the effective vertex:
\begin{align}
 \label{eq:lolli_v3_3}
 % Contraction 1
 \contraction{\tr \big(}{\pt_I}{ [A^3, [\pcl_I, }{A_3}
 % Contraction 2
 \contraction{
 \tr \big( \pt_I [ A^3, [\pcl_I, A_3]] \big)
 = \tr \big(}{\pt_I}{[}{A^3}
 % Equation
 \tr \big( \pt_I [ A^3, [\pcl_I, A_3]] \big)
 = \tr \big( \pt_I [ A^3, [\pcl_I, A_3]] \big)
 = 0.
\end{align}

\paragraph{Fermions in the loop}

The action contains three cubic vertices including fermions. The first one is
\begin{align}
 \label{eq:lolli_ferm_1}
 % Contraction
 \contraction{V_{\mathrm{F},1} = \frac{1}{2} \sum_{i = 1}^3 \tr \big(}{\bar \psi_j}{G^i_{jk} [\pt_i, }
             {\psi_k}
 % Equation
 V_{\mathrm{F},1} = \frac{1}{2} \sum_{i = 1}^3 \tr \big( \bar \psi_j G^i_{jk} [\pt_i, \psi_k] \big)
 = N n_\psi \tr \left( t^{(1)}_i \pt^{(1)}_i \right) \tr \tilde{K}_F^{(1)},
\end{align}
the second vertex gives a similar result,
\begin{align}
 \label{eq:lolli_ferm_2}
 % Contraction
 \contraction{V_{\mathrm{F},2} = \frac{1}{2} \sum_{i = 4}^6 \tr \big(}{\bar \psi_j}
             {G^i_{jk} [\pt_i, \gamma_5}{\psi_k}
 % Equation
 V_{\mathrm{F},2} = \frac{1}{2} \sum_{i = 4}^6 \tr \big( \bar \psi_j G^i_{jk} [\pt_i, \gamma_5 \psi_k] \big)
 = - N n_\psi \tr \left( t^{(2)}_i \pt^{(2)}_i \right) 
   \tr \left(i \gamma_5 \tilde{K}_F^{(2)} \right),
\end{align}
and the last contraction vanishes,
\begin{align}
 \label{eq:lolli_ferm_3}
 % Contraction
 \contraction{\frac{1}{2} \tr \big(}{\bar \psi_j}{ \gamma_\mu [A^\mu, }{\psi_k}
 % Equation
 \frac{1}{2} \tr \big( \bar \psi_j \gamma^\mu [A_\mu, \psi_j] \big) = 0.
\end{align}
It is important to remember that when the fermion propagators are being regulated one has to use~\eqref{eq:trace-chiral-rotation} and~\eqref{eq:fermion-prop-regularized}.
The combinations of propagators~\eqref{eq:lolli_ferm_1} and~\eqref{eq:lolli_ferm_2} are
\begin{equation}
 \begin{aligned}
 \tilde{K}_F^{(1)}
 &= \frac{1}{(2 \ell_1 + 1) (2 \ell_2 + 1)} \left[
 (\ell_2 + 1) \Big(\tilde{K}_F^{-\ell_1, \ell_2} - \tilde{K}_F^{\ell_1 + 1, \ell_2}\Big) +
 \ell_2 \Big(\tilde{K}_F^{-\ell_1, -\ell_2 - 1} - \tilde{K}_F^{\ell_1 + 1, - \ell_2 - 1} \Big)
 \right]\!,
 \\
 \tilde{K}_F^{(2)}
 &= \frac{1}{(2 \ell_1 + 1) (2 \ell_2 + 1)} \left[
 (\ell_1 + 1) \Big(\tilde{K}_F^{-\ell_1, \ell_2} - \tilde{K}_F^{-\ell_1, -\ell_2 - 1}\Big) +
 \ell_1 \Big(\tilde{K}_F^{\ell_1 + 1, \ell_2} - \tilde{K}_F^{\ell_1 + 1, - \ell_2 - 1} \Big)
 \right]\!,
\end{aligned}
\end{equation}
and the replacement~\eqref{eq:replacement-off-diagonal-block-so3-so3} is understood.

\paragraph{Summing up all vertices}
\label{sec:sum-all-vertices}

The full effective vertex is the sum of all the contractions calculated in the previous subsection.
We also have to remember to restore the overall prefactor of $\tfrac{2}{\gym^2}$ of the action, i.e.\
\begin{align}
 V_{\mathrm{eff}} = \frac{2}{\gym^2} \left(V_{\mathrm{G}} + V_1 + V_2 + V_3 + V_4 + V_5 + V_6 + V_7 + V_8 + V_{\mathrm{F},1} + V_{\mathrm{F},2} \right).
\end{align}
Inserting the expressions from the previous paragraphs, we see that the vertex contains a part that depends on the regulator terms $f_\epsilon(y) = -\tfrac{1}{\epsilon} - \log(4 \pi) + \gamma_{\mathrm{E}} - 2 \log(y_3) - 1$ and a part that is finite as $\epsilon \rightarrow 0$,
\begin{align}
 V_{\mathrm{eff}} = V_{\mathrm{eff},\epsilon} + V_{\mathrm{eff,fin}}.
\end{align}
The $\epsilon$-dependent part is
\begin{align}
 \label{eq:effective-vertex-divergent}
\begin{split}
 V_{\mathrm{eff},\epsilon}(y; k_1, k_2)
 &=
 \frac{- N}{32 \pi^2 y_3^3} f_\epsilon(y) \Big[
  (k_1^2 + k_2^2) (n_{\mathrm{c}} + 2 n_{\psi} - n_{\phi,{(1)}} - n_{\phi,{(2)}} - n_{\mathrm{A,easy}}) \\
 & \quad
 - 2 (n_{\mathrm{c}} + 2 n_{\psi} + 5 n_{\phi,{(1)}} - n_{\phi,{(2)}} - n_{\mathrm{A,easy}} - 18)
 \Big] \tr\left(\pt_i^{(1)} t_i^{(1)}\right) + (1 \leftrightarrow 2).
\end{split}
\end{align}
This is zero for $n_{\mathrm{A,easy}} = 3, n_{\mathrm{c}} = 1, n_{\psi} = 4$ and $n_{\phi,{(1)}} = n_{\phi,{(2)}} = 3$.
Note that here we are using that we can keep $n_{\mathrm{A,easy}} = 3$ in dimensional reduction, cf.\ the discussion after~\eqref{eq:lolli_v3_1}.
The finite part is
\begin{align}
 \label{eq:effective-vertex-finite}
 \begin{split}
 V_{\mathrm{eff,fin}}(y; k_1, k_2)
 &=
 \frac{-N}{2 \pi^2 y_3^3} \left(
 W^{(1)}(k_1, k_2) \tr\left(\pt_i^{(1)} t_i^{(1)}\right) + W^{(2)}(k_1, k_2) \tr\left(\pt_i^{(2)} t_i^{(2)}\right) \right),
 \end{split}
\end{align}
where the functions $W^{(1)}(k_1, k_2)$ and $W^{(2)}(k_1, k_2)$ are given in~\eqref{eq:effective-vertex-W1} and~\eqref{eq:effective-vertex-W2} in the main text.
This result is exact, i.e.\ we have not expanded for large $k_1$ and $k_2$.

\subsection{Contraction of the stick}
\label{sec:stick-lollipop}

Now we proceed to contract the external field with the effective vertex. 
The traces $\tr( \pt_j^{(b)} t_j^{(b)} )$ with $b = 1, 2$ coming from the effective vertex will be contracted with an external field $\pt^{(a)}_i$. 
For simplicity, let us consider the case where we are contracting fields from the first sector, i.e.\ the case $a = b = 1$.
Notice that $t_j^{(1)}$ is a matrix in the $k_1 k_2 \times k_1 k_2$ block padded with zeros. 
Thus, when we multiply it with $\pt_j^{(1)}$ only the $k_1 k_2 \times k_1 k_2$ block survives when taking the trace. 
Expanding $\pt_i^{(1)}$ and $\pt_j^{(1)}$ in this block in terms of fuzzy spherical harmonics and their Hermitian conjugates, we obtain
\begin{equation}
  \label{eq:contraction-vertex-trace}
\begin{aligned}
 \contraction{}{\, \pt_i^{(1)}}{\tr }{\pt_j^{(1)}}
 \pt_i^{(1)} \, \tr( \pt_j^{(1)} t_j^{(1)} )
 &= 
 \left\langle 
   (\pt_i^{(1)})     _{\ell_1 , m_1 ; \ell_2 , m_2 } 
   (\pt_j^{(1)})^\dag_{\ell_1', m_1'; \ell_2', m_2'} 
 \right\rangle\\&\qquad\qquad
 \hat{Y}^{m_1}_{\ell_1} \otimes \hat{Y}^{m_2}_{\ell_2} \,
 \tr \left[ 
   \left( \hat{Y}^{m_1'}_{\ell_1'} \otimes \hat{Y}^{m_2'}_{\ell_2'} \right)^\dagger 
   \left( t_j^{k_1} \otimes \mathds{1}_{k_2} \right)
 \right].
\end{aligned}
\end{equation}
In the previous expression, the trace can be simplified further. 
We start by expanding the matrices $t_i$ and $\mathds{1}$ in terms of the fuzzy spherical harmonics $\hat{Y}^{m}_{\ell}$ as
\begin{align}
 t_i^{k_1} = \sum_{m_1 = \pm 1, 0} (c_i)_{m_1} \hat{Y}^{m_1}_{\ell_1 = 1}, \quad
 \mathds{1}_{k_2} = (-1)^{k_2+1} \sqrt{k_2} \, \hat{Y}^{m_2 = 0}_{\ell_2 = 0}.
\end{align}
The explicit coefficients $(c_i)_{m_1}$ can be obtained from~\eqref{eq:relation-t-Ylm-1} and~\eqref{eq:relation-t-Ylm-2} in Appendix~\ref{sec:conventions-su2-and-Ylm}.
Using that the $\hat{Y}_\ell^m$ are traceless for $\ell > 0$ and proportional to the identity for $\ell = 0$, we obtain
\begin{align}
 \begin{aligned}
 \tr 
  \left[ 
    \left( \hat{Y}^{m_1'}_{\ell_1'} \otimes \hat{Y}^{m_2'}_{\ell_2'} \right)^\dag 
    (t_j^{k_1} \otimes \mathds{1}_{k_2}) 
  \right]
 & =
  \tr \left[ \left( \hat{Y}^{m_1'}_{\ell_1'} \right)^\dag t_j^{k_1} \right]
  \tr \left[ \left( \hat{Y}^{m_2'}_{\ell_2'} \right)^\dag \right] \\
 & =
 (-1)^{k_2 + 1}
 \sqrt{k_2} \;
 \delta_{\ell_1',1} \;
 \delta_{\ell_2',0} \; 
 \delta_{m_2',0} \;
 (c_j)_{m_1'}.
 \end{aligned}
\end{align}
Inserting this into~\eqref{eq:contraction-vertex-trace}, we find that the propagator between the scalars has to be evaluated for $\ell_2 = \ell_2' = m_2 = m_2' = 0$ and $\ell_1 = \ell_1' = 1$. The explicit form of the propagator is
\begin{align}
 \begin{aligned}
 \langle (\pt_i)_{1, m_1} (\pt_j)^\dag_{1, m_1'} \rangle 
 & = 
  \delta_{ij} \delta_{m_1, m_1'} 
  \left(
    \tfrac{2}{3} K^{m^2=0} +
    \tfrac{1}{3} K^{m^2=6} 
  \right) \\
 &\phaneq - \tfrac{i}{3} \epsilon_{ijk} [t^{k_1 = 3}_k]_{2 - m_1, 2 - m_1'} 
  \left( K^{m^2 = 0} - K^{m^2 = 6} \right).
 \end{aligned}
\end{align}
Combining this propagator with the explicit form of $c_j^m$ and the $3 \times 3$ matrices $t^{k_1 = 3}_{i}$, we obtain
\begin{align}
 \label{eq:contractions-Veff-nonzero}
 \contraction{}{\pt_i}{{}^{(1)}\tr }{\pt}
 \pt_i^{(1)} \tr( \pt^{(1)}_j t^{(1)}_j )
 =
 (t_i^{k_1} \otimes \mathds{1}_{k_2}) K^{m^2 = 6},
 \quad
 \contraction{}{\pt_i}{{}^{(2)}\tr }{\pt}
 \pt_i^{(2)} \tr( \pt^{(2)}_j t^{(2)}_j )
 =
 (\mathds{1}_{k_1} \otimes t_i^{k_2}) K^{m^2 = 6}.
\end{align}
The contractions where the external field and the one inside the trace are from different sectors vanish,
\begin{align}
 \contraction{}{\pt_i}{{}^{(1)}\tr }{\pt}
 \pt_i^{(1)} \tr( \pt^{(2)}_j t^{(2)}_j )
 =
 \contraction{}{\pt_i}{{}^{(2)}\tr }{\pt}
 \pt_i^{(2)} \tr( \pt^{(1)}_j t^{(1)}_j )
 = 0.
\end{align}
The contraction of the easy components $A_0$, $A_1$ and $A_2$ of the gauge field with the vertex vanishes because the propagator between them and the scalars is zero.
Furthermore, we find
\begin{align}
 \contraction{}{\tilde A}{{}_3\tr }{\pt_j}
 A_3 \tr( \pt^{(1)}_j t^{(1)}_j )
 =
 \contraction{}{\tilde A}{{}_3\tr }{\pt_j}
 A_3 \tr( \pt^{(2)}_j t^{(2)}_j )
 = 0.
\end{align}
This shows that only the vevs of the scalars receive one-loop corrections.

\subsection{Spacetime integral}
\label{sec:spacetime-integral}

In order to evaluate the correction to the scalar vevs~\eqref{eq:one-loop-corr-vevs}, we are only missing the calculation of the integral over $y$. The propagator in the integral has mass $m^2 = 6$, or equivalently $\nu = \tfrac{5}{2}$, and it can be expressed in terms of elementary functions,
\begin{align}
\begin{split}
 K^{\nu = \tfrac{5}{2}}(x, y)
 &= \frac{\gym^2}{2} \frac{\xi(x, y)^4}{10 \pi^2 x_3 y_3} \; 
    {}_2F_1 \left(2, \tfrac{5}{2}; \tfrac{7}{2}; \xi(x, y)^2 \right) \\
 &=\frac{\gym^2}{2}  \frac{1}{4 \pi^2 x_3 y_3} \left(
 \frac{2 \xi^2 - 3}{\xi^2 - 1} - \frac{3 \, \arctanh(\xi)}{\xi}
 \right).
\end{split}
\end{align}
In the second equality, we have dropped the explicit dependence of $\xi$ on $x$ and $y$ to simplify the notation. In the integral, this propagator will be multiplied by a factor of $1 / (y_3)^3$ that comes from the effective vertex.
Thus, the integral is
\begin{align}
\label{eq:Veff-spacetime-integral}
\begin{split}
 \int \diff^4 y \frac{1}{y_3^3} K^{\nu = \tfrac{5}{2}}(x, y)
 &= \frac{\gym^2}{2} \frac{1}{4 \pi^2}
 \int_{0}^{\infty} \diff y_3 \int_0^{\infty} \diff r \int \diff \Omega \;
 \frac{r^2}{x_3 y_3^4} \left( \frac{2 \xi^2 - 3}{\xi^2 - 1} - \frac{3 \, \arctanh(\xi)}{\xi} \right) \\
 &= \frac{\gym^2}{2} \frac{1}{5} \int_0^{\infty} \diff y_3
 \left.\begin{cases}
  (x_3)^{-2} & \text{for }0 \leq y_3 < x_3 \\
  (x_3)^3 (y_3)^{-5} & \text{for }0 \leq x_3 < y_3
 \end{cases}\right\}
 =\frac{\gym^2}{2} \frac{1}{4 x_3},
\end{split}
\end{align}
where we have used spherical coordinates defined by $r^2 = (x_0 - y_0)^2 + (x_1 - y_1)^2 + (x_2 - y_2)^2$ and we are working in Euclidean signature as anticipated when we discussed the spacetime part of the scalar propagator.

One can combine the effective vertex~\eqref{eq:effective-vertex-finite}, the contractions~\eqref{eq:contractions-Veff-nonzero} and the spacetime integral~\eqref{eq:Veff-spacetime-integral} to obtain the correction to the vevs given in \eqref{eq:correction-vevs} of the main text.

\section{Color traces}
\label{sec:color-traces-so3-so3}

For the calculation of $\langle \tr\, Z^L \rangle$ to one-loop order in Section~\ref{sec:one-loop-trZL}, we need expressions for the color traces.
More precisely, we need to calculate traces where $(Z^{\mathrm{cl}})^L$ is multiplied with a number of \su{2} generators $t_i$ from each sector.

It was shown in~\cite{Buhl-Mortensen:2016jqo} that
\begin{align}
 \displaystyle \tr\Big[(t_3^{k})^L\Big]
  = (-1)^{L+1} \frac{2}{L+1} B_{L+1}\left( \tfrac{1-k}{2} \right)
  = \frac{k^{L+1}}{2^L (L + 1)} + \mathcal O (k^L), &
\end{align}
for  $L$ even while 
$\tr\Big[(t_3^{k})^L\Big] = 0 $ for $L$ odd.
Here $B_{L+1}(k)$ denotes the Bernoulli polynomial of degree $L+1$.
In this paper, the most general trace that we will evaluate is
\begin{align}
 \tr \left[ 
   (Z^{\mathrm{cl}})^L 
   \left( t_3^{k_1} \right)^{n_1} 
   \otimes
   \left( t_3^{k_2} \right)^{n_2} 
 \right]
 = \frac{(-1)^L}{x_3^L} 
 \sum_{n = 0}^L \binom{L}{n} i^{L-n} 
 \tr \left[ \left( t_3^{k_1} \right)^{n + n_1} \right]
 \tr \left[ \left( t_3^{k_2} \right)^{L + n_2 - n} \right] .
\end{align}
A particular term in this sum will not vanish if $n + n_1$ is even and  $L + n_2 - n$ is even. In order for the entire sum not to vanish we need that $n_1$ and $L + n_2$ have the same parity, or equivalently, we need that $n_1$ and $L + n_2$ are both even or both odd. In either case, only half of the terms in the sum will contribute to the result.

When $n_1$ is even and $L + n_2$ is even, only the terms with $n$ even contribute. Thus, we must sum over a new variable $m$ such that $n = 2m$ and $m = 0, \ldots, \lfloor \frac{L}{2} \rfloor$. If we expand for large $k_1$ and $k_2$, we obtain
\begin{align}
  \frac{(-1)^L}{2^{L + n_1 + n_2} x_3^L} 
  \sum_{m = 0}^{\lfloor \frac{L}{2} \rfloor} \binom{L}{2m} i^{L-2m}
  \frac{k_1^{2m + n_1 + 1}}{(2m + n_1 + 1)}
  \frac{k_2^{L + n_2 - 2m + 1}}{(L + n_2 - 2m + 1)}
  + \mathcal O(k^{L + n_1 + n_2 + 1}).
\end{align}
Here $\mathcal{O}(k^\ell)$ stands for terms where the combined powers of $k_1$ and $k_2$ are less than or equal to $\ell$.

When $n_1$ is odd and $L + n_2$ is odd, only the terms with $n$ odd contribute. Thus, we must sum over a new variable $m$ such that $n = 2m + 1$ and $m = 0, \ldots, \lfloor \frac{L-1}{2} \rfloor$. If we expand for large $k_1$ and $k_2$, we obtain
\begin{align}
  \frac{(-1)^L}{2^{L + n_1 + n_2} x_3^L} 
  \sum_{m = 0}^{\lfloor \frac{L-1}{2} \rfloor} \binom{L}{2m + 1} i^{L-2m-1}
  \frac{k_1^{2m + n_1 + 2}}{(2m + n_1 + 2)}
  \frac{k_2^{L + n_2 - 2m}}{(L + n_2 - 2m)}
  + \mathcal O(k^{L + n_1 + n_2 + 2}).
\end{align}

The above sums can be carried out explicitly for particular values of $n_1$ and $n_2$. In all cases of interest for us, the traces will vanish for $L$ odd, so we will assume that $L$ is even in the rest of this section. It will also be convenient to express the results in terms of the angle 
\begin{align}
 \label{eq:def-psi0}
 \psi_0 \equiv \arctan \left( \frac{k_1}{k_2} \right).
\end{align}

In the following results, the symbol $\simeq$ means that the right-hand side only contains the leading-order term in $k_1$ and $k_2$. The trace for $n_1 = 0$ and $n_2 = 0$ is
\begin{align}
 \tr \left[ 
   \left( Z^{\mathrm{cl}} \right)^L
 \right]
 \simeq 
   \frac{(-i)^L (k_1^2 + k_2^2 )^{\frac{L}{2}+1} 
         \sin \left[ (L+2) \psi_0 \right]}
        {2^{L} x_3^L (L + 1)(L + 2)}.
\end{align}
When $(n_1, n_2) = (1, 0)$ or $(n_1, n_2) = (0, 1)$, we find
\begin{align}
 \label{eq:color-traces-one-t}
 \begin{aligned}
  \tr \Big[ \left( Z^{\mathrm{cl}} \right)^{L-1} t_3^{k_1}\otimes \mathds{1}_{k_2} \Big]
  \simeq \frac{(-i)^L ( k_1^2 + k_2^2 )^{\frac{L}{2}}}
  {2^L x_3^{L-1} L (L+1) (L+2)}
  \Big[
&   - k_1 k_2 L \cos \left( L \psi_0 \right) \\
&   + \left[k_2^2 + k_1^2 (L+1) \right] \sin \left( L \psi_0 \right)
  \Big], \\
  \tr \Big[ \left( Z^{\mathrm{cl}} \right)^{L-1} \mathds{1}_{k_1} \otimes t_3^{k_2} \Big]
  \simeq \frac{(-i)^{L-1} ( k_1^2 + k_2^2 )^{\frac{L}{2}}}
  {2^L x_3^{L-1} L (L+1) (L+2)}
  \Big[ 
&   + k_1 k_2 L \cos \left( L \psi_0 \right) \\
&   + \left[k_1^2 + k_2^2 (L+1) \right]
    \sin \left( L \psi_0 \right)
  \Big].
 \end{aligned}
\end{align}
For the case $n_1 = n_2 = 1$, the trace gives
\begin{align}
 \label{eq:color-traces-t-t}
 \begin{aligned}
  \tr \Big[ 
     \left( Z^{\mathrm{cl}} \right)^{L-2} t_3^{k_1} \otimes t_3^{k_2} 
  \Big]
  &\simeq \frac{(-i)^{L+1} ( k_1^2 + k_2^2 )^{\frac{L}{2}}}
  {2^L x_3^{L-2} L (L+2) (L-1)}
  \Big[
   + k_1 k_2 L \cos \left( L \psi_0 \right) \\
   & \hspace{3cm}
   + (k_1 - k_2)(k_1 + k_2) \sin \left( L \psi_0 \right)
  \Big].
 \end{aligned}
\end{align}
Finally, for the cases $(n_1, n_2) = (2, 0)$ and $(n_1, n_2) = (0, 2)$ the traces evaluate to
\begin{align}
\label{eq:color-traces-two-t}
\begin{aligned}
  \tr \Big[ 
    \left( Z^{\mathrm{cl}} \right)^{L-2} 
    \left( t_3^{k_1} \right)^{2}\otimes \mathds{1}_{k_2}
  \Big] 
& \simeq
 -\frac{(-i)^L ( k_1^2 + k_2^2 )^{\frac{L}{2}}}
       {2^L x_3^{L-2} (L-1) L (L+1) (L+2)}
  \Big[
   + 2 k_1 k_2 L \cos \left( L \psi_0 \right) \\
&  \hspace{3cm}
   + \left( -2 k_2^2 + k_1^2 L(L+1) \right) 
   \sin \left( L \psi_0 \right)
  \Big], \\
  \tr \Big[ 
   \left( Z^{\mathrm{cl}} \right)^{L-2} 
  \mathds{1}_{k_1} \otimes \left( t_3^{k_2} \right)^{2}
  \Big] 
& \simeq
  \frac{(-i)^L ( k_1^2 + k_2^2 )^{\frac{L}{2}}}
       {2^L x_3^{L-2} (L-1) L (L+1) (L+2)}
  \Big[
   + 2 k_1 k_2 L \cos \left( L \psi_0 \right) \\
&  \hspace{3cm}
   + \left( 2 k_1^2 - k_2^2 L(L+1) \right) 
   \sin \left( L \psi_0 \right)
  \Big].
\end{aligned}
\end{align}

% \newpage
\bibliographystyle{utphys2}
% \enlargethispage{1\baselineskip}
\bibliography{references}

\end{document}